\documentclass[twocolumn,aps,prx,superscriptaddress]{revtex4-2}

\usepackage{amsmath}
\usepackage{graphicx}
\usepackage{MnSymbol}
\usepackage{upgreek}

\begin{document}

\title{Granite sliding on granite: friction, wear rates, surface topography, and the scale-dependence of rate-state effects}

\author{Sergey V. Sukhomlinov}
\affiliation{
 Dept. of Materials Science and Engineering,
  Saarland University, 66123 Saarbrücken, Germany
}

\author{Martin H. Müser}
\affiliation{
 Dept. of Materials Science and Engineering,
  Saarland University, 66123 Saarbrücken, Germany
}

\author{B.N.J. Persson}
\email{B.Persson@fz-juelich.de}
\affiliation{State Key Laboratory of Solid Lubrication, Lanzhou Institute of Chemical Physics, Chinese Academy of Sciences, 730000 Lanzhou, China}
\affiliation{Peter Grünberg Institute (PGI-1), Forschungszentrum Jülich, 52425 Jülich, Germany}
\affiliation{MultiscaleConsulting, Wolfshovener Str. 2, 52428 Jülich, Germany}

\begin{abstract}
We study tribological granite–granite contacts as a model for tectonic faulting, combining experiments, theory, and molecular dynamics simulations.
The high friction in this system is not dominated by particulate wear or plowing, as frequently assumed, but by cold welding within plastically deformed asperity junctions.
We base this conclusion on the observation that wear is repeatedly high after cleaning contacts but decreases as gouge accumulates, while friction shows the opposite trend.
Moreover, adding water reduces wear by a factor of ten but barely decreases friction. 
Thermal and rate-dependent effects—central to most earthquake models—are negligible: friction remains unchanged between 
$-40^\circ$C and $20^\circ$C, across abrupt velocity steps, and after hours of stationary contact.
The absence of rate-state effects in our macroscopic samples is rationalized by the scale-dependence of pre-slip.
The evolution of surface topography shows that quartz grains become locally smooth, with height spectra isotropic for wavelength below 
10 microns but anisotropic at longer wavelengths, similar to natural faults.
The resulting gouge particles have the usual characteristic sizes near 100~nm.
Molecular dynamics simulations of a rigid, amorphous silica tip sliding on $\alpha$-quartz reproduce not only similar friction coefficients near unity but also other experimentally observed features, including stress-introduced transitions to phases observed in post-mortem faults, as well as theoretical estimates of local flash temperatures.
Additionally, they reveal a marked decrease of interfacial shear strength above 600$^\circ$C. 
\end{abstract}

\maketitle

\section{Introduction}

Friction between rock surfaces controls the stability of faults and thereby the dynamics of earthquakes and slow-slip events~\cite{Byerlee1978PAG,Scholz1998N,Faulkner2010JSG}.
Despite decades of laboratory and field studies, the dominant microscopic origin of rock friction remains uncertain:
whether resistance to sliding is governed primarily by plastic deformation~\cite{Goldsby2002GRL,Beeler2008JGR}, adhesion~\cite{Li2011N,Li2014TL}, or particulate wear at asperity junctions~\cite{Aghababaei2022MRSB} is still debated.
Understanding these mechanisms is essential for linking phenomenological descriptions---such as rate-and-state friction laws~\cite{Dieterich1979JGRSE,Baumberger2006AP}---to the atomic and microscale physics that ultimately control fault strength.

Granite, a dominant constituent of the continental crust, provides a natural model system for studying these processes under controlled laboratory conditions.
Previous experiments on granite have mostly been performed at very high nominal pressures, typically of order $100~{\rm MPa}$~\cite{Ishibashi2018WRR,Aharonov2019JGRSE}, conditions representative of deep crustal faults. 
Such studies have yielded valuable constraints on macroscopic friction laws but leave open how interfacial bonding, wear, and plasticity at the asperity scale act together across the scales to produce friction.

At the same time, a growing body of work has revealed that the physics of friction and wear at mineral contacts~\cite{Li2011N,Goldsby2002GRL}, or more generally, tetrahedrally coordinated solids~\cite{Pastewka2010NM,Li2014TL,Moras2018PRM,Atila2025PRL}, differs markedly from that of metallic or polymeric systems.
Silicate interfaces, in particular, can sustain large normal stresses~\cite{Whitney2007AM,Li2014TL,Lei2022G} and develop interfacial bonds that persist over geological timescales.
Under shear, these contacts undergo plastic flow, local amorphization~\cite{Sharma1996PMS}, and nanometer-scale wear, producing debris that evolves into the gouge layers ubiquitous along mature faults~\cite{Rice2006JGR}.
Yet it remains unclear whether the high friction observed in rocks arises mainly from the plowing and abrasion of these particles, or from the intrinsic strength of plastically deformed, cold-welded asperity junctions.
Addressing this question requires experiments, including simulations, that isolate the elementary mechanisms under controlled conditions and theories that link the microscopic and macroscopic observations.

Linking microscopic and macroscopic observations also allows one to address the question of to what degree the 
large discrepancy between the static and kinetic friction coefficients observed at small scales can be reduced at larger scales.
To understand this, first note that contact between two solids objects usually strengthen during stationary contact, e.g., due to an
increase in the contact area (due to slow thermally activated processes, creep), or due to formation of strong bonds across the interface, e.g., Si-O-Si bonds for silica in contact with silica, which is also a thermally activated process.
For elastic solids, if the elastic modulus is small or the solids have large nominal contact area, when an external shear force is applied the asperity contact regions may break at different times and if this pre-slip result in breaking of all the contact regions before full slip occur, then the effective static friction coefficient equal the kinetic friction coefficient.
Stated differently, if all the contact have been broken (and renewed)  when the applied force approach the kinetic friction force $F_{\rm k} = \mu_{\rm k} F_{\rm N}$, then the static or breakloose friction force equal the kinetic friction force.
(Note: We prefer the notation ``breakloose friction force''
rather than the ``static friction force'' as many scientists associate with it a well defined quantity, while in fact it depends on how the external forces are applied and on the sliding history. 
Thus, static friction coefficients are often tabulated without stating how they were obtained.)

It must be kept in mind that the roughness of fault lines spans from nanometers, the scale of atomic-force microscopy experiments, to kilometers~\cite{Brodsky2011EPSL}.
Hence, for large enough length scale, the breakloose friction force will equal the kinetic friction force. 
This pre-slip effect has been studied  
for a 1D model \cite{Lorenz2012JPCM}, which can be considered as an extension of the Burridge-Knopoff model~\cite{Burridge1967BSSA}, 
where the driving force is applied on one of the elastic block.
However, pre-slip will also occur in more realistic situations~\cite{Liang2025TL}, and is supported by several 
experiments where $\Delta \mu = \mu_{\rm s} - \mu_{\rm k} \approx 0$ in spite of the
fact that a strong strengthening of the contact occurs during the time of stationary contact.
Thus it has been observed in rubber friction experiments that $\Delta \mu  \approx 0$, although the true contact area continues to grow over time due to viscoelastic relaxation~\cite{Tuononen2016SR,Tada2025JCP,Persson2025JCP_a}.
In the context of earthquakes, McClimon et al.~\cite{McClimon2024TL} have shown that for nanoscale silica-silica (amorphous ${\rm SiO_2}$) contacts, $\Delta \mu/\mu_{\rm k}$ can be very large, on the order of $\approx 4$.
However, in macroscopic contact experiments on granite, which consists mainly of quartz (crystalline ${\rm SiO_2}$), $\Delta \mu/\mu_{\rm k}$ is very small, typically $0.01$–$0.03$, and in the study presented below we observe no difference between the breakloose friction force and the kinetic friction force.
In another study Li et al\cite{Li2020PRL} observed that for silica-silica contacts $\Delta \mu/\mu_{\rm k} \approx 3$ at the nanoscale and $\Delta \mu/\mu_{\rm k} \approx 0.03$ at the millimeter length scale. In a related hard-matter contact, the ratio 
$\mu_{\textrm{s}}/\mu_{\textrm{k}}$ was close to two at loads small enough to have contact merely in a single asperity, 
but approached unity with increasing loads~\cite{Peng2025PRL}. This load-dependency is also predicted by the model in Ref.~\cite{Lorenz2012JPCM}.
We do note, however, that the study in~\cite{Ji2022EA} shows that the friction dynamics at meter-scale is nearly the same as at cm-scale~\cite{Ji2022EA}. In this study a thick
gauge separated the surfaces and the slip was localized to micrometer-sized shear bands in the gauge.

In this work, we aim to present a meaningful link between microscopic and macroscopic friction and wear of rocks.
To this end, we conduct experiments and molecular dynamics simulations on silicate materials.
The results are connected and interpreted using advanced contact-mechanics theories~\cite{Persson2001JCP,Persson2001PRL,Almqvist2011JMPS} for randomly rough surfaces, which have successfully passed systematic tests~\cite{Dapp2014JPCM,Muser2017TL} by numerically rigorous methods—unlike the still-popular bearing-area models, which fail such tests because they ignore long-range elastic deformation~\cite{Afferrante2018TL,Persson2022MRS}.
The experiments are conducted using granite as the primary model system for rock~\cite{Marone1990JGRSE,Scholz1998N,Kanamori2001PT,Kanamori2004RPP,Scholz2018Book}.
They allow us to correlate friction and wear and to scrutinize the importance of the produced gouge, by repeatedly removing it from the contact.
The microscopic interpretations of the experiments are assessed in molecular dynamics simulations, which enable the measurement of local stresses as a function of temperature and sliding velocity with nanometer-scale resolution.
Here, we focus on contacts between a hard, adhesive indenter and quartz, the load-bearing constituent of granite.
Finally, the results are interpreted using Persson’s contact-mechanics theory as well as a continuum variant~\cite{Lorenz2012JPCM} of the Burridge–Knopoff model~\cite{Burridge1967BSSA}.

\section{Methods}

\subsection{Experimental methods}

\begin{figure}[htbp]
\includegraphics[width=0.45\textwidth,angle=0]{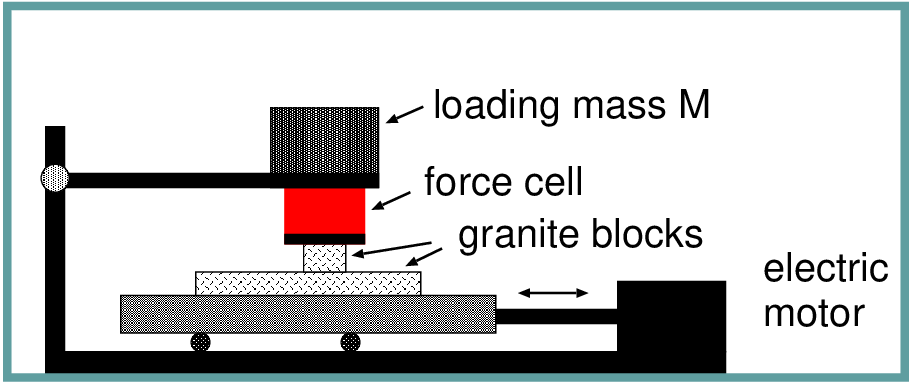}
\caption{
Low-temperature linear friction slider used for the experiments. 
The instrument allows precise control of temperature ($-40^\circ{\rm C}$ to $+20^\circ{\rm C}$), sliding speed ($1~{\rm \upmu m/s}$ to $1~{\rm cm/s}$), and normal load ($250$–$1000~{\rm N}$). 
}
\label{SetUp.eps}
\end{figure}

We have performed friction measurements using our low-temperature linear friction slider shown in Fig.~\ref{SetUp.eps}. 
With the present set-up we can vary the temperature from $T=-40^\circ{\rm C}$ to $+20^\circ{\rm C}$ and the sliding speed from $v=1~{\rm \upmu m/s}$ to $1~{\rm cm/s}$. 
The normal load can be changed from $\approx 250~{\rm N}$ to $1000~{\rm N}$ by adding calibrated masses on top of the force sensor.
Most experiments were performed in the normal atmosphere during a $\sim 1$ month time period. The air humidity was not controlled
and may have differed on different days.
The experiments reported on in Fig. \ref{1x.2FxoverFz.last.eps} and \ref{1time.2Temp.eps} were performed in a closed chamber 
containing humidity absorbing salt where the relative humidity was $\approx 0.2$.

The force sensor has a resolution (sensitivity) of about $1 \ {\rm N}$. The friction results presented below are very reproducible.
This can be seen in, e.g. Fig. 8 where the lines for the 100-runs consist of 100 data points, where
each data point is the average of the friction force during once sliding cycle ($40 \ {\rm cm}$ sliding distance). 
Note that the fluctuations (noise) in the lines are extremely small.

A granite block with the square cross-section area $A_0=4~{\rm cm}\times4~{\rm cm}$ is glued to an aluminum plate, which gets attached to the force cell (see Fig.~\ref{SetUp.eps}). 
The granite specimen can move vertically with the carriage to adapt to the substrate profile. 
The substrate is also a granite block approximately $10~{\rm cm}$ wide and $30~{\rm cm}$ long. 
Both the sliding block and the substrate are $3~{\rm cm}$ thick. 
The substrate sample is attached to the machine table, which is moved by a servo drive via a gearbox in a translational manner. 

We control the relative velocity between the granite specimen and the substrate sample, while the force cell acquires the normal and friction forces. 
During each run the substrate is displaced $20~{\rm cm}$ forward and $20~{\rm cm}$ backward to return to the initial position. 

The friction slider collects forces at a frequency of $10~{\rm Hz}$. 
Thus, as we increase the sliding speed from $0$ to $1~{\rm \upmu m/s}$, the first friction data are obtained at a sliding distance of $0.1~{\rm \upmu m}$ at that sliding velocity. 
The theory in Sec.~\ref{sec:contact_mechanics} finds in agreement with established~\cite{Dieterich1994PAG} and more recent~\cite{Hayward2019JGR} experimental work that the diameter of typical meso-scale contact regions is of order $10~{\rm \upmu m}$; hence, the static-friction peak occurring within sliding distances of that magnitude can be resolved up to sliding velocities of $v = 100~\upmu$m/s.

The wear rate was determined from the mass $m$ of the granite wear particles produced after a given sliding period. 
Our experiments show that the wear mass $m$ is proportional to the normal force $F_{\rm N}$ and to the sliding distance
$L$ when sliding on clean surfaces (no wear particles), and we will give the wear rate as $k_{\rm w} = m/F_{\rm N} L$.
This is consistent with the Archard wear equation~\cite{Archard1953JAP}. 
The wear particles were transferred by a soft brush from the granite surface to a paper surface of known mass, and the total mass was measured using a Mettler Toledo analytical balance (model MS104TS/00) with a sensitivity of $0.1~{\rm mg}$. 
Two main protocols were used: a 1-run, where particles were removed after each $0.4~{\rm m}$ sliding cycle, and a 100-run, where particles were removed after 100 cycles ($40~{\rm m}$ total). 
However, other $n$-runs were also conducted. 

The surface roughness of the granite samples was measured using a Mitutoyo Surftest SJ-410 profilometer equipped with a diamond tip of radius $R=1~{\rm \upmu m}$ and a tip–substrate force $F_{\rm N}=0.75~{\rm mN}$. 
Measurements were taken with a step length of $0.5~{\rm \upmu m}$, scan length $L=25~{\rm mm}$, and tip speed $v=50~{\rm \upmu m/s}$. 
The size and morphology of the wear particles were studied using a JEOL Scanning Electron Microscope (SEM). 

\begin{figure}[htbp]
\includegraphics[width=0.45\textwidth,angle=0]{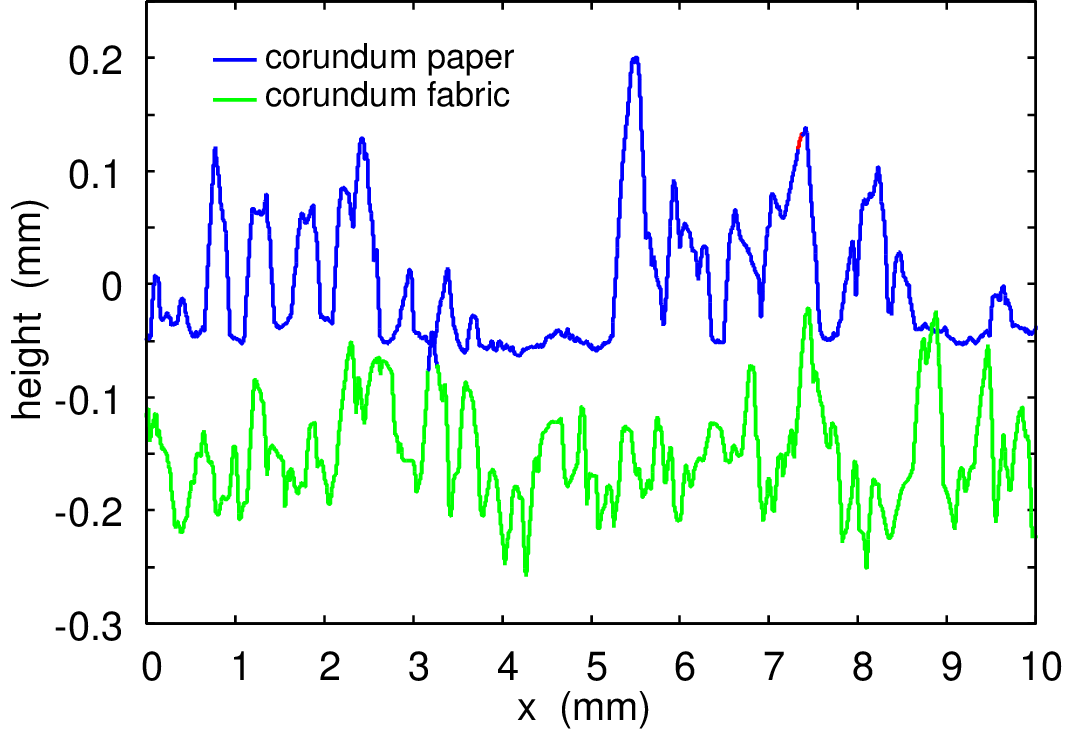}
\caption{
The surface height topography of the corundum abrasive paper and fabric.
}
\label{SANDPAPER.1x.2height.eps}
\end{figure}

\begin{figure}[hbtp]
\includegraphics[width=0.45\textwidth,angle=0]{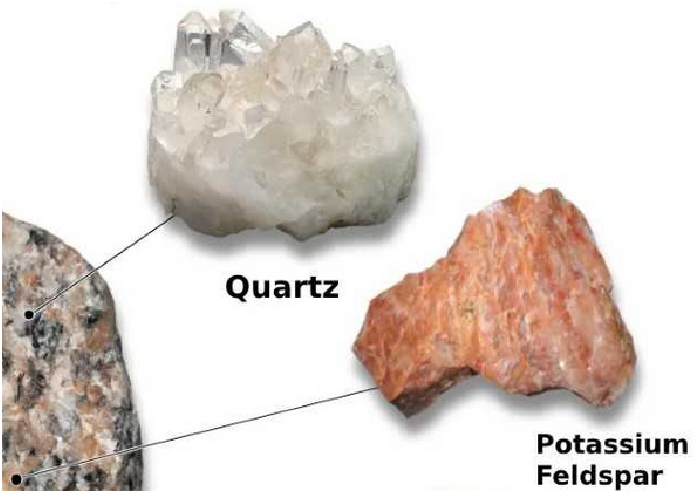}
\caption{
Granite consist of several minerals but the largest fractions are quartz and feldspar.
Our granite looks visually similar to granite whose composition has been studied in detail, with $\approx 72\%$ quartz and $\approx 15\%$ feldspar.
}
\label{Gratite.compo.eps}
\end{figure}

Experiments involving corundum-coated paper and fabric were used as references to compare the relative hardness and wear behavior of the constituent minerals.
Their linescans are shown in Fig.~\ref{SANDPAPER.1x.2height.eps}.
Corundum (aluminum oxide) is a very hard mineral with a penetration hardness $\sigma_{\rm P} \sim 30~{\rm GPa}$ and a 
Mohs hardness $\kappa \approx 9$ compared to the main constituent of granite, which is silica with $\sigma_{\rm P} \sim  15~{\rm GPa}$, $\kappa \approx 7$. 
Feldspar, another constituent of granite, is even softer with $\sigma_{\rm P} \lesssim  3~{\rm GPa}$, $\kappa \approx 6$.
Fig.~\ref{Gratite.compo.eps} shows optical images of granite and its two main constituents.  

\subsection{Molecular Dynamics Simulations}

Molecular dynamics simulations were conducted to substantiate claims made in the experimental part of the work with a particular focus on non-flooded contacts.
To this end, the force-field proposed by van Beest, Kramer, and van Santen (BKS) was used~\cite{vanBeest1990PRL}.
Owing to its simple form, the potential is computationally cheap and allows for the simulation of relatively large systems.
At the same time, it has successfully reproduced the properties of liquid~\cite{Vollmayr1996PRB,Horbach1999PRB}, amorphous~\cite{Vollmayr1996PRB,Horbach2001EPJB} and crystalline~\cite{Herzbach2005JCP} silica alike, including pressure-induced amorphization~\cite{Tse1991PRL} and crystal-crystal phase transformations of quartz~\cite{Campana2004PRB}.
In some cases, parts of the simulations were conducted or repeated with ReaxFF~\cite{vanDuin2003JPC} parameterized for silica by Noaki~\textit{et. al.}~\cite{Noaki2023NCM}.
This allowed us, among other things, to ensure that the trends observed in the simulation were not exclusive to BKS.
However, this could only be done selectively, as the simulations run about a factor 10 times more slowly with ReaxFF than with BKS.

While a thin contamination or water layer may exist in non-flooded contacts, high pressures will squeeze them out rapidly and siloxane (Si-O-Si) bridges have been argued to occur over time even in the presence of extra water~\cite{Li2014TL,McClimon2024TL}.
Since the squeeze-out might not be complete at the high simulation sliding velocities of $0.1-10$~m/s, we expect them to be complete at the lower experimental velocities, which we attempt to target with our simulation.

Our set-up consists of an initially fully crystalline substrate ($\alpha$-quartz) and a rigid amorphous indenter made of silica.
The dimensions of the substrate at room temperature are $39.69\times2.58\times13.42$~nm in $x$, $y$, and $z$ directions, respectively.
Periodic boundary conditions are employed in the $xy$ plane.
All atoms within the substrate interact via the BKS potential.
The atoms in the six bottom layers, i.e., two layers of silicon and four of oxygen atoms, are confined to their crystallographic positions in the in-plane directions and coupled with a spring of stiffness $k = 0.5$~N/m to their ideal lattice sites in the normal direction.
The coupling constitutes a compromise between zero strain and constant stress conditions at the lower surface.
The value of the spring stiffness was chosen so that a normal surface undulation of wavelength $L_x$ experiences approximately the same restoring force as it would in a semi-infinite solid.

The indenter consists of a cylindrical shell with an outer radius of $13.5$~nm.
This value was chosen during the initial stages of the project, while exploring the possibility to use crystalline indenters.
The given radius was the smallest size at which the BKS indenter did not spontaneously amorphize.
Ultimately, it was decided to use an amorphous tip in order to reduce the dependence of the results on the crystal orientation.
Atoms of the indenter are kept fixed to its center of mass.
In most simulations, the indenter and substrate atoms interact through BKS interactions.
Adhesion and cold welding were suppressed in selected complementary simulations by retaining only the short-range exponential two-body repulsion in the indenter–substrate interaction.
This effectively passivates the contact and makes the indenter behave like a rigid, slightly corrugated wall.
These simulations will be labeled as \textit{non-adhesive}.

In the simulations, the quartz surface is indented by a rigid amorphous indenter as a modeling simplification. While macroscopic granite contacts involve two deformable bodies, at the asperity scale deformation is expected to localize preferentially in one of the two opposing asperities due to differences in curvature, local structure, or prior damage. The rigid indenter therefore serves as a geometrically well-defined stress-imposing boundary condition, allowing us to isolate and analyze pressure-driven deformation mechanisms in $\alpha$-quartz without additional complexities associated with simultaneous plasticity of the opposing asperity. The indenter may be viewed as representing a heavily damaged or amorphized silica-rich counter-asperity.

Since the initial cleaved quartz surface is quite unstable when using the BKS potential, it was subjected to an initial equilibration using ReaxFF for 30~ps.
The resulting equilibrated structure, which can be characterized by an approximately 5~\AA~thick amorphous surface layer, proved stable in continuation runs using BKS.
We chose the (001) quartz surface with middle-O termination, thus exposing a single oxygen atom per unit cell of $\alpha$-quartz.
This termination was shown to be the most stable~\cite{Wang2018M}.

In our simulations, the lack of a passivation layer means that direct Si–O bonding occurs more readily than in a real-laboratory experiment.
Thus, the interface cold welds rather quickly in the simulations.
This acceleration is desirable to mitigate the large gap between the sliding velocities imposed in the simulation, 10~m/s by default, and the experimental velocities, which are three or more orders of magnitude slower.

Finally, a time step of 1~fs was used throughout this work.
Temperature was maintained at $T = 300$~K in most simulations using a Grønbech-Jensen thermostat~\cite{GronbechJensen2019MP} with
a thermostat time constant of $\tau = 100$~fs.
Only atoms in layers 7 through 12 (six next to those connected to springs) were thermostatted.
All simulations were run using the LAMMPS simulation package~\cite{Thompson2022CPC}.
The snapshots were prepared using OVITO~\cite{Stukowski2009MSMSE}.

\subsubsection{Determination of contact area and interfacial stresses}

The task of determining interfacial stress fields and real contact areas from atomistic data — in a form that can be meaningfully compared to continuum models — deserves its own short section.

Comparing simulations with continuum-mechanics-based theories requires:
(a) assessments of the real contact area in atomistic simulations so that mean contact pressures can be determined;
(b) spatially resolved contact stresses for detailed comparison.
This is non-trivial for the following ~\cite{Cheng2010TL}:
(a) different reasonable contact criteria (e.g., distance- vs. stress-based) can yield noticeably different contact areas; this is particularly tricky for nominally repulsive contacts, where the interfacial gap between a Hertzian indenter and a linear elastic half-space opens only with $\Delta u \sim a_c (\Delta r/a_c)^{3/2}$, with $a_c$ the contact radius;
(b) real, even nominally repulsive, interactions have finite range, i.e., they are body forces rather than surface stresses.
Nonetheless, results from atomistic/discrete simulations (which produce sharply localized stress peaks) and continuum theory can be compared meaningfully when both are processed with the same coarse-graining filter~\cite{Muser2019TL}.

Before describing our procedure, we note that Gaussian smoothing centered at the origin does not shift the location of maximum slope for an ideal step, which illustrates that smoothing does relatively little harm when stress fields are already flat across the load-bearing region — as they are during slow indentation past the onset of plasticity.
At the same time, one should not assume that an automated criterion based on a chosen smoothing width can robustly define contact edges without human inspection.

Despite these caveats, the problem becomes manageable if uncertainties on the order of 10\% are acceptable.
The data shown in the Results section were processed as follows:
The atomistic contact stress is defined as the sum of forces exerted by the substrate atoms on those indenter atoms whose $(x,y)$-coordinates fall into an areal bin of size $\approx 1$~Å~$\times$~1~Å.
For 1D stress profiles, the values are averaged along the $y$-direction and smoothed with a Gaussian of width $\sigma \approx 2.5$~Å.
This produces smoothly varying stress profiles.
Nonetheless, human inspection remains required to locate the contact line. 
This necessity arises because weak adhesive necks appear at the contact edge for small indentation depths, but disappear at higher loads.
This makes it difficult to define a single, universally reliable quantitative measure for the contact line's position.

\subsubsection{Local von Mises strains}
Local atomic strains were calculated using the best-fit local affine deformation method following Falk and Langer~\cite{Falk1998PRE}.
For each atom $i$, neighbors within a cutoff distance 10~\AA~ were identified in a reference configuration, which was chosen to be the undeformed sample where the indenter and the substrate form a contact.
The local deformation gradient tensor $\mathbf{F}_i$ is then determined from the minimization of the mean-square difference between the actual relative displacements of the central atom and its neighbors and the relative displacements that they would have if the region within the cutoff distance were strained uniformly.
This procedure yields the local best affine transformation that maps the deformed atomic neighborhood onto the reference one.
The local Lagrangian strain tensor $\boldsymbol{\eta}_i$ is then obtained as $\boldsymbol{\eta}_i = 1/2\left(\mathbf{F}_i^\mathrm{T}\mathbf{F}_i-1\right)$.
The local von Mises strains were calculated from the deviatoric part of $\boldsymbol{\eta}_i$.
In this work, OVITO~\cite{Stukowski2009MSMSE} was used to calculate local strains.

\subsubsection{Classification of stress-induced crystal phase}

During the sliding simulation, a new crystalline phase emerged underneath the leading contact edge.
Here, we describe how the phase was characterized.
An apparent unit cell was carved out of the simulation box and repeated $6\times4\times4$.
The resulting supercell was then subjected to a stability test: 
a 50~ps long NVT simulation keeping the sheared unit cell was performed at $T=300$~K.
Next, the cell was quenched to zero temperature using a cooling rate of $5\times10^{12}$~K/s, one time allowing  the lattice constants to change while keeping the box angles and one time with full optimization. 
Both resulting supercells were then analyzed to determine their crystallographic space group symmetry from the reduced atomic positions and lattice parameters.
This symmetry analysis was carried out using the spglib library~\cite{Togo2024STAMM} which identifies symmetry operations and assigns the corresponding space group.
A symmetry tolerance (symprec parameter) of 0.05~\AA~was used, meaning that atoms within this distance after application of a symmetry operation were considered symmetry-equivalent.
Such a choice allows for small relaxation-induced distortions to be ignored while preserving physically-meaningful symmetry-breaking.

\section{Results}

\subsection{Experimental results}

Granite is a polymineralic composite typically dominated by hard quartz and softer feldspar grains, see the optical image of granite and its two main constituents in Fig.~\ref{Gratite.compo.eps}.
Because of the hardness contrast, quartz having a Mohs hardness of $\kappa \approx 7$ and feldspar $\kappa \approx 6$, it is widely inferred~\cite{Chester1986PAG,Mair1999JGR} that granite surfaces wear non-uniformly so that quartz grains extend above the average surface plane while feldspar remains below.
Thus, one expects mature interfaces to exhibit comparatively flat, quartz-dominated load-bearing plateaus.
In this section, we report and correlate results on wear, friction, and topography changes in granite–granite (dry and in water), as well as dry granite–corundum contacts, where corundum with $\kappa \approx 9$ is even harder than quartz.

\begin{figure}[hbtp]
\includegraphics[width=0.4\textwidth,angle=0]{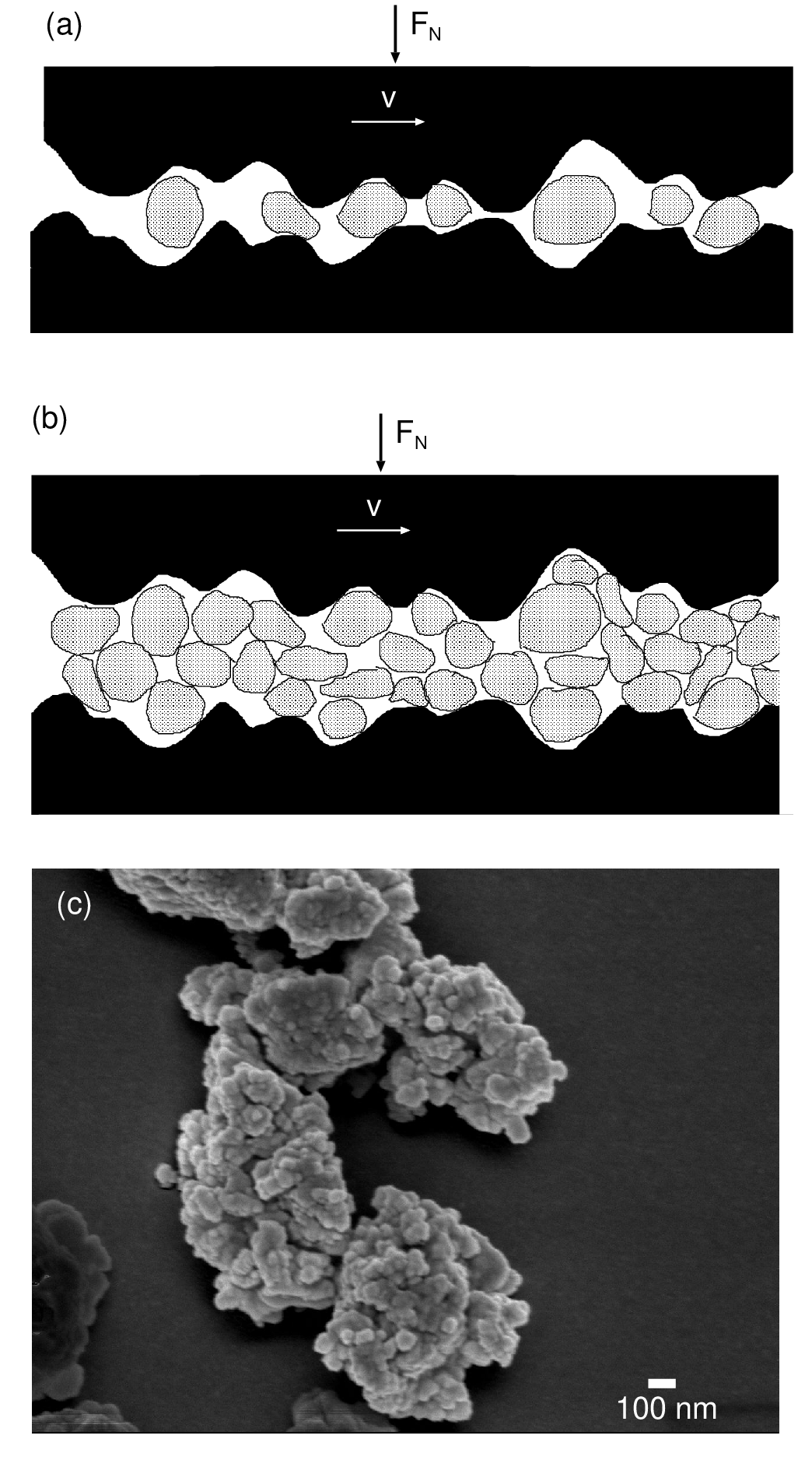}
\caption{
Schematics of a granite--granite contact with (a) few versus (b) many gouge particles.
(c) SEM image of wear-particle clusters.
}
\label{fig:gouge}
\end{figure}

\subsubsection{Wear}

\begin{figure}[tbp]
\includegraphics[width=0.45\textwidth,angle=0]{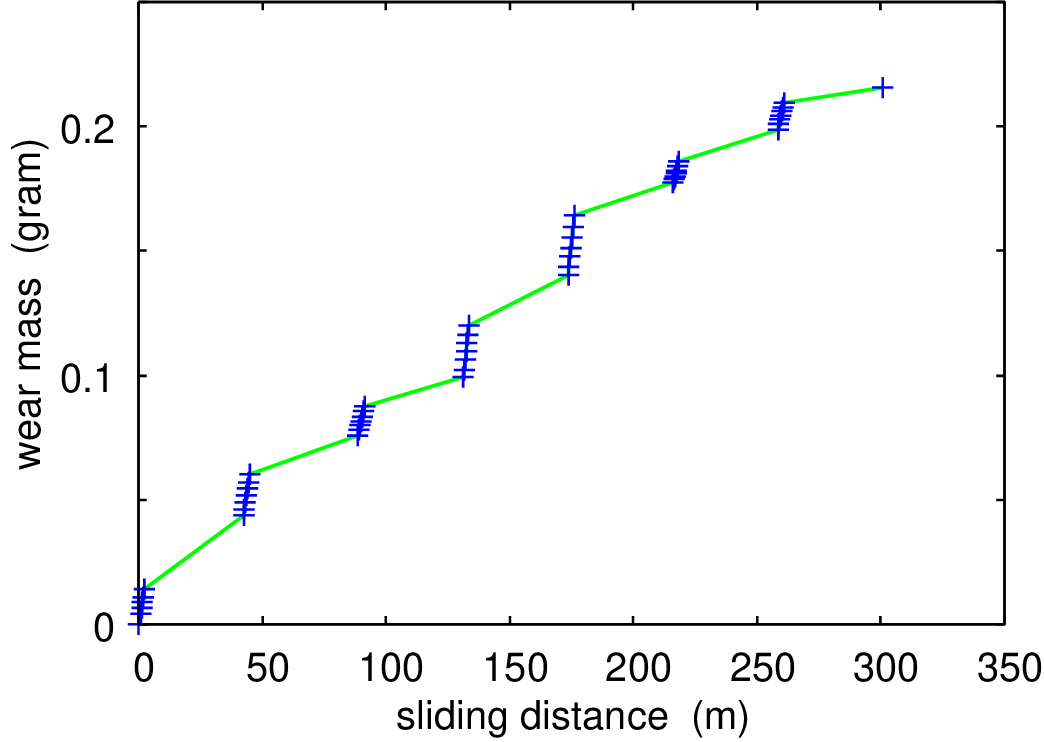}
\caption{
Cumulative wear mass as a function of sliding distance for a normal load of $F_\mathrm{N} = 250\,\mathrm{N}$ and a sliding velocity of $v = 3\,\mathrm{mm/s}$.
The symbols correspond to 1-run measurements, each consisting of 0.2~m forward and 0.2~m backward motion, after which the accumulated wear  is removed.
Green lines connecting the symbols show the mean wear obtained from 100-run measurements, in which wear accumulates over a total sliding distance of 40~m.
}
\label{1distance.2wearmass.eps}
\end{figure}

We begin by corroborating the established result that the presence of gouge dramatically suppresses the further production of gouge in granite--granite contacts~\cite{Chester1986PAG, Mair1999JGR}.

The standard interpretation is that once gouge is present, it shields direct contact between the opposing surfaces, 
and shear is accommodated in the deformable granular layer rather than by continued comminution of intact asperities~\cite{Chester1986PAG, Mair1999JGR, Rathbun2013JGRSE}.
A qualitative sketch of this shielding is shown in Fig.~\ref{fig:gouge}(a--b).
The quantitative effect is analyzed in Fig. \ref{1distance.2wearmass.eps}: the cumulative wear mass increases steeply 
with sliding distance when debris is removed after each cycle (“1-run”), whereas the mean wear rate is 
strongly reduced when sliding proceeds through a pre-existing debris layer for 100 cycles before removal (“100-run”).
A characteristic gouge cluster is shown in the inset of Fig.~\ref{fig:gouge}(c). 

The particles shown in Fig. 4(c) are from a 100-run. Most particles have diameters around $100 \ {\rm nm}$ 
but form clusters as indicated in the figure.
We also observed larger particles with diameter $\sim 0.1-1  \ {\rm mm}$, in particular when using new surfaces.      
The big particles are most likely removed by brittle fracture, while the smaller particles may    
originate from a different process, maybe related to the formation of cold welded junctions, 
or from big particles trapped for some time at the sliding interface
where they could break up into smaller fragments. 
While the particle may spend some time in the asperity contact regions, most of the particles after a 100-run accumulate
at the turn-around edges of the sliding track. 

Analyzing the data shown in Fig. \ref{1distance.2wearmass.eps} quantitatively reveals that the wear rate for 1-runs 
is roughly 13 times higher than for 100-runs, which illustrates the protective effect of the accumulated gouge layer.
In detail, one sliding cycle consists of 0.2~m forward and 0.2~m backward motion.
We always first perform a sequence of six 1-runs, which is followed by a single 100-run, yielding a full sequence.

The wear mass during the seven 100-runs totaled $m=0.12 \ {\rm g}$ over a distance of $L = 280 \ {\rm m}$, yielding (with $F_\mathrm{N} = 250$~N) a wear rate of $k_\mathrm{W} = 0.00171 \ {\rm mg/Nm}$.
In contrast, the cumulative wear from all 1-runs was $m=(0.2156-0.1200) \ {\rm g}$ over $L=16.8 \ {\rm m}$, giving a rate of $k_\mathrm{W} = 0.0223 \ {\rm mg/Nm}$.
We note that the deduced thickness of the removed layer $d = \Delta V/A_0 \approx 51 \ {\rm \upmu m}$, where $\Delta V = m/\rho$ is the removed volume, agrees closely with changes in topography pictures, which will be analyzed further below. 

We also performed a 100-run experiment when the granite surface was covered by a $\sim 1 \ {\rm mm}$ thick water film.
This resulted in much smaller wear, $m = 0.0018 \ {\rm g}$, corresponding to a wear rate of $0.00018 \ {\rm mg/Nm}$, approximately ten times smaller than for the dry surface.

In addition, we performed measurements where the granite block was slid on two different P100 abrasive surfaces obtained from different suppliers.
In both cases, corundum particles are bound to the substrate surface (in one case paper and in the other a fabric) with an acrylic resin.
The measurements consist of three cycles, where the granite block moves again 20~cm forward and 20~cm backward in each cycle, repeated twice on different substrate areas.

For granite sliding on the corundum paper, the wear rates are $0.701 \ {\rm mg/Nm}$ and $0.803 \ {\rm mg/Nm}$ for the first and second 3-run, and for the two runs using corundum fabric $0.369 \ {\rm mg/Nm}$ and $0.338 \ {\rm mg/Nm}$.
These rates are 17--34 times higher than in self-mated contacts.
The sand paper has sharp and hard asperities, which can penetrate deeper into the feldspar than into the quartz. Hence, when approaching the quartz grain it may contact the quartz grain from the side rather than from the top. This may result in the much larger (by a factor of $\sim 30$; see Table I) wear rate for sandpaper on granite than for the same sandpaper on a flat quartz surface.

The higher wear rate of the corundum paper versus the corundum fabric supposedly reflects the taller asperities and more open structure of the corundum paper.
Pertinent height topographies are shown in Fig.~\ref{SANDPAPER.1x.2height.eps}.

It is interesting to compare these wear rates with the wear rate we found in an earlier study~\cite{Xu2025TL} for quartz blocks with smooth surfaces sliding on the P100 corundum paper (1-run): $0.024 \ {\rm mg/Nm}$.
This wear rate is roughly 30 times smaller than for the 1-run of granite on the same surface.
The much larger wear rate for granite is attributed to its composite nature, with hard quartz crystals surrounded by a smaller amount of the softer feldspar. 

Table~\ref{TABLE1} summarizes all measured wear rates, including results for two silica glass surfaces sliding on P100 corundum (from Ref.~\cite{Xu2025TL}).
It also lists approximate friction coefficients for all cases, the detailed acquisition of which is described further below.
Since the friction coefficient for the abrasive corundum counterface is smaller than for granite 1-runs ($\sim 0.6$ compared to $\sim 0.9$), while the wear rate is 17–34 times higher, we can safely conclude that plowing contributes negligibly to the friction for granite sliding on granite.
Overall, no (positive) correlation between wear rate and friction coefficients can be identified from Table~\ref{TABLE1}.

\begin{table}[hbtp]
\renewcommand{\arraystretch}{1.8}
\centering
      \noindent\makebox[\linewidth][c]{%
      \begin{tabular}{@{}|l||c|c|@{}}
\hline
              system & $m/F_{\rm N}L$ (mg/Nm) & $\upmu$ \\
\hline
\hline
              granite–granite 100-run & 0.0017 & 0.95 \\
\hline
              100-run in water & 0.00018 & 0.7 \\
\hline
              granite–granite 1-run & 0.022 & 0.9 \\
\hline
              granite–P100 paper & 0.75 & 0.6 \\
\hline
              granite–P100 fabric & 0.35 & 0.7 \\
\hline
              quartz–P100 paper & 0.024 & 0.32 \\
\hline
              borosilica–P100 paper & 0.058 & 0.34 \\
\hline
              soda-lime–silica–P100 paper & 0.14 & 0.38 \\
\hline
\end{tabular}
}
\caption{
Summary of wear rates and friction coefficients for different sliding systems. 
The sliding speed was $v=3~{\rm mm/s}$ and the nominal contact pressures were $\sim0.16~{\rm MPa}$. 
P100 paper and fabric are corundum abrasives with an average particle size of $\approx160~{\rm \upmu m}$. 
}
\label{TABLE1}
\end{table}

Even though the wear rate is small for granite on granite in 100-runs, the friction coefficient is close to unity.
Assuming that the points of contact are predominantly in plastic flow --- an assumption that will be supported by both molecular dynamics simulations and the theoretical treatment presented further below --- one may conclude that the interfacial shear stresses are of similar order of magnitude as the penetration hardness, which is about 10--15~GPa. 
We suggest that such large stresses can only be obtained when covalent or ionic bonds (e.g., Si–O–Si) form across the interface, 
leading to cold-welded-like contacts. Cold welding between gauge particles, and between the particles and the walls, may also occur.

The granite blocks used in our studies were kept in the normal atmosphere and may have water and organic contamination on the
surfaces. We believe that even if a thin contamination film occurs on the granite surface, it is removed (or penetrated) by the high normal and shear stresses acting in the contact regions. Thus, the contamination film may act as a weak barrier against the formation of cold welded junctions~\cite{Li2014TL}, but since the local stresses in the contact regions
are so large as to plastically deform the quartz, a thin contamination film is unlikely to prevent direct contact between the
${\rm SiO_2}$ quartz grains. For metals, which often have thick oxide coatings, measurements using radioactive tracer atoms
have shown that cold welded junctions and metal transfer occurs~\cite{Rabinowicz1995Book}.
The plastic yield stress and hence the asperity
contact pressure are lower for metals than that for quartz, and quartz has no oxide coating, so the barrier
against formation of cold welded junctions is likely to be smaller for quartz (or granite) than for metals.
In a contact flooded with water, silica surfaces will be continuously re-hydrogenated and re-hydroxylated.
The resulting termination apparently acts similar to a sacrificial, protective layer.
Breaking the terminating bonds costs energy, which is why friction remains relatively high, but the transfer of silicon atoms or silicon-containing clusters is strongly reduced~\cite{Li2014TL,McClimon2024TL}.

\begin{figure}[tbp]
\includegraphics[width=0.45\textwidth,angle=0]{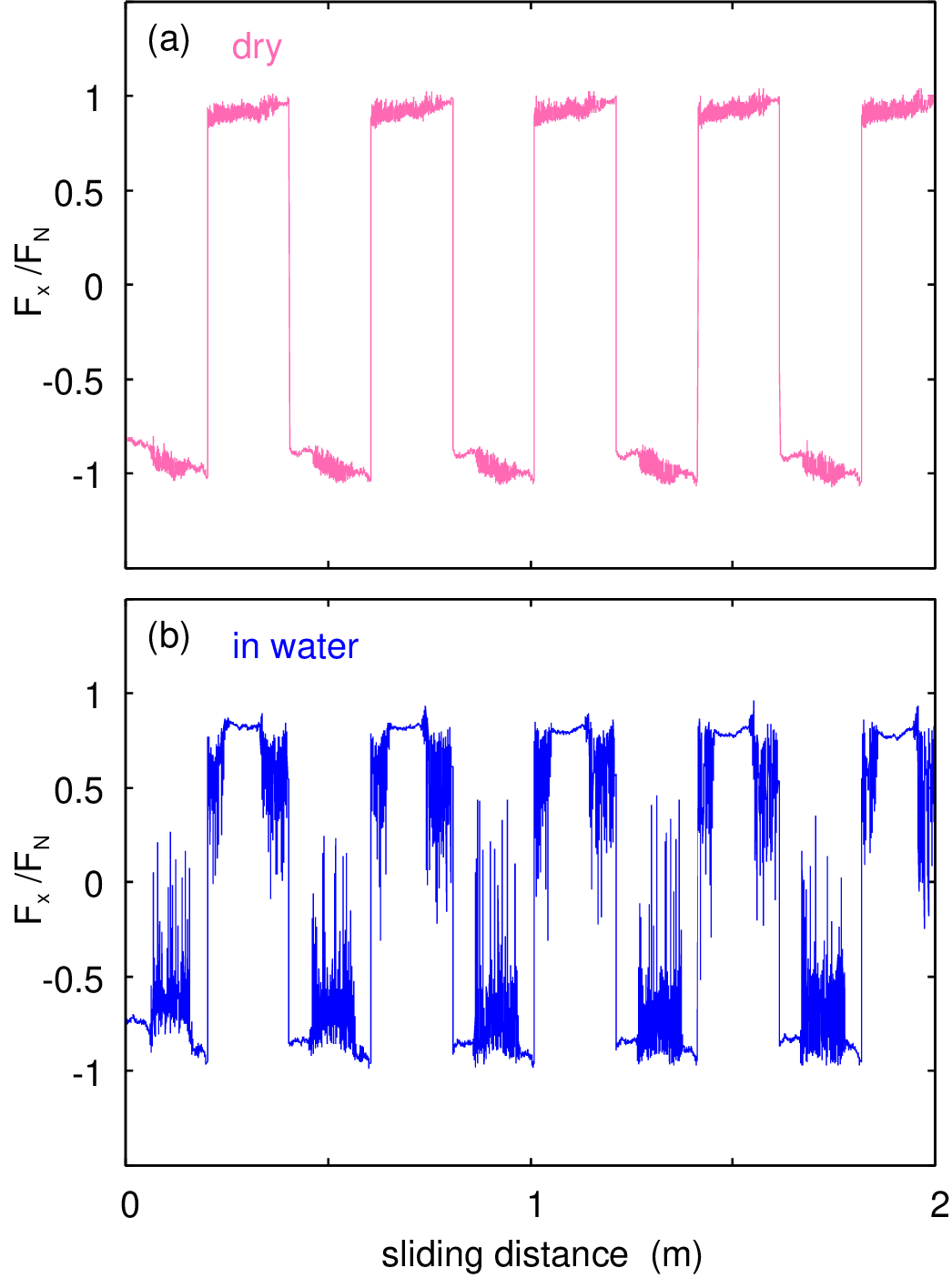}
\caption{
The ratio of tangential and normal force $F_x/F_{\rm N}$ as a function of the sliding distance during (a) the first 5 sliding cycles during the
last 100-run in the dry state, and (b) the first 5 sliding cycles during the 100-run in water.
In both cases the wear particles are not removed from the surface.
The sliding speed $v=3 \ {\rm mm/s}$ and the normal force $F_{\rm N} = 250 \ {\rm N}$.
}
\label{1x.2mu.dry.last1.eps}
\end{figure}

\begin{figure}[tbp]
\includegraphics[width=0.45\textwidth,angle=0]{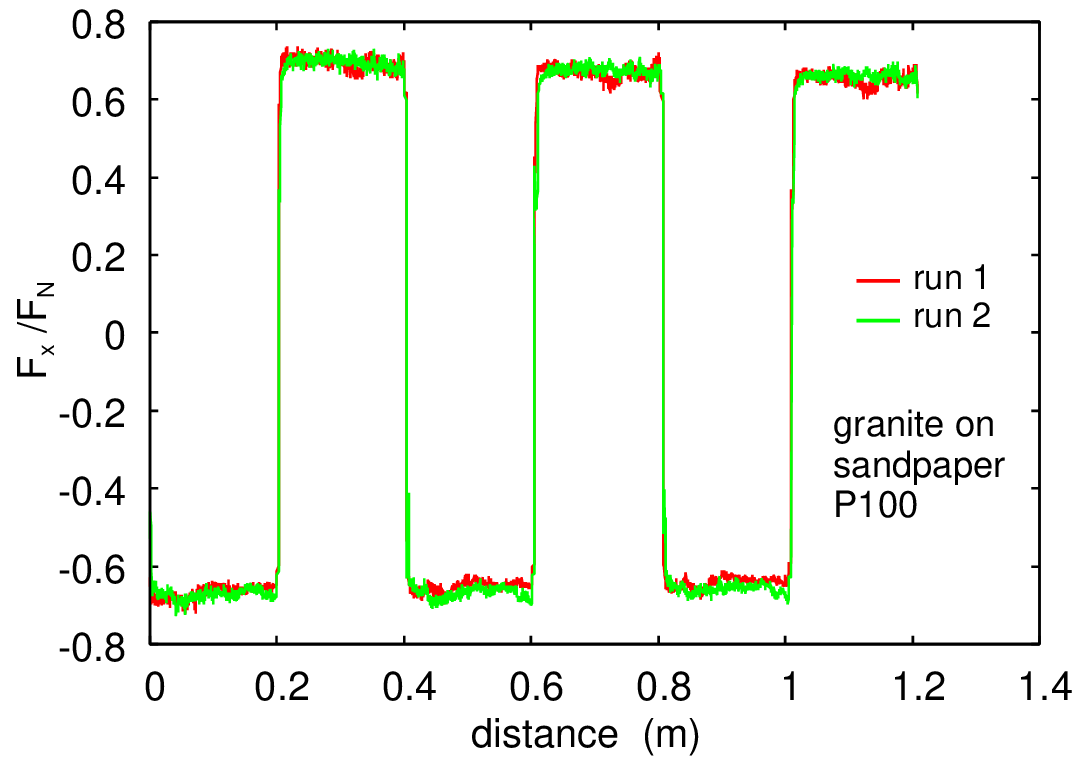}
\caption{
The ratio of the tangential and normal forces $F_x/F_{\rm N}$ as a function of sliding distance for granite on the corundum fabric P100 surface.
Runs~1 and~2 consisted of 3-cycles, performed on different surface regions of the corundum surface.
}
\label{1x.2mu.second.granite.P100fabric.eps}
\end{figure}

\subsubsection{Sliding friction}

We investigated three representative sliding configurations to disentangle the effects of wear and friction.
These include (i) granite sliding on granite under dry conditions, (ii) granite sliding on granite in water, and (iii) granite sliding on corundum abrasives.
Representative results for these cases are summarized in Fig. \ref{1x.2mu.dry.last1.eps} and Fig. \ref{1x.2mu.second.granite.P100fabric.eps}.

Fig.~\ref{1x.2mu.dry.last1.eps} shows that water does not substantially reduce the friction compared to the dry case, although it dramatically reduces the wear rate.
A clearly negative correlation between wear and friction is even evident from Fig.~\ref{1x.2mu.second.granite.P100fabric.eps}.
It reveals a reduced kinetic friction of corundum against granite compared to self-mated granite contacts
($\approx 0.7$ as compared to $0.95$ for granite on granite), despite the excessive wear produced by corundum. 
Thus, if the high wear rates on corundum are attributed to plowing by the sharp and stiff corundum asperities, then plowing can only give a marginal contribution to the friction for granite sliding on granite. 
This is further supported by the observation that the wear rate during the 100-run of granite on granite is ~13 times smaller than for the 1-run, but the friction coefficient is slightly higher, about 0.95.

Some details about the results shown in Fig. \ref{1x.2mu.dry.last1.eps} and \ref{1x.2mu.second.granite.P100fabric.eps} are in place.
The data in Fig. \ref{1x.2mu.dry.last1.eps}(a) represent the first five sliding cycles during the last 100-run in the dry state using $v = 3$~mm/s.
Fig. \ref{1x.2mu.dry.last1.eps}(b) shows the same for the experiment in water.
In either case, the kinetic friction matches the static friction.
Large stick–slip oscillations occur in the presence of water, with a frequency much higher than the instrumental resolution $\Delta x = 0.3$~mm, obtained  
at the sliding speed $3 \ {\rm mm/s}$ and $10 \ {\rm Hz}$ measurement rate.
Fig. \ref{1x.2mu.second.granite.P100fabric.eps} shows two different 3-runs of granite on P100 sandpaper, conducted at different locations on the corundum surface.

\begin{figure}[tbp]
\includegraphics[width=0.45\textwidth,angle=0]{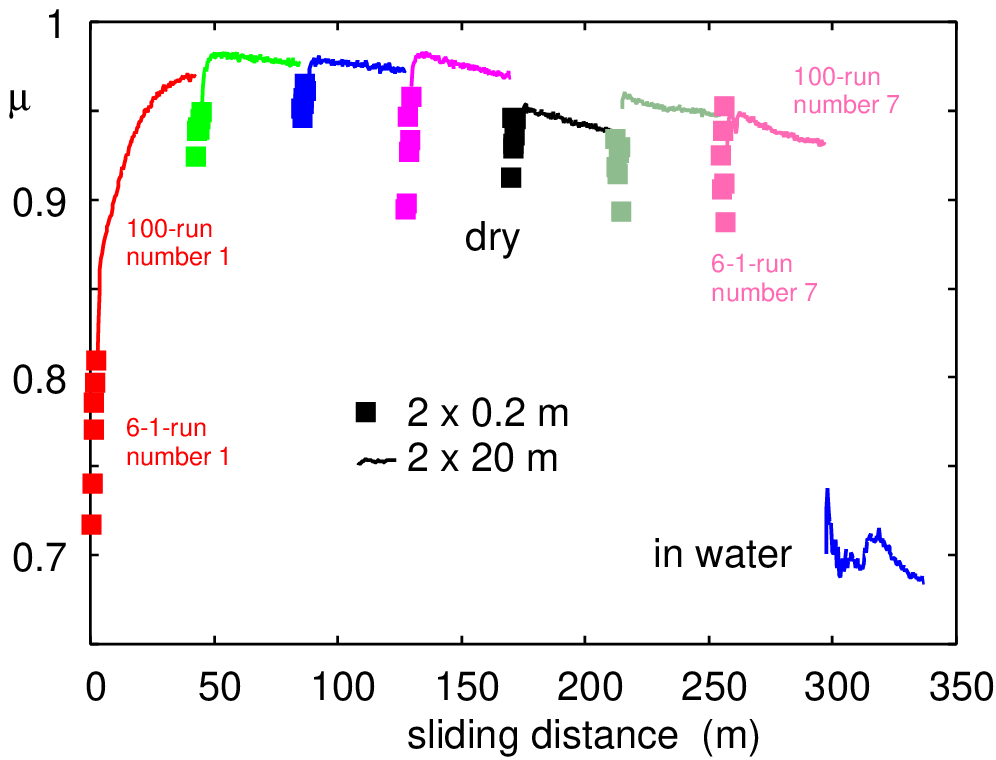}
\caption{
Friction coefficient for the self-mated contact as a function of the accrued sliding distance at a sliding velocity of $v = 3$~mm/s.
Results of the 1-runs are shown by square symbols, while those of the 100-runs are shown by continuous lines.
Water was added after 300~m of dry sliding.
The subsequent 100-run (blue line), which showed stick-slip motion as in Fig. \ref{1x.2mu.dry.last1.eps}(b), gave a lower sliding friction coefficient of $\approx 0.7$ versus $\approx 0.95$ of the dry contact. 
Lines represent running averages over whole cycles of 0.4~m sliding distance.  
}
\label{1distance.2friction.with.water.eps}
\end{figure}

\begin{figure}[tbp]
\includegraphics[width=0.45\textwidth,angle=0]{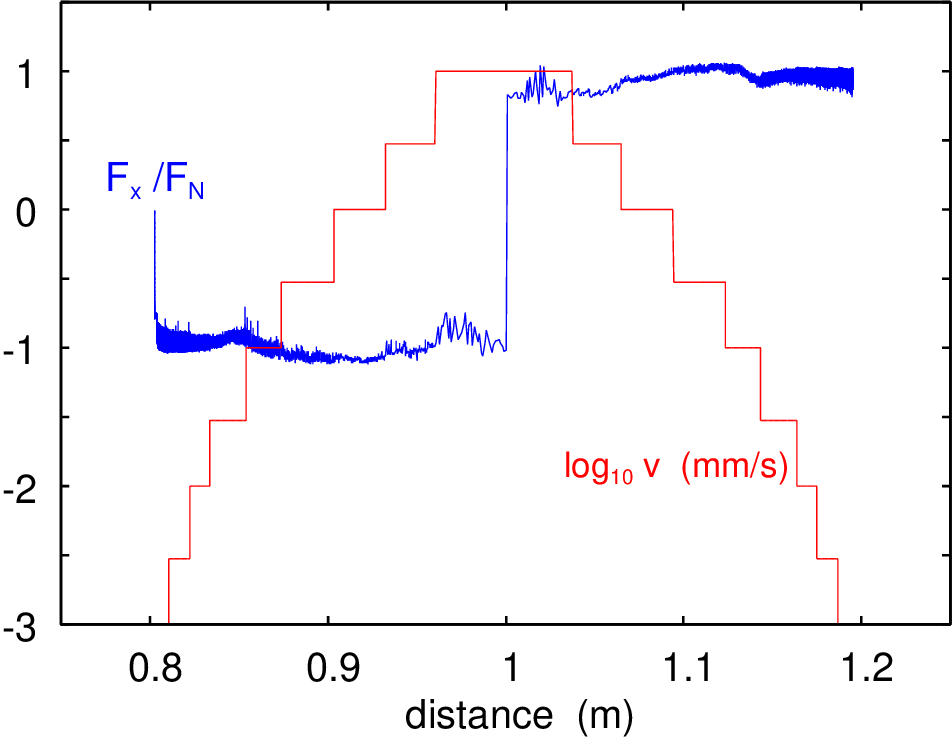}
\caption{
Blue curve: The normalized lateral force $F_x/F_{\rm N}$ as a function of the sliding distance
for the second (repeat) experiment at $T=22^\circ {\rm C}$.
Red curve: The logarithm of the sliding speed (in mm/s). The speed was changed in steps as follows: $1 \ {\rm cm}$ sliding distance at the speed
$1 \ {\rm \upmu m/s}$, $1 \ {\rm cm}$ at $3 \ {\rm \upmu m/s}$ , $1 \ {\rm cm}$ at $10 \ {\rm \upmu m/s}$ , $2 \ {\rm cm}$ at $30 \ {\rm \upmu m/s}$,
$2 \ {\rm cm}$ at $100 \ {\rm \upmu m/s}$, $3 \ {\rm cm}$ at $300 \ {\rm \upmu m/s}$, $3 \ {\rm cm}$ at $1 \ {\rm mm/s}$, $3 \ {\rm cm}$ at $3 \ {\rm mm/s}$,
$4 \ {\rm cm}$ at $10 \ {\rm mm/s}$. 
The direction of motion is inverted and the velocity protocol is reversed after a sliding distance of 1~m.
}
\label{1x.2FxoverFz.last.eps}
\end{figure}

\begin{figure}[tbp]
\includegraphics[width=0.45\textwidth,angle=0]{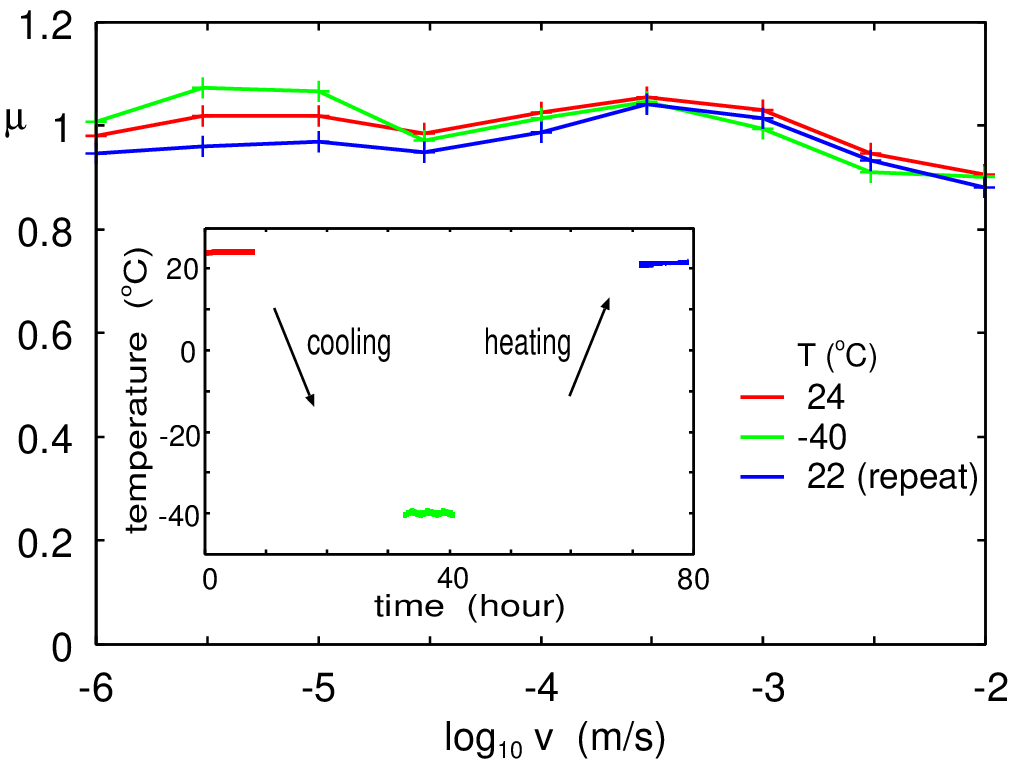}
\caption{
The friction coefficient obtained from data of the form shown in Fig.~\ref{1x.2FxoverFz.last.eps} by averaging over the sliding distance for each sliding speed and temperature.
The measurements were first performed at $T = 24^\circ\mathrm{C}$ (red curves), next at $T = -40^\circ\mathrm{C}$ (green curve), and finally at $T = 22^\circ\mathrm{C}$ (blue curve).
Inset: The temperature in the sliding friction container. The first experiment was done at $T=24^\circ {\rm C}$.
After the system was cooled to $-40^\circ {\rm C}$ where the second experiment was performed. After the system was
heated up to $T=22^\circ {\rm C}$ where the third experiment was done.
}
\label{1time.2Temp.eps}
\end{figure}

The friction coefficient as a function of sliding distance for self-mated granite is shown in Fig.~\ref{1distance.2friction.with.water.eps}.
It contains the seven full sequences of the dry contact and the 100-run in water. 
The data reveal that friction is always the smallest in the first 1-run after debris was removed from the previous 100-run.
This underlines the negative correlation of friction and wear in self-mated granite contacts. 

The friction values we have found for granite sliding on granite are similar to the friction values
found for many other types of rocks. Thus Byerlee found (see Fig. 5 in Ref.~\cite{Byerlee1978PAG}) that the maximum friction coefficient, 
that is, the friction at the onset of sliding,
where the sliding velocity and temperature effects may be very small, is about $0.85$ for a large number of different rocks with different roughness
and for nominal pressures between $1$ and $100 \ {\rm MPa}$.
The natural explanation for this is that because of the surface roughness, the solids make contact only in a small fraction of the
nominal contact area and that all contacts have yielded plastically. When the solids yield plastically it is likely that
any contamination film, which may be formed in the normal atmosphere, will be penetrated so that cold-welded junctions can be formed. In this case,
the friction coefficient will be determined by the ratio $\tau /\sigma_{\rm P}$ of the yield stress in shear $\tau$ 
and the penetration hardness $\sigma_{\rm P}$; for rocks $\tau /\sigma_{\rm P}$ may be of order 1.
The importance of cold-welded junctions and the equation $\mu \approx \tau /\sigma_{\rm P}$, was first proposed 
by  Bowden and Tabor~\cite{Bowden2001Book} for metals. 

Fig.~\ref{1x.2FxoverFz.last.eps} highlights an unexpected insensitivity of the kinetic friction force to abrupt velocity changes.
For these measurements, the system was first ``run-in'' by 10-runs at room temperature. 
After, measurements were performed, in which the velocity $v$ was increased in discontinuous steps from 1~$\upmu$m/s to 1~cm/s, reversed and then decreased by decreasing the speed again in similar steps.
Specifically, the sliding distance per velocity were: 
1~cm at the speed of 1, 3, and 10~$\upmu$m/s, 
2~cm at 30 and 100 $\upmu$m/s,
3~cm at 0.3, 1, and 3~mm/s, and 
4~cm at 10 mm/s.
Fig. \ref{1x.2FxoverFz.last.eps} shows that the friction coefficient does not show anomalies at the points of time, where velocity changes abruptly. 

To analyze the velocity dependence of friction further and to add temperature to the picture, 
the friction was measured near room temperature and at $T = -40^\circ$C using the same protocol as that just described. 
To this end (see inset in Fig. \ref{1time.2Temp.eps}), 
the first set of experiments was conducted at $T = 24^\circ$C at different $v$ over eight hours.
The experimental apparatus was then cooled down to $T = -40^\circ$C and allowed to equilibrate for one day.
Next, the velocity sweeps were repeated for the duration of eight hours.
Finally, the apparatus was heated back to high temperature, this time to $T = 22^\circ$C, again over the duration of one day to ensure full thermal equilibration in the system. 
The result for the repeat at $T = 22^\circ$C is depicted in Fig.~\ref{1x.2FxoverFz.last.eps}.
Despite the relatively long waiting times, no stiction peak is apparent during the onset of sliding. 
This result is surprising because peaks during the onset of sliding after long relaxation times have been observed in many cases and motivate the basis for rate-and-state friction laws, which are frequently used in modeling earthquakes.

Fig. \ref{1time.2Temp.eps} shows that there is only a very small dependency of the friction on the temperature and on the sliding speed for $v < 1 \ {\rm mm/s}$. 
For  $v > 1 \ {\rm mm/s}$ the friction coefficient decreases  slightly, which also shows up in the friction force as a weak stick-slip noise for the highest sliding speed, see the blue curve in Fig.~\ref{1x.2FxoverFz.last.eps}
for $v=1 \ {\rm cm/s}$. 
The fact that the friction coefficient is the same for $20^\circ {\rm C}$ and $-40^\circ {\rm C}$ indicates that 
thermal excitation is unimportant for the friction and wear processes for temperatures up to room temperature. 

\begin{figure}[tbp]
\includegraphics[width=0.45\textwidth,angle=0]{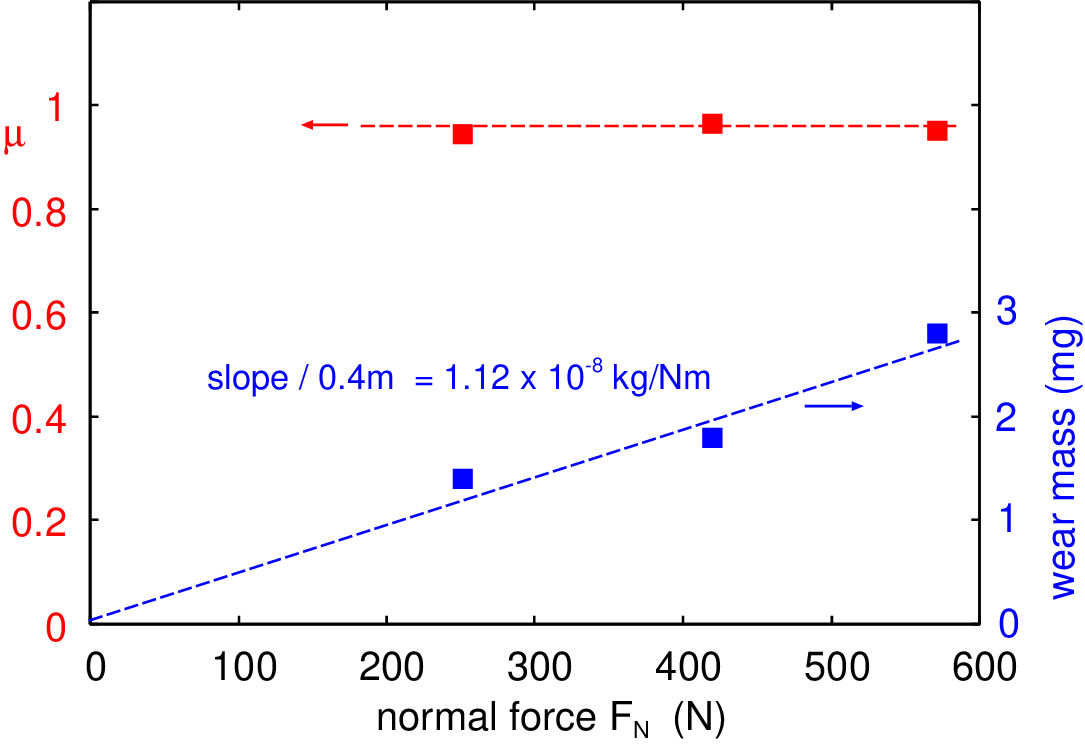}
\caption{
The friction coefficient (red data points) and the wear rate (blue) as a function on the normal load
for 1-runs at $T=20^\circ {\rm C}$ and sliding speed $v=3 \ {\rm mm/s}$.
}
\label{1FN.2mu.and.wear2.eps}
\end{figure}

As a side remark, we note that the friction force and the wear rate both increase roughly linearly with the normal force $F_\mathrm{N}$ whenever tested, e.g., for 1-runs at $T = 20^\circ$C and $v = 3$~mm/s, see Fig.~\ref{1FN.2mu.and.wear2.eps}.
This is expected if increasing the load increases the area of real contact proportional to $F_\mathrm{N}$, while leaving typical contact-patch sizes approximately unchanged.
The former assumption is well established from a wide range of theories and simulations, while the latter 
is also true~\cite{Muser2018L}, at least for the system and pressure range of interest in this study (see Theory section).
The dashed line in Fig.~\ref{1FN.2mu.and.wear2.eps} corresponds to a wear rate of 
$k_\mathrm{W}\approx1.12\times10^{-2}$~mg/Nm, which is smaller than the average wear rate for 1-runs during the run-in over the first 300.8~m sliding distance, $2.47\times10^{-2}$~mg/Nm.

\begin{figure}[tbp]
\includegraphics[width=0.45\textwidth,angle=0]{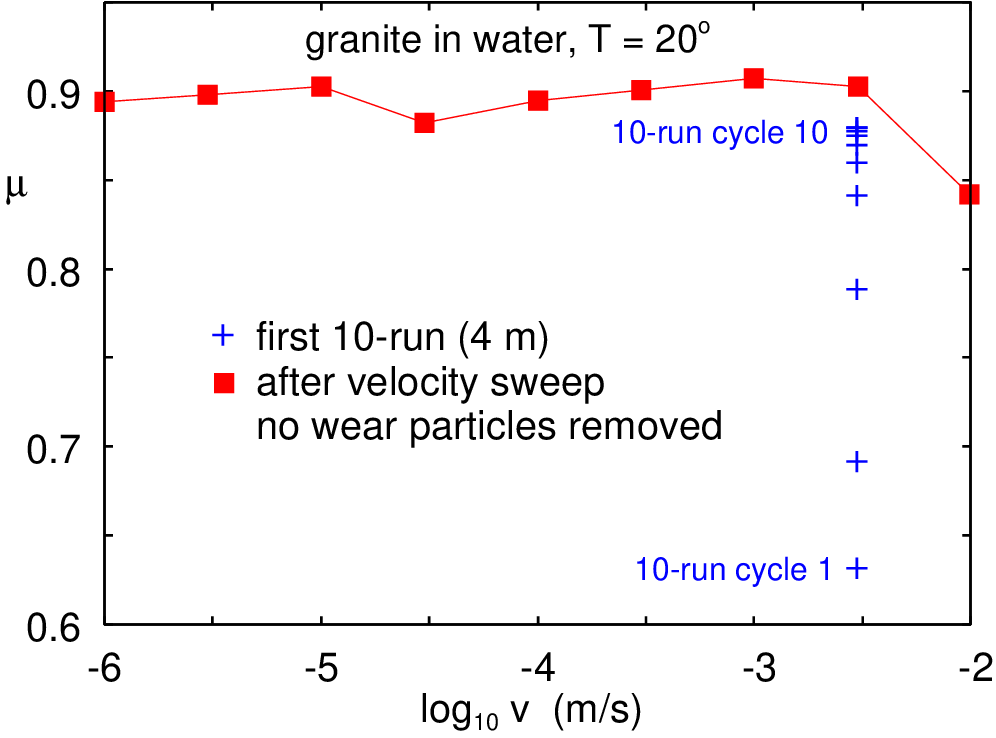}
\caption{
Friction coefficient $\mu$ as a function of sliding speed $v$ of self-mated granite in water for a normal force of $F_{\rm N} = 250\,\mathrm{N}$ and a temperature of $T = 20^\circ\mathrm{C}$.
The blue + symbols show averages over individual cycles in the initial 10-run.
The red squares show the results of a velocity sweep.
No wear particles were removed during the experiment.
}
\label{1logv.2mu.in.water.many.v.eps}
\end{figure}

The running‐in of a self‐mated granite contact in water and the velocity dependence of the friction coefficient are shown in Fig. \ref{1logv.2mu.in.water.many.v.eps} 
for a normal force of $F_\mathrm{N} = 250$~N and at a temperature of $T = 20^\circ$C.
The friction coefficient first increases monotonically from 0.63 to 0.88 during a 10-run at 3~mm/s.
Without removing the debris, the same velocity sweep explored earlier in Fig.~\ref{1x.2FxoverFz.last.eps} was conducted next.
The friction coefficient remained approximately constant from $v = 1~\upmu$m/s to 3~mm/s, but then decreased from its small velocity value of $\mu \approx 0.9$ to $\mu \approx 0.85$ at $v = 1$~cm/s.
This drop explains the stick–slip dynamics reported earlier in Fig. \ref{1x.2mu.dry.last1.eps}(b), since decreasing $\mu(v)$ relationships unavoidably lead to unsteady motion when the system is driven at constant velocity using a finite (machine/granite-block) compliance and small intrinsic damping. 

Similar experiments were repeated using two additional loads at a velocity of 3~mm/s, which shows, as in the dry system, a negligible dependence of the friction coefficient on the normal force when $F_{\rm N}$ 
increases from $250 \ {\rm N}$ to $520 \ {\rm N}$. 
In all experiments in water, wear particles were not removed manually, but may have been partially flushed from the track by the water flow, at least at the highest sliding speeds. 
This is one possible — albeit speculative — explanation for why the friction drops in water at the highest velocities.

\begin{figure}[tbp]
\includegraphics[width=0.45\textwidth,angle=0]{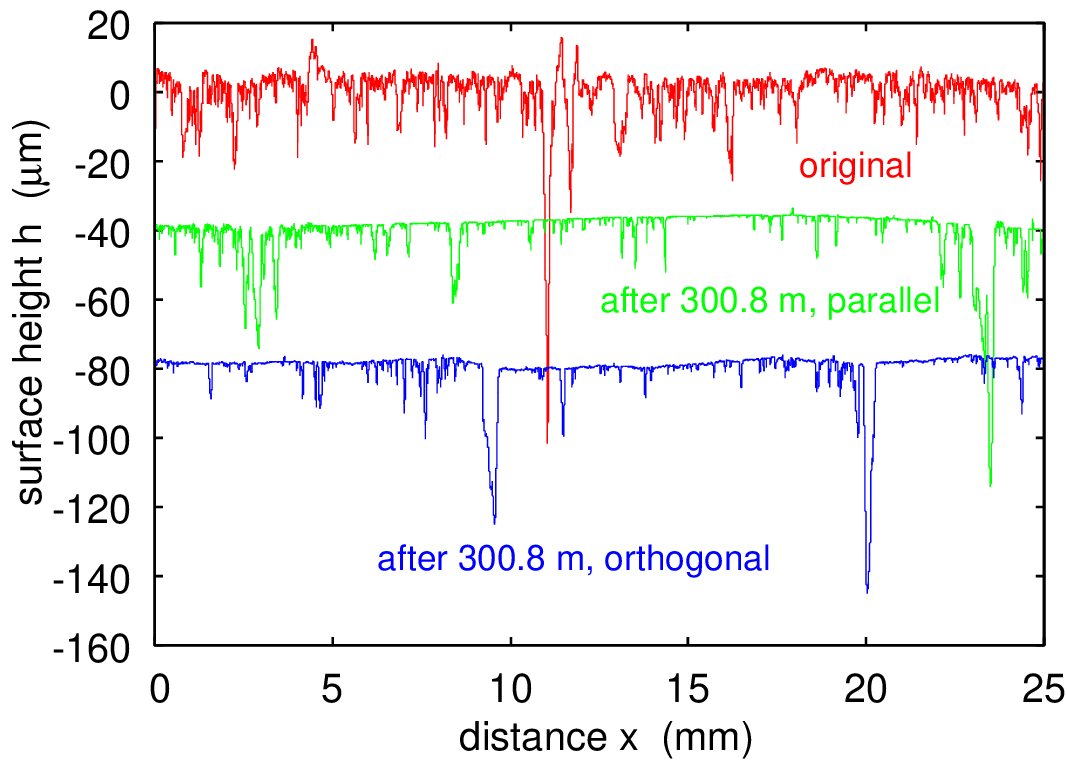}
\caption{
The line topography of the original surface (red line), and after sliding $ 300.8 \ {\rm m}$ (green and blue lines). 
The green line is the topography along the sliding direction and the blue line orthogonal to the sliding direction.
}
\label{1x.2h.all3.eps}
\end{figure}

\begin{figure}[tbp]
\includegraphics[width=0.45\textwidth,angle=0]{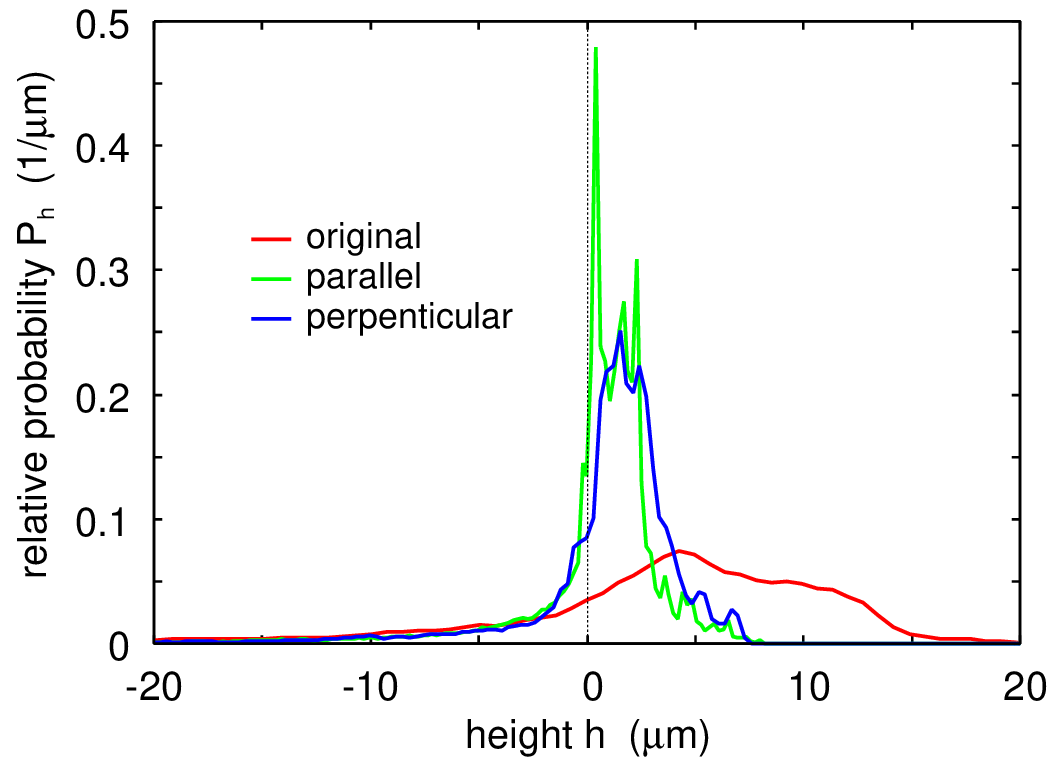}
\caption{
The surface height probability distributions deduced from the line scans shown in Fig.~\ref{1x.2h.all3.eps}.
}
\label{1h.2Ph.eps}
\end{figure}

\begin{figure}[tbp]
\includegraphics[width=0.45\textwidth,angle=0]{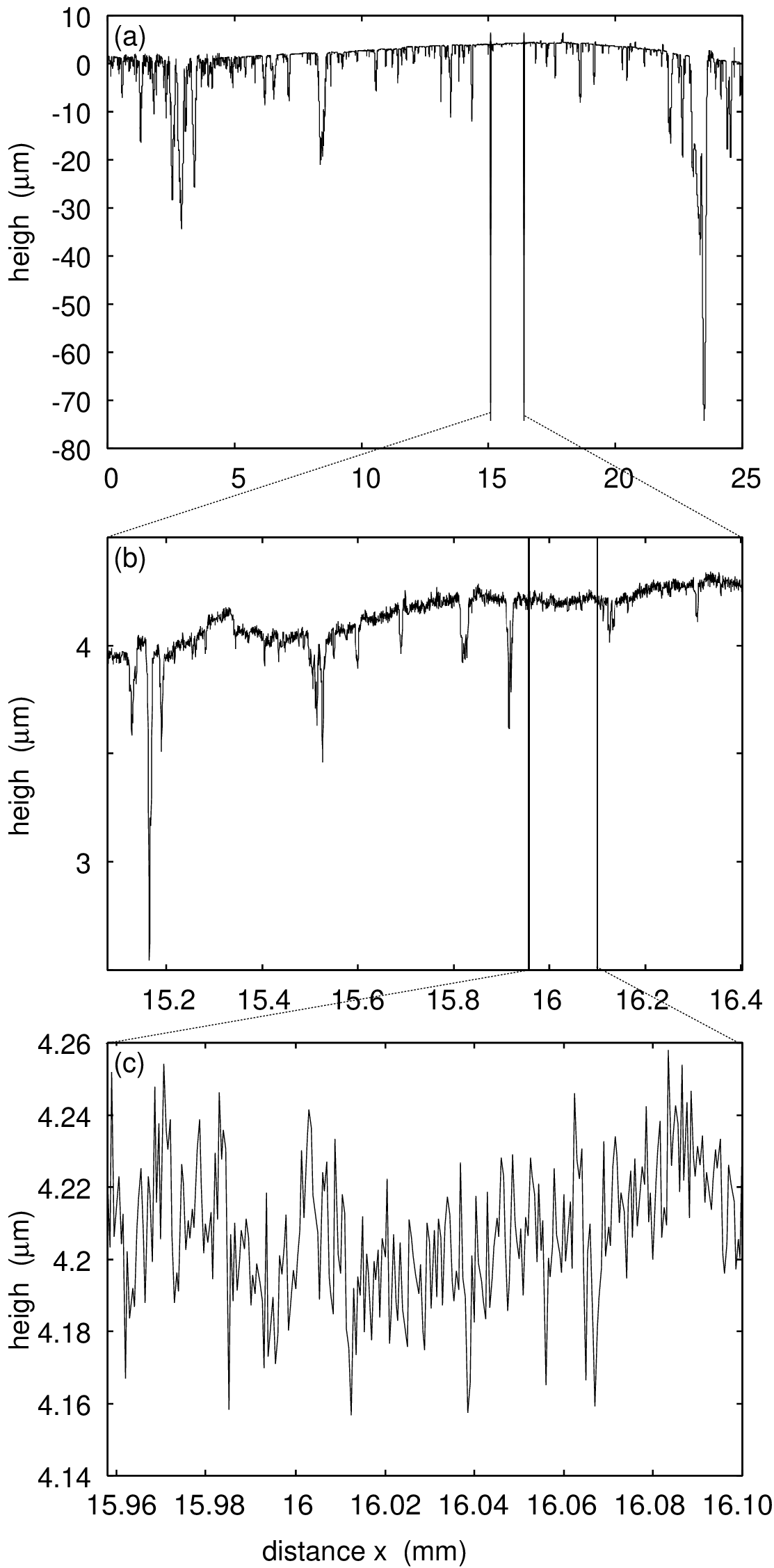}
\caption{
(a) The line topography along the sliding direction after sliding $ 300.8 \ {\rm m}$
(green line in Fig. \ref{1x.2h.all3.eps}).
(b) Magnified segment from (a). (c) magnified segment from (b). Note that the roughness amplitude
in (c) is of order $10 \ {\rm nm}$.
}
\label{parallel0.eps}
\end{figure}

\begin{figure}[tbp]
\includegraphics[width=0.45\textwidth,angle=0]{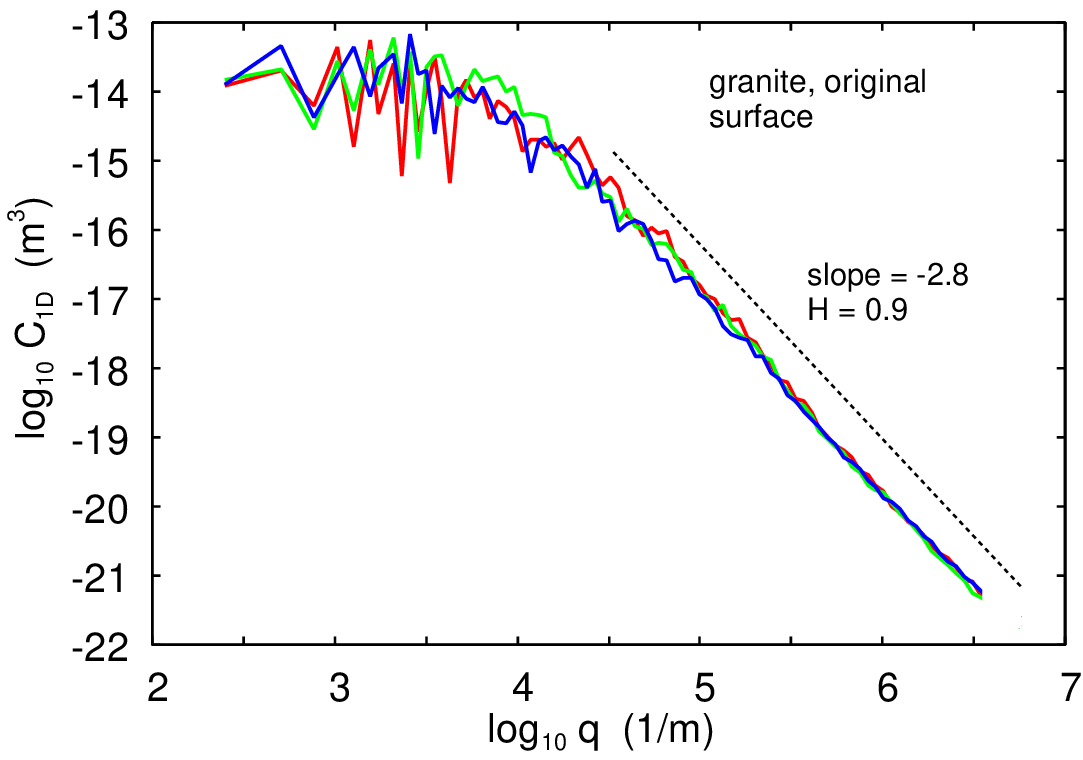}
\caption{
The surface-roughness power spectra $C_{\mathrm{1D}}(q)$ as a function of wavevector $q$ for the original surface, evaluated along three different tracks at different angular orientations.
The surface exhibits isotropic and translational invariant statistical properties.
}
\label{ORIGINAL.1logq.2logC.eps}
\end{figure}

\begin{figure}[tbp]
\includegraphics[width=0.45\textwidth,angle=0]{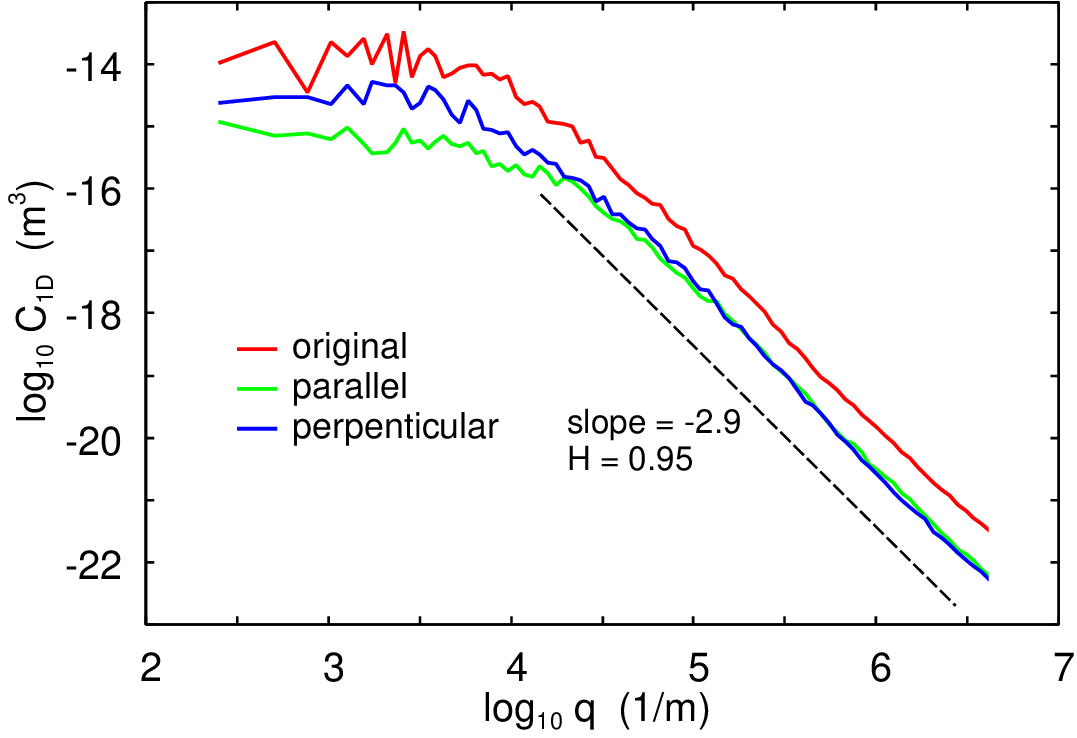}
\caption{
The surface roughness power spectra of the original surface (red line), and after sliding $300.8 \ {\rm m}$
(green and blue lines). The green line is the power spectrum along of the roughness profile along the sliding direction
and the blue line orthogonal to the sliding direction. The root-mean-square roughness are $16.8$, $7.1$ and $5.1 \ {\rm \upmu m}$
for the red, blue and green case, respectively.
}
\label{1logq.2logC.eps}
\end{figure}

\begin{figure}[tbp]
\includegraphics[width=0.45\textwidth,angle=0]{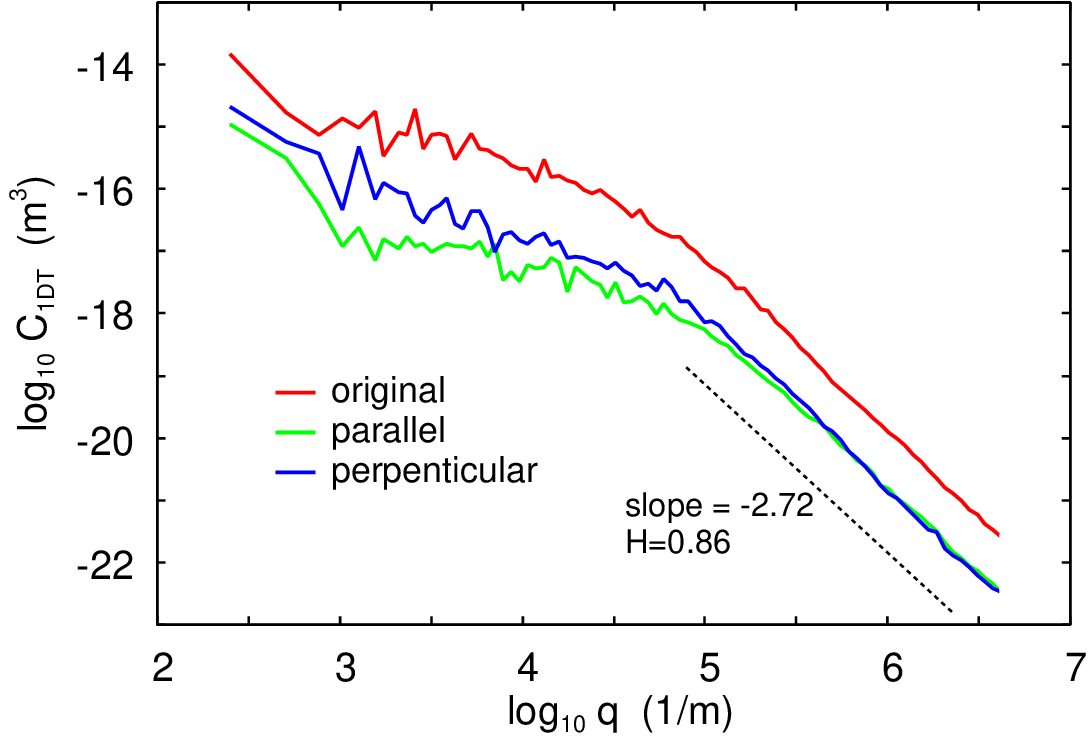}
\caption{
Similarly to Fig.~\ref{1logq.2logC.eps}, but this time the top, rather than the full, power spectra are shown.
For the top spectra, the root-mean-square roughness values are $6.47$, $2.37$, and $1.88\,\upmu\mathrm{m}$ for the red (original), blue (perpendicular), and green (parallel) cases, respectively.
}
\label{1logq.2logC1D.top.eps}
\end{figure}

\subsubsection{Line topography and power spectra}

We have measured the line topography on the 0.2~m long granite sliding track before and after sliding.
Fig.~\ref{1x.2h.all3.eps} shows the line topography of the original surface (red line), and after having slid the distance of 300.8~m (green and blue lines). 
The top of the surface is much smoothened through the friction experiments, however,  the thickness of the removed layer of granite (about 5 µm) is too small to remove the deep wells or cavities of the original surface.
In fact, when overlaying the height histograms resulting from the respective curves, which is done in Fig.~\ref{1h.2Ph.eps}, 
the valleys superimpose nicely for all cases, while the tail associated with large height is substantially reduced after the friction experiments.
Thus, differences between them reveal predominantly stochastic differences, but also some deterministic differences, which arise from (normal) pressures not being perfectly isotropic in the given geometry/loading condition.
Note that Fig.~\ref{1x.2h.all3.eps} is not true to scale so that slopes are magnified by a factor of $\approx 100$.

An interesting observation in Fig.~\ref{1x.2h.all3.eps} is that about 10-15\% of the scan line can be associated with deep valleys.
This is roughly the fraction of feldspar in granite.
Thus, it seems clear that the feldspar is worn off, while the silica 
particles standing proud of the mean surface get polished. 

Although the multi-scale nature of tribological surfaces is well known~\cite{Majumdar1990Wear,Jacobs2017STMP,Persson2014TL}, including that of tectonic plates~\cite{Brown1985JGR, Power1987GRL, Candela2012JGR}, we show the height scan of our default granite surface at different resolution in Fig.~\ref{parallel0.eps} covering the lateral range of (a) 25~mm, (b) 1.3~mm, and (c) 0.14~mm. 
The scan of the worn surface parallel to the sliding direction in panel (a) reveals that the smooth areas are even smoother in the center of the scan than at the edges.
The roughness on the smooth zone called out in panels (b) and (c) extends over a lateral range a little more than 1~mm, on which it shows standard deviations of about $\lesssim 20$~nm. 
This is approximately a third of the typical size of the wear particles shown in Fig.~\ref{fig:gouge}(c).

As argued earlier, it is likely that the granite surface wears non-uniformly, with hard quartz grains extending
above the average surface plane while feldspar recedes.
Thus, regions with very smooth topography (rms roughness of order $10$--$30$~nm), such as in Fig.~\ref{parallel0.eps}, are likely to consist of a “polished” quartz grain rather than a mixture of the two minerals. (The high polishing of the zoomed-in region in Fig. \ref{parallel0.eps}
is due to a slight curvature of the surface giving rise to the highest contact pressure in this region of the substrate.)
Such roughness amplitudes are far below the wavelength of visible light, while the lateral extent of the flat patches --- quartz grains being typically a few mm --- is orders of magnitude larger.
In the absence of strong subsurface scattering, this geometric scale separation explains why the surface patches~\cite{Fondriest2013G} look optically shiny/specular.

Figs.~\ref{ORIGINAL.1logq.2logC.eps}, \ref{1logq.2logC.eps}, and \ref{1logq.2logC1D.top.eps} contain surface roughness power spectra pertaining to the line scans shown in Fig.~\ref{1x.2h.all3.eps}.
Specifically, Fig. ~\ref{ORIGINAL.1logq.2logC.eps} shows the surface roughness power spectrum of the original surface along three sliding directions, 
Fig.~\ref{1logq.2logC.eps} the average of those three spectra (red line) and parallel (green line) as well as orthogonal (blue line) to the sliding direction after sliding 300.8~m.
Values of the root-mean-square roughness are 16.8, 7.1, and 5.1~µm for the original, the orthogonal, and the parallel direction, respectively.
The power spectra along the sliding direction and orthogonal to the sliding direction are similar for 
$q > 1.6 \times 10^4~\rm{m}^{-1}$, 
but for smaller wavenumbers, the roughness orthogonal to the sliding direction is larger than along the sliding direction. 
Similar results have been observed for earthquake faults by Brodsky and coworkers~\cite{Brodsky2011EPSL}.

Fig.~ \ref{1logq.2logC1D.top.eps} shows the surface roughness top power spectrum of the original surface (red line) and after sliding, along the orthogonal (blue line) and parallel (green line) sliding directions. 
The associated root-mean-square roughnesses are 6.47, 2.37 and 1.88~µm for the red, blue and green cases, respectively.
Top power spectra are obtained by considering only the roughness profile above the average plane, while replacing below-plane roughness with values having the same statistical properties as those above. 
This spectrum is appropriate for use in the Persson contact mechanics theory, because for materials as stiff and hard as granite, contact occurs exclusively at asperities above the centroid plane even for elevated nominal contact pressures, while valleys below do not contribute.

\subsection{Simulation results}
\label{sec:simulation}

In our simulations of quartz sliding against a rigid counterbody, contact was first established by moving the indenter downward with a velocity of $v_z = -10$~m/s.
The primary property of interest in indentation experiments is the normal force as a function of displacement $d$, or, in continuum mechanics, as a function of overlap.
In large systems, displacement and overlap are identical up to an additive constant, which can be eliminated by subtracting this constant from the displacement.
In our finite, periodically repeated system, we define both quantities to be zero when the normal position of the rigid indenter is 1.5~Å---roughly the equilibrium separation between silicon and oxygen atoms---above the undeformed quartz surface.
This configuration defines a reference distance between the tip ($z_\mathrm{t}$) and the bottom layer of quartz ($z_\mathrm{b}$), which is subtracted from subsequent values of $z_\mathrm{t} - z_\mathrm{b}$.
Our procedure accounts for the fact that the elastic coupling between the surface and the bottom layer of quartz is macroscopically large in the experiments but finite in a finite, periodically repeated simulation cell.

\begin{figure}[hbtp]
\includegraphics[width=0.45\textwidth]{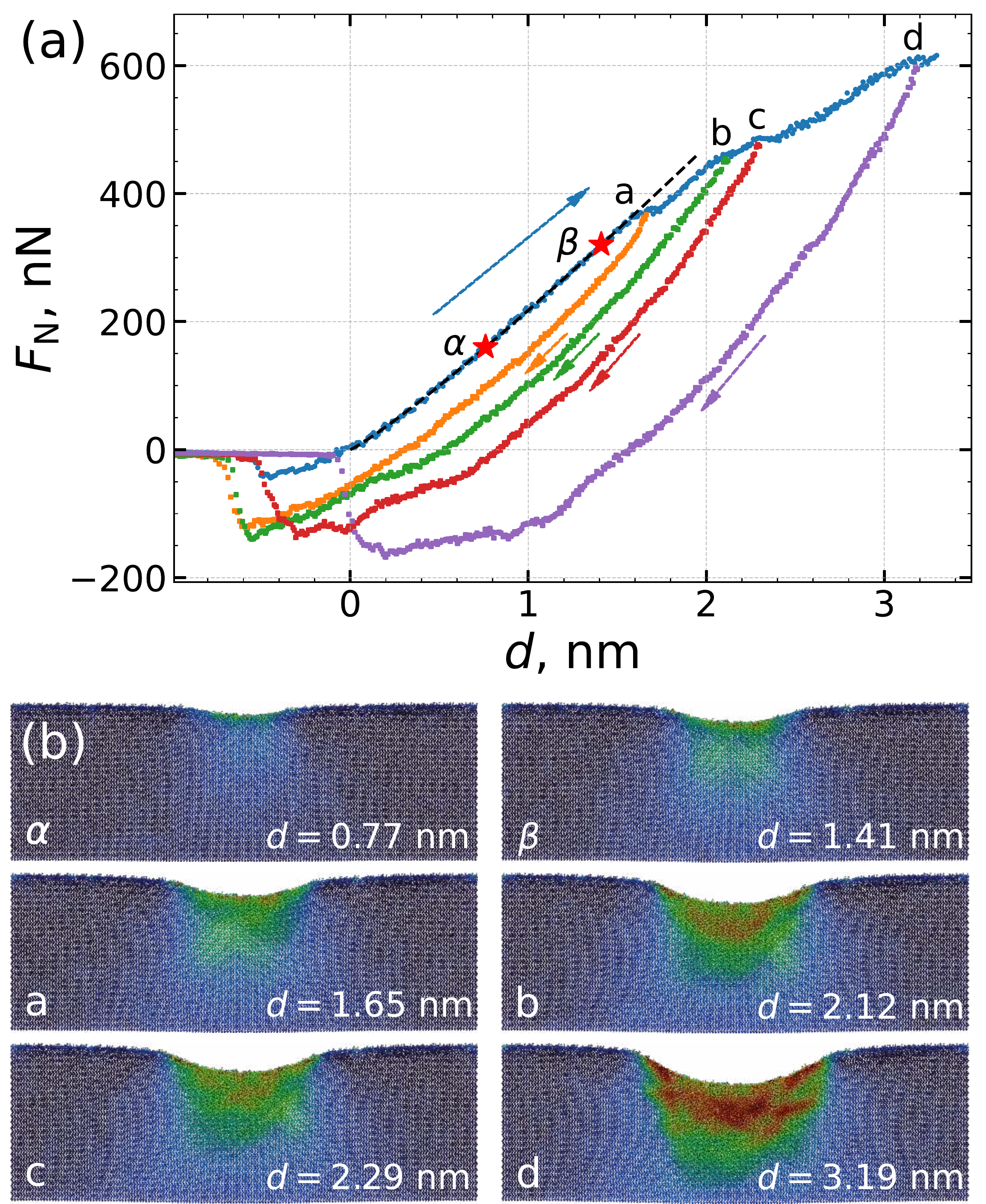}
\caption{\label{fig:indentation}
(a) Nominal force $F_\mathrm{N}$ as a function of indentation depth $d$ for one approach curve (blue dots) and several retraction curves, each starting from a point labeled a (orange triangles), b (green diamonds), c (red squares), or d (purple stars). The black dashed line is a power-law fit, $F_\mathrm{N}/\mathrm{nN} = 217.5~(d/\mathrm{nm})^{1.124}$, to the initial elastic regime of the approach curve.
(b) Configurations corresponding to the labeled points along the approach curve in (a). Atoms are colored according to their von Mises strain, using the undeformed sample as reference. 
}
\end{figure}

Fig.~\ref{fig:indentation}(a) shows the normal force as a function of the indentation depth.
The initial part is well fit by a power law $F_\mathrm{N}/\mathrm{nN} = 217.5\times(d/\mathrm{nm})^{1.124}$ up to $d_\mathrm{a}=16.5$ Å.
At that point there is a first apparent discontinuity, followed by a stronger second discontinuity at $d_\mathrm{c}=22.9$ Å.
The character of the relation $F(d)$ changes between these two discontinuities at $d_\mathrm{b}=21.2$ Å.
The maximum indentation is reached at $d_\mathrm{d}=31.9$ Å.
A brief comment on the non-integer exponent is in place:
because a cylinder, rather than a sphere, indents the surface, the force–indentation relation in a finite simulation cell contains quasi-logarithmic terms that do not admit a simple closed form.
They are effectively emulated by small corrections to an exponent near unity.

Selected configurations of the indentation process are displayed in Fig.~\ref{fig:indentation}(b).
For each configuration, the atoms are colored corresponding to their von Mises strain, and the reference configuration is chosen to be the undeformed solid.
The load-displacement relation shown in Fig.~\ref{fig:indentation}(a) indicates that there is no significant plasticity in the initial regime, e.g., for the points labeled ''$\alpha$'' and ''$\beta$''.
At the point marked ''a'', the sample experiences the first significant plastic deformation.
As the penetration depth increases, the plastic deformation also penetrates deeper into the $\alpha$-quartz block.
The following mean contact stresses were identified at the marked points of penetration:
$\alpha$: 7.6, $\beta$: 11.4, a: 11.6, b: 13.0, c: 12.7, d: 14.0~GPa.
Thus, noticeable plasticity sets in at 11.6~GPa (point a) and massive plasticity at about 14.0~GPa (point d) upon pure indentation. 
These results are in excellent agreement with the established literature values in the range of 10--12~GPa for quartz~\cite{Whitney2007AM,Strozewski2021JGR}, despite our relatively high indentation rates. 

\begin{figure}
    \includegraphics[width=0.4\textwidth]{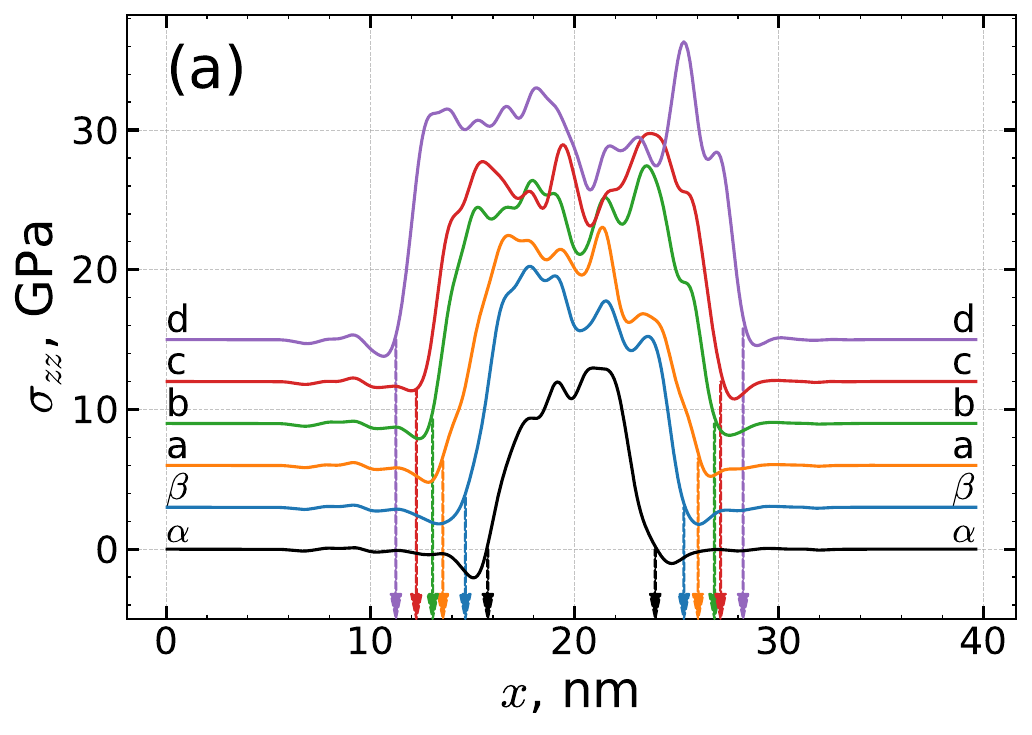}
    \includegraphics[width=0.4\textwidth]{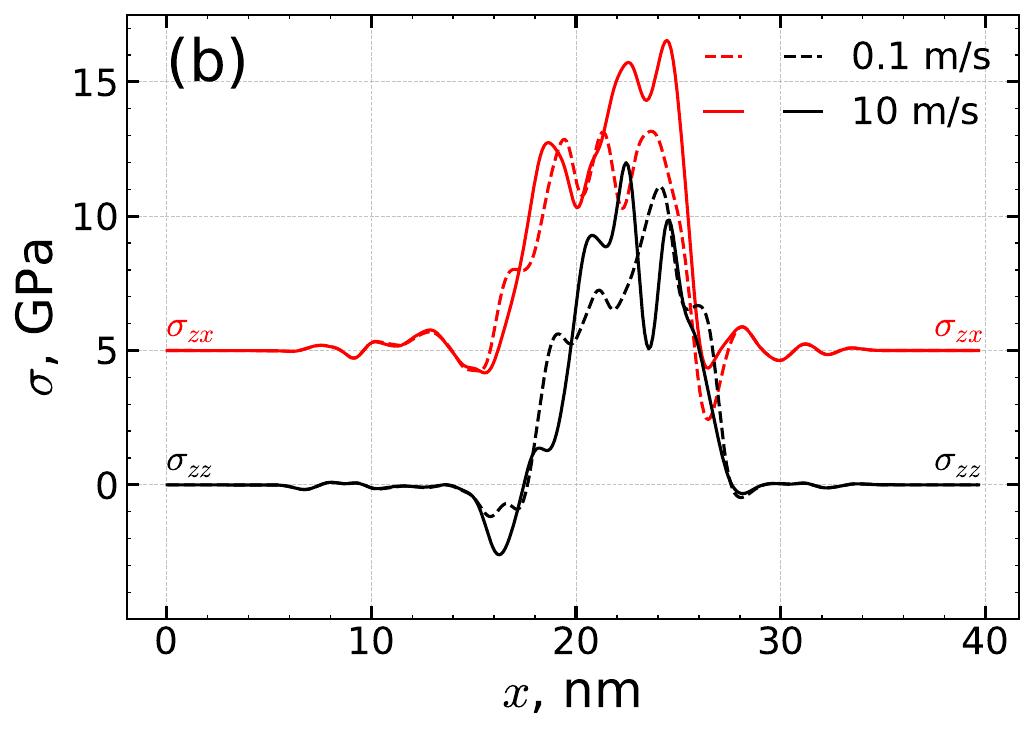}
    \caption{\label{fig:stress_profiles}
    (a) Normal stress $\sigma_{zz}$ profiles for different indentation depths with labels introduced in Fig.~\ref{fig:indentation}.
    The estimated locations of contact edges are marked by arrows.
    The zero-line for stress is shifted by 3~GPa from one curve to the next.
    (b) Normal stress $\sigma_{zz}$ (black lines) and shear stress $\sigma_{zx}$ (red lines) profiles for sliding simulations at $s = 16$~nm for $v_x = 10$~m/s (solid lines) and $v_x = 0.1$~m/s (dashed lines) at a normal load $F_\mathrm{N} = 160$~nN. The zero-line for shear stress is shifted up 5~GPa for both velocities.
}
\end{figure}

Spatially resolved stresses during nanoindentation are shown in Fig.~\ref{fig:stress_profiles}(a).
At the lowest investigated normal force, the shape of the stress profile is reminiscent of a Hertzian contact.
Fluctuations are relatively large due to the small size and because individual oxygen atoms above the quartz surfaces leave large signals. 
Even before the onset of appreciable plasticity, i.e., at the point marked by $\beta$, a plateau near 15 GPa develops in the center of the contact.
At the highest investigated force, this plateau is fully developed.
The large fluctuations reflect the small system size and arise from individual oxygen atoms that protruded above the original surface before contact with the rigid indenter.

\begin{figure*}
    \includegraphics[width=0.95\textwidth]{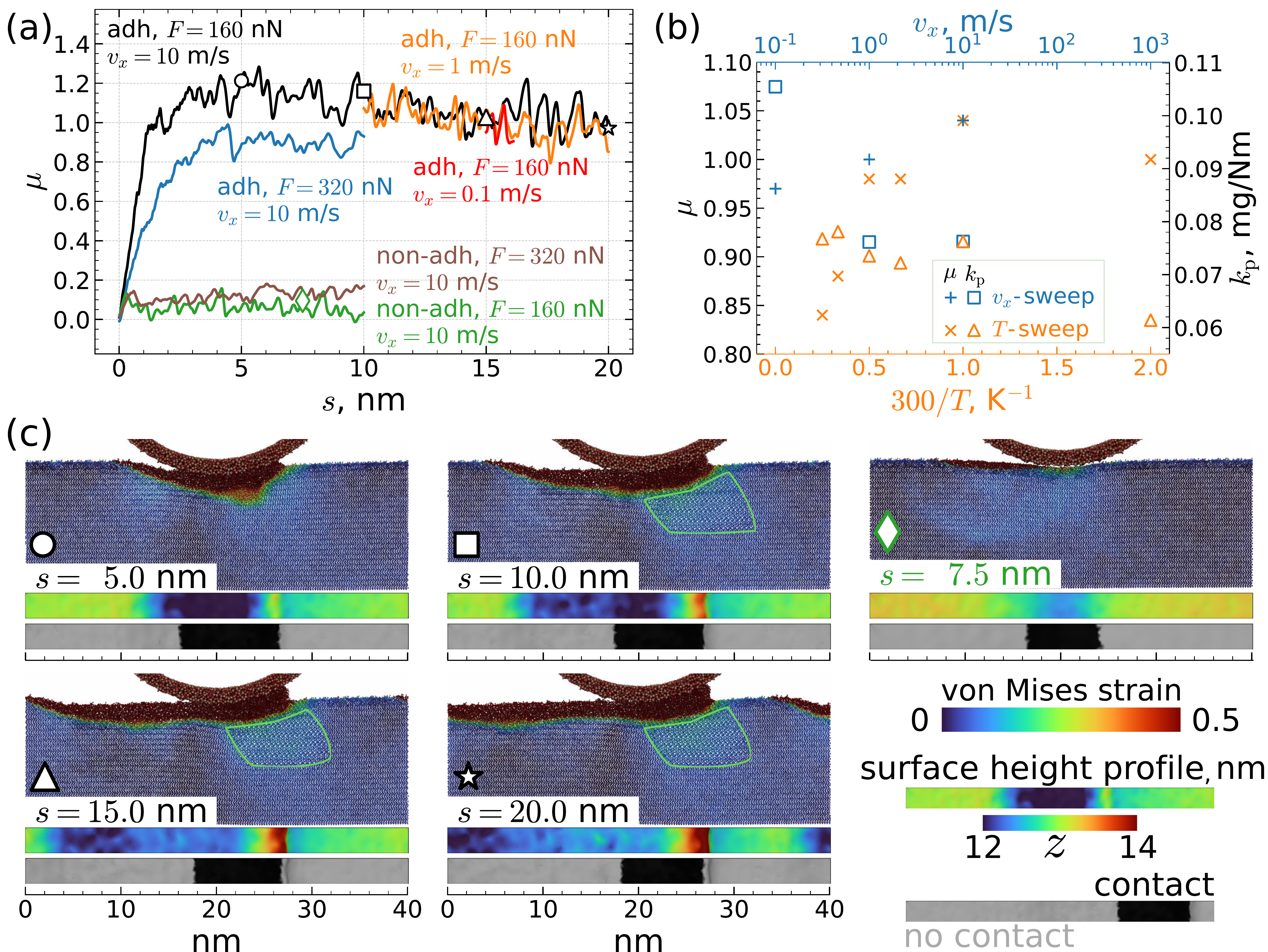}
    \caption{\label{fig:friction_coefficient}
    (a) Friction coefficient $\mu(s)$ for quartz sliding at $v_x=10$~m/s under $F_N=160$~nN (black). 
    Orange and red: continuation runs with reduced $v_x=1$ and $0.1$~m/s, respectively, at the same $F_N$.
    Blue: $v_x=10$~m/s at $F_\mathrm{N}=320$~nN.
    Green and brown: non-adhesive contacts at $v_x=10$~m/s with $F_N=160$ and $320$~nN, respectively.
    All curves were smoothed using a Gaussian filter of width $\Delta s = 0.4$~\AA.
    (b) Left $y$-axis: average friction coefficient as a function of sliding velocity (blue pluses) and as a function of inverse temperature scales to 300~K (orange crosses). 
    Right $y$-axis: plastification rate $k_\mathrm{p}$ as a function of sliding velocity (blue squares) and as a function of inverse temperature scales to 300~K (orange triangles). 
    The velocity sweep (blue symbols) is performed at a fixed temperature $T=300$~K and the temperature sweep (orange symbols) at a fixed sliding velocity $v_x=10$~m/s.
    (c) Snapshots at selected $s$ with surface height profiles and contact/no-contact maps (distance criterion). 
    Zones boxed in green mark regions where a (reversible) stress-induced phase transformation occurred.
}
\end{figure*}

We next run simulations at 300~K and at the load of the point marked by $\alpha$ but at a constant sliding velocity of $v_x = 10$~m/s with an initial configuration, in which the tip and the substrate barely touched.
This set-up will be referred to as \emph{default} in the following.
Fig.~\ref{fig:friction_coefficient}(a) reveals friction coefficients close to $\mu = 1$, which is similar to those found experimentally.
When reducing the velocity by a factor of ten to $v_x = 1$~m/s or even by a factor of one hundred to $v_x = 0.1$~m/s, the friction coefficient remains almost unchanged, as evidenced in Fig.~\ref{fig:friction_coefficient}(a,b).
However, it drops by a factor of ten when switching off the adhesive interactions between the tip and the substrate.
This result suggests that the adhesive interactions induce cold-welding and subsequent plasticity causing significant instabilities and thus dissipation. 

Although the net friction force changes little between $v_x = 0.1$~m/s and 10~m/s, the local stress profiles, see Fig.~\ref{fig:stress_profiles}(b), differ markedly.
At the lower sliding velocity, the normal stress has evolved toward an approximately constant plateau, whereas at the higher velocity a slightly skewed, Hertz-like pressure profile is found.
This behavior can be rationalized by noting that quartz is not given enough time to yield beneath the rapidly moving indenter, which is consistent with the experimental observation that wear rates are greater at small than at large velocities.
Moreover, the shear stress almost mirrors the normal stress as in extended silica-silica interfaces~\cite{Li2014TL}.
This behavior is reminiscent of that predicted by generic models of boundary lubricants~\cite{He1999S,Muser2001PRL}.

To further illuminate the processes leading to the observed frictional behavior, Fig.~\ref{fig:friction_coefficient}(c) depicts configurations at selected sliding stages.
For adhesive contact, the upper layers of the block amorphize in front and under the sliding indenter, and a substantial amount of the amorphized material is dragged in front of the tip.
The amount of piled-up material at the leading edge of the indenter increases with sliding distance, as evident from the red regions in the surface-height profiles, which become progressively larger and more intense. 
This process creates a ``wear track''.
For the non-adhesive contact at the same load, the plastically deformed zone is almost negligible, and the contact is substantially more symmetric than in the adhesive case.
This is evident from the configuration marked with a green diamond in Fig.~\ref{fig:friction_coefficient}(c), which exhibits a relatively symmetric von Mises strain and height profile.
We thus conclude that the main contribution to friction in our simulations must come from cold-welding sites, which lead to significant plasticity at the surface.

In sliding simulations with adhesive contact, we observe a zone where $\alpha$-quartz transforms into a different crystalline phase, which was identified as tridymite, a polymorph of silica. 
Specifically, the symmetry of the sheared structure turned out to be P$2_1$ (space group 4), while allowing the angles to relax (in bulk simulations of this phase) leads to a higher symmetry, P$2_12_12_1$ (space group 19).
It is often referred to as low-temperature OP tridymite, where OP stands for orthorhombic-primitive.
Experimentally, this phase has a stability range of $110--150^\circ$C at ambient pressures~\cite{Kihara1977ZKCM, Graetsch1998AM}.

The region that underwent the phase transformation is boxed in green in Fig.~\ref{fig:friction_coefficient}(c).
It moves along with the indenter, slightly in front of it while maintaining nearly constant size as sliding progresses.
This zone is dynamic, and thus some energy must be spent to transform $\alpha$-quartz to tridymite at the leading edge of the zone and a certain amount of energy is released as heat during the reverse transformation. 
As a result, this hysteresis contributes to friction.
However, the creation of such a zone also strengthens the material against shear-induced amorphization, leading to a smaller contact area through reduction of the built-up material at the leading edge.
We note that the size of the tridymite zone grows with normal load and temperature.
It disappears below a critical load, which happens to be slightly below the used default value of $F_\mathrm{N} = 160$~nN.

\begin{figure}
    \centering
    \includegraphics[width=0.95\linewidth]{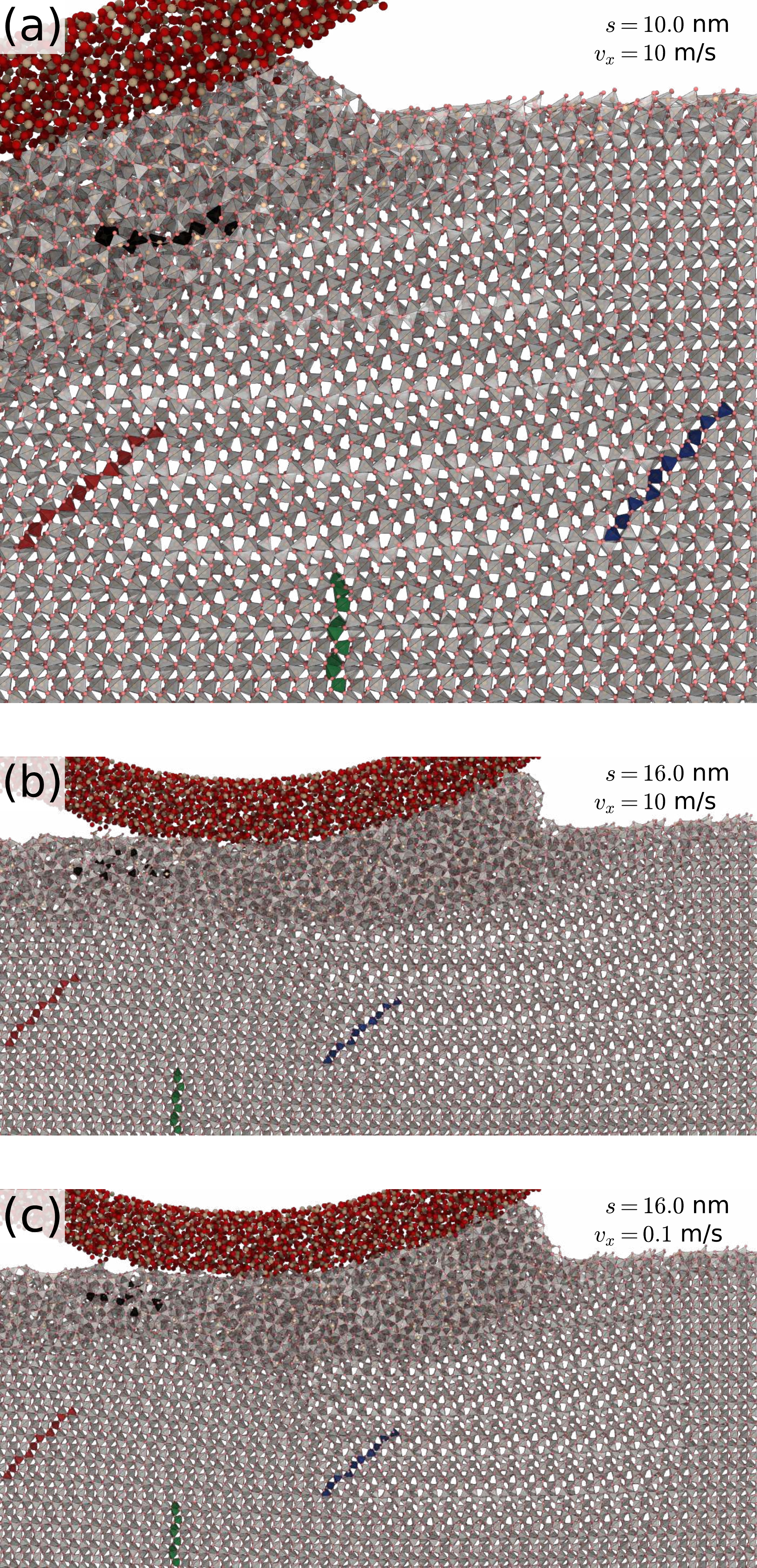}
    \caption{Zoom into the structure of the default configuration at a sliding distance of (a) $s = 10$~nm and (b) at $s = 16$~nm and (c) at a reduced velocity of $v_x = 0.1$~m/s again at $s = 16$~nm. For lines of ten (initially) connected tetrahedra are marked in color allowing their motion to be tracked.}
    \label{fig:config}
\end{figure}

A zoom into the atomistic structure of the simulated system is shown in Fig.~\ref{fig:config}.
Atoms belonging to the amorphous tip are represented as simple spheres, while coordination tetrahedra (SiO$_4$ units) belonging to the substrate are depicted as tetrahedra.
Four rows of initially connected tetrahedra, ten in total, are highlighted in color.
These rows remain connected during the conversion from $\alpha$-quartz to tridymite and back to $\alpha$-quartz.
In contrast, the single row of tetrahedra that becomes part of the amorphized region ends up disconnected.

An animation of the dynamics is available in the supplementary information.
It shows that tridymite begins to nucleate beneath the near-surface region without passing through any other phase (including $\beta$-quartz) and then grows inward into the substrate.
At the moment of its first appearance, both the potential energy and the instantaneous friction coefficient exhibit a small dip.
The observed dynamics suggest a shear-driven displacive transition path between $\alpha$-quartz and OP tridymite that bypasses the intermediate phases typically accessed during purely thermal transitions between these structures.
This finding strongly supports the idea that processes often interpreted as ``frictional melting'' or high-temperature phase transitions of silicates can be driven by shear at substantially lower temperatures than those required under isotropic stress~\cite{Lee2017NG,Woo2023SR}.  
Thus, our finding contrasts with the classical explanation, which attributes the presence of tridymite and other phases in post-mortem faults~\cite{Jackson2010AM} or crushed $\alpha$-quartz~\cite{Martinelli2020M} to high-temperature phase transitions during extreme stress.

Another structural feature worth discussing is the difference in the geometry of amorphized silica at $v = 10$~m/s and 0.1~m/s sliding velocity.
At the higher velocity, a bulge forms with a shape somewhat reminiscent of a compressed elastomer, see Fig.~\ref{fig:config}(b).
This bulge is no longer visible at the lower velocity in panel (c), as if the silica had enough time to ``flow'' and reduce the surface energy.
This result is surprising because the imposed temperature was 300~K, while flash temperatures (discussed further below) are only about 100~K higher.

We also varied the simulation temperature $T$ in a range from 150 to 1,200~K by conducting similar continuation runs as those just discussed.
The results presented in Figure~\ref{fig:friction_coefficient}(c) reveal a rather weak $\mu(T)$ dependence up to 600~K. 
This insensitivity can be rationalized by the argument that amorphization is driven mechanically rather than thermally, as in other tetrahedral network formers far below the melting temperature~\cite{Pastewka2010NM,Moras2018PRM,Atila2025PRL} so that an interfacial disordered zone can be rapidly generated despite being far away from melting point conditions.
The newly generated amorphous silica is essentially solid even at 600~K with extremely small crystallization rates and no (experimentally accessible) viscous flow.
Its plastic deformation is thus expected to be in the strongly shear-thinning regime, that is, it proceeds via localized structural rearrangements over high energy barriers so that thermal activation and thus temperature dependence are weak.
Thus, not only the generation of amorphous material but also the shear response of the amorphized zone are insensitive to temperature, unless shear rates are  extremely small.

At temperatures exceeding 600~K, above the $\alpha-\beta$-quartz phase-transformation temperature, which is about $T^* = 846$~K for real silica~\cite{Carpenter1998AM} and 750~K for \textit{in-silico}-BKS silica~\cite{Muser2001PCM}, there is a marked decline of the friction with increasing temperature.
A possible reason for the decrease in friction could be that the more highly symmetric $\beta$-quartz is stiffer than $\alpha$-quartz, making it less prone to deformation.  
In addition, thermal activation will eventually start to matter at elevated temperature allowing the amorphous zone to be sheared more easily.

While frictional heating at the interface increases the temperature by about 100~K in our default set-up, as discussed further below, the system would remain below $T^*$ when sliding with our default system, for which $T = 300$~K, $v = 10$~m/s, and $F_N = 160$~nN.
Even if temperature went above $T^*$, the $\beta$-phase would have to nucleate during a relatively brief moment of increased temperature.

\begin{figure}
    \includegraphics[width=0.45\textwidth]{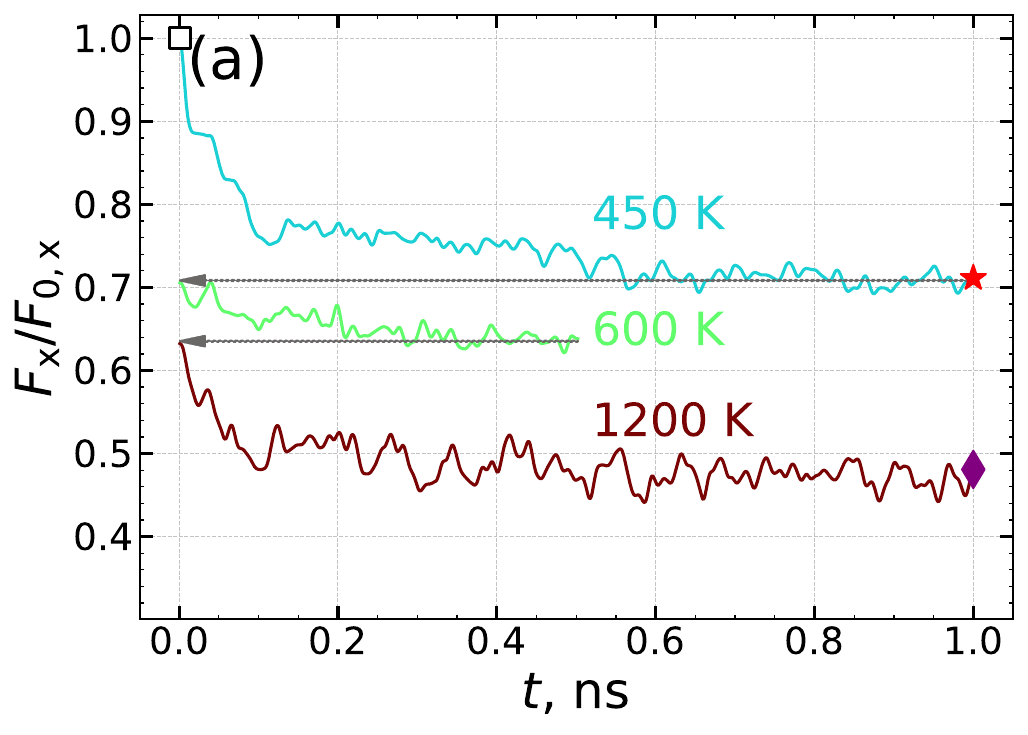}
    \includegraphics[width=0.45\textwidth]{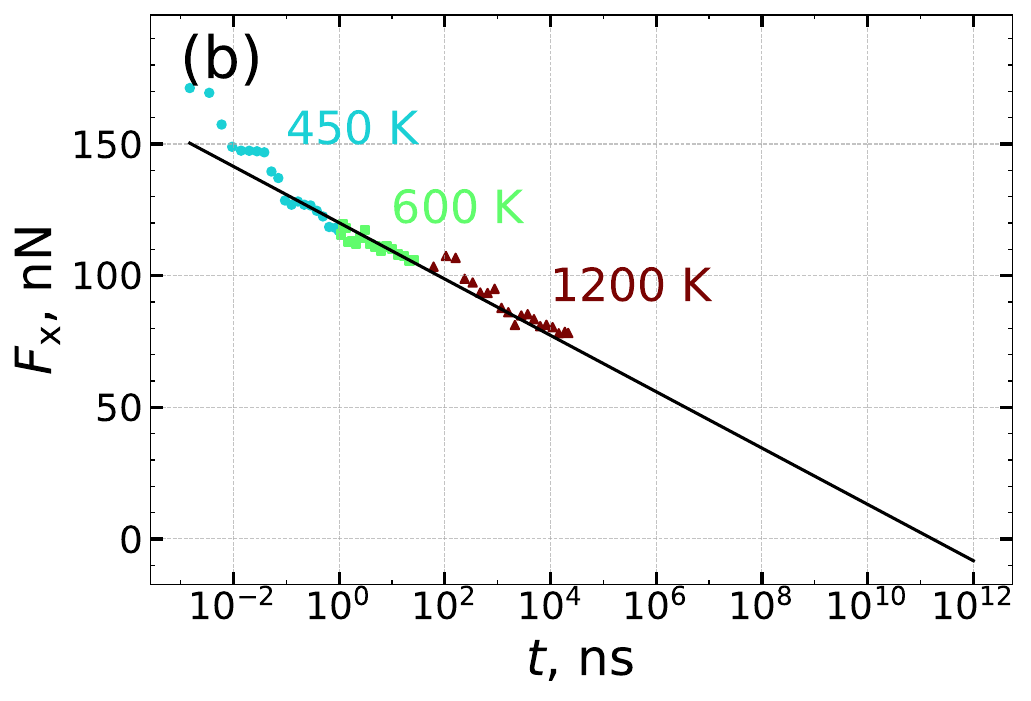}
    \includegraphics[width=0.45\textwidth]{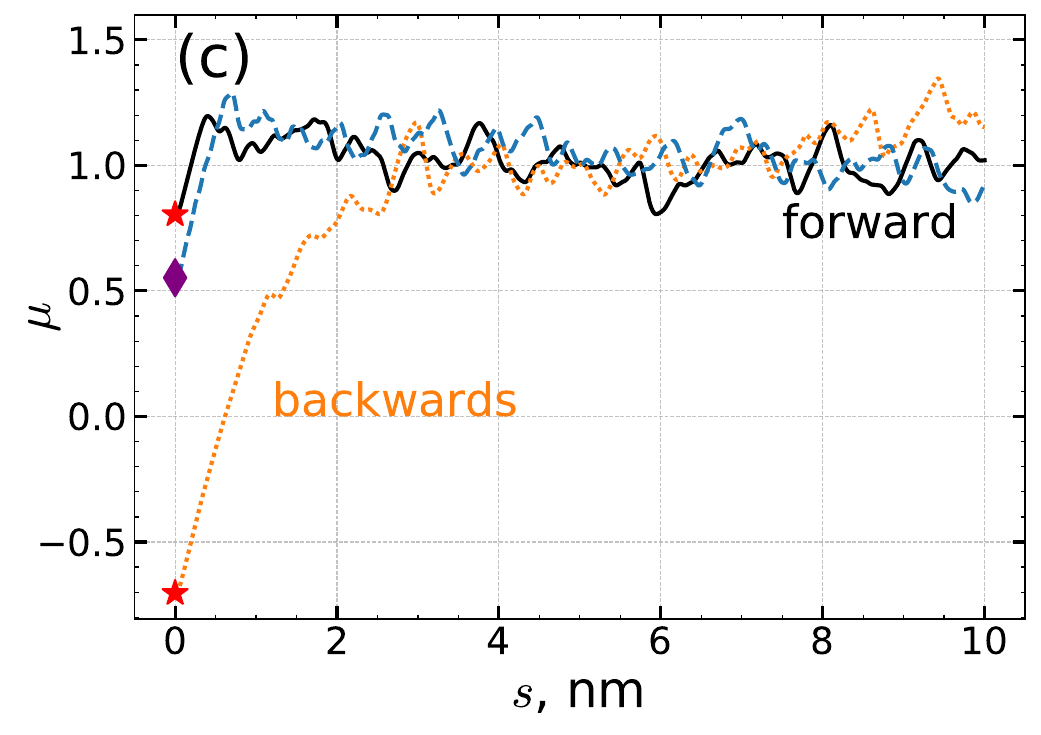}
    \caption{\label{fig:annealing_relaxation}
    (a) Lateral force $F_\mathrm{x}$ relaxation while starting from a configuration that corresponds to the black square in Fig.~\ref{fig:friction_coefficient}(a) (adhesive contact at sliding velocity $v_x = 10$~m/s under a normal load $F_\mathrm{N} = 160$~nN). The annealing was performed in three steps each time increasing temperature.
    (b) Same data as in (a), however, represented as a continuous run, where a time--temperature equivalence principle based on an activation energy of $\Delta E = 0.64$~eV was applied.
    (c) Friction coefficient $\mu$ after reinitiation of sliding at $T = 300$~K starting from the configuration marked by a red star in (a), one time in the same direction as previously (forward, black solid line), one time in the opposite direction (backwards, orange dotted line). Additionally, friction coefficient after reinitiation of sliding at $T = 300$~K starting from configuration marked by a purple diamond in (a) (blue dashed line).
    All curves in (a) and (c) were smoothed using a Gaussian filter of width $\Delta s = 0.4$~\AA.
}
\end{figure}

The insensitivity of the friction to temperature and velocity does not imply that thermal or structural relaxation is absent in a sliding or resting contact up to 600~K.
In fact, besides the structural relaxation of the bulge / built-up material at low sliding velocity, the shear stress relaxes noticeably when the motion is suddenly halted, even on the short time scales of the simulations, particularly at elevated temperatures, as evident in Fig.~\ref{fig:annealing_relaxation}(a) and even more clearly in its redrawn version, Fig.~\ref{fig:annealing_relaxation}(b).
For the latter, a time--temperature principle was applied such that the progression of time is sped up at high temperatures relative to the reference annealing temperature of $T_{\rm ref} = 450$~K by a factor of $\exp[-(\beta - \beta_{\rm ref})\Delta E]$.
When choosing $\Delta E = 0.64$~eV, the data can be adjusted to relax approximately logarithmically in time.
Using this activation energy, one would infer that the shear stress in the nano-scale tip relaxes after roughly one week when annealed at 450~K, and thus after 75~years annealing time at room temperature.

The above estimate, however, is highly uncertain, as (a) a relatively broad range of $\Delta E$ values around 0.5~eV yields an apparently logarithmic relaxation, (b) it is unclear whether the relaxation remains logarithmic over more than 12 decades in time, and (c) the effective barrier is far below activation energies reported for liquid silica at elevated temperatures, which are about 5.0~eV experimentally~\cite{Doremus2002JAP} and roughly 10\% smaller for BKS-silica~\cite{Horbach1999PRB}.
While stress may well reduce $\Delta E$ for the structural reorganization of disordered silica, an order-of-magnitude reduction would be difficult to justify.
We therefore expect that stress relaxation in our nano-scale contact would take substantially longer than estimated from our oversimplified, yet instructive, time--temperature-equivalence analysis of the accrued relaxation data.

When restarting the simulations from the annealing run at 1,200~K, which relaxes the shear stress to half its original value, a small but noticeable overshoot in the friction coefficient is revealed under the given constant sliding velocity in Fig.~\ref{fig:annealing_relaxation}(c). 
This peak might be substantially larger if sliding were imposed by a weak spring of stiffness $k = 20$~N/m, which is not atypical for atomic-force microscope cantilevers. 

\begin{figure}[hbtp]
\includegraphics[width=0.45\textwidth]{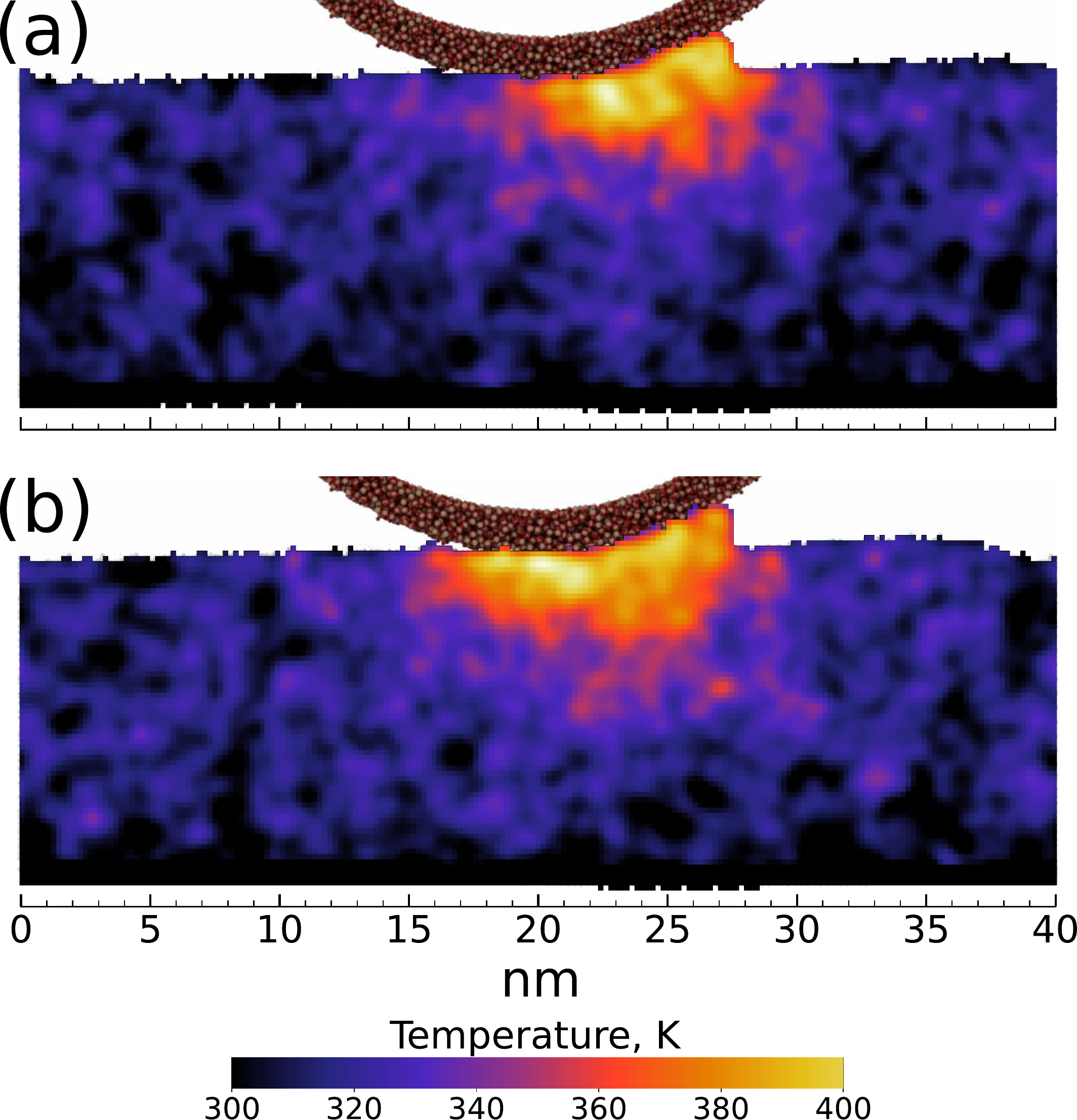}
\caption{\label{fig:local_heating}
Temperature profile in a sliding contact at slid distance (a) $s$ = 17.5~nm, and (b) $s$ = 20.0~nm, both at $v_x$ = 10~m/s, $T$ = 300~K.
}
\end{figure}

Besides friction, we also monitored the amount of material having undergone (massive) plastic deformation.
All atoms being assigned a von Mises strain exceeding 0.5 are counted toward such material.
This allows us to define a plastification rate $k_\textrm{p}$ as the mass of material having undergone plastic deformation normalized by the product of normal load and sliding distance, which has the same unit as the wear rate $k_\textrm{w}$. 
Fig.~\ref{fig:friction_coefficient}(b) reveals that $k_p$ increases with decreasing velocity, which correlates with the experimentally observed trend on the wear rate $k_w$.
The simulated $k_p$ numbers are a little less than an order of magnitude greater than the experimentally measured $k_p$ values. 
However, plastification rates in the simulations and experimental wear rates are expected to differ since the  underlying mechanisms are scale-dependent, as discussed next. 
Nonetheless, our plastification rates correlate nicely with the wear rates obtained in simulations of two colliding $\alpha$-quartz hemispheres having the rather small value of 3.4~nm radius of curvature~\cite{Li2025GRL}.

A recently developed theory~\cite{Aghababaei2016NM} places the transition from atom-by-atom plastification to debris formation at a critical junction size $d^* = \lambda G \Delta\gamma / \sigma_y^2$, where $\lambda$ is a geometric factor ($\lambda \approx 8/\pi$ for our geometry), $G \approx 60$~GPa is the shear modulus, taken as the rough average of $C_{33}$ and $C_{44}$, and $\sigma_y \approx H/3 \approx 3$~GPa is the yield strength, approximated as one third of the indentation hardness $H$.
This gives a critical junction size of $d^* \approx 60$~nm, above which debris forms, which is clearly larger than our simulated contacts and thus consistent with our findings.
For the larger experimental contacts, however, debris formation should dominate.

The sliding-induced instabilities causing Coulomb like friction are the processes that are responsible for local heating and ultimately for local flash temperatures.
They are difficult to visualize because they suffer from large scatter.
This is because instabilities happen relatively rarely on the time-scale of MD simulations and moreover kinetic energies, which are used to determine local temperature via the equipartition theorem, are already quite noisy even without instabilities.
This is why temperature maps in a contact averaged over 1,000 time steps and run through a 5~Å Gaussian blurring filter can look quite different at different moments of time, e.g, depending on whether the last ``major'' instability was bond breaking at the trailing edge or plasticity at the leading edge of the contact. 
As the shape of the contact is non-stationary even in the moving frame of the tip, averaging temperature maps over long times is not an ideal option either.
Nonetheless, we find that the typical temperature raise $\Delta T$ at the very tip-quartz interface at a sliding velocity of 10~m/s is about 100~K, as shown for two representative situations in Fig.~\ref{fig:local_heating}.
The zone where quartz undergoes reversible phase-transformations appears to be always part of the warmest regions. 
As argued in Sect.~\ref{sec:contact_mechanics}, $\Delta T$ is proportional to the product of contact radius and sliding velocity--given an approximately constant shear stress.
This product is roughly similar for the highest velocity used in the experiments and in MD, i.e, $v_{\rm max} = 10$~m/s and $a_c = \mathcal{O}( 10~{\rm nm})$ and in the experiments, $v_{\rm max} = 1$~cm/s and $a_c = \mathcal{O}(10~\textrm{µm})$. 

We conclude the simulation section by supporting our claim that the simulation results are not exclusive to the employed potential.
To this end, we performed continuation runs for the default system ($v=10$~m/s, $s=10$~nm), where ReaxFF~\cite{vanDuin2003JPC,Noaki2023NCM} replaced the BKS~\cite{vanBeest1990PRL} potential (i) without and (ii) with a 200~ps relaxation before further sliding of about 5~nm.
The friction coefficient fluctuated around 0.8 and 1.0, which is $\lesssim 10$\% lower than with BKS and thus still of similar magnitude.
Moreover, the phase-transformed zone did not convert back to quartz.
Instead, we observed an even slightly increased propensity for a phase transformation to occur below the tip, although temperature did not raise compared to the BKS simulation. 

\section{Theory}
\label{sec:theory}

\subsection{Contact mechanics}
\label{sec:contact_mechanics}

In this section, the Persson contact-mechanics theory~\cite{Persson2001PRL,Xu2024IJSS} for elasto-plastic contact~\cite{Persson2001PRL} and interfacial separation~\cite{Almqvist2011JMPS} 
is applied to the experimentally studied systems for clean surfaces, 
with the goal of estimating the true contact area, interfacial separation, typical contact patch sizes, and ultimately local flash temperatures.
Since granite is stiff and hard, two granite blocks will typically only make contact above the average surface plane, which is why the top rather than the full power spectrum should be used in the theory. 
It has in fact been shown to predict highly accurate stresses and interfacial separations in the elastic limit when the relevant stochastic properties of the surface were sampled over the true contact, not only for random~\cite{Almqvist2011JMPS,Dapp2012PRL}, but even for deterministic height profiles~\cite{Muser2021TL}.

The main idea behind the theory is to study the contact at different magnifications $\zeta$ starting at a coarse scale, where microscopic roughness is not yet resolved, the effect of roughness is then added gradually by increasing the magnification $\zeta$, i.e., by including height variations at scales $L/\zeta$~\cite{Persson2001PRL,Almqvist2011JMPS}. 
(Here, spatial features of wavelength $L = 2.5$~cm are resolved for $\zeta = 1$.) 
As $\zeta$ increases, regions that appeared to be in elastic contact at coarse resolution are resolved into smaller patches which may either fall out of contact or enter the plastically deformed part of the interface, and the apparent interfacial separation increases. 
The procedure stops once roughness features have been resolved down to the smallest, i.e., atomic scale. 
For the calculations reported below, the measured power spectrum was multiplied by a factor of two to account for two uncorrelated rough surfaces.

\begin{figure}[htbp]
\includegraphics[width=0.45\textwidth,angle=0]{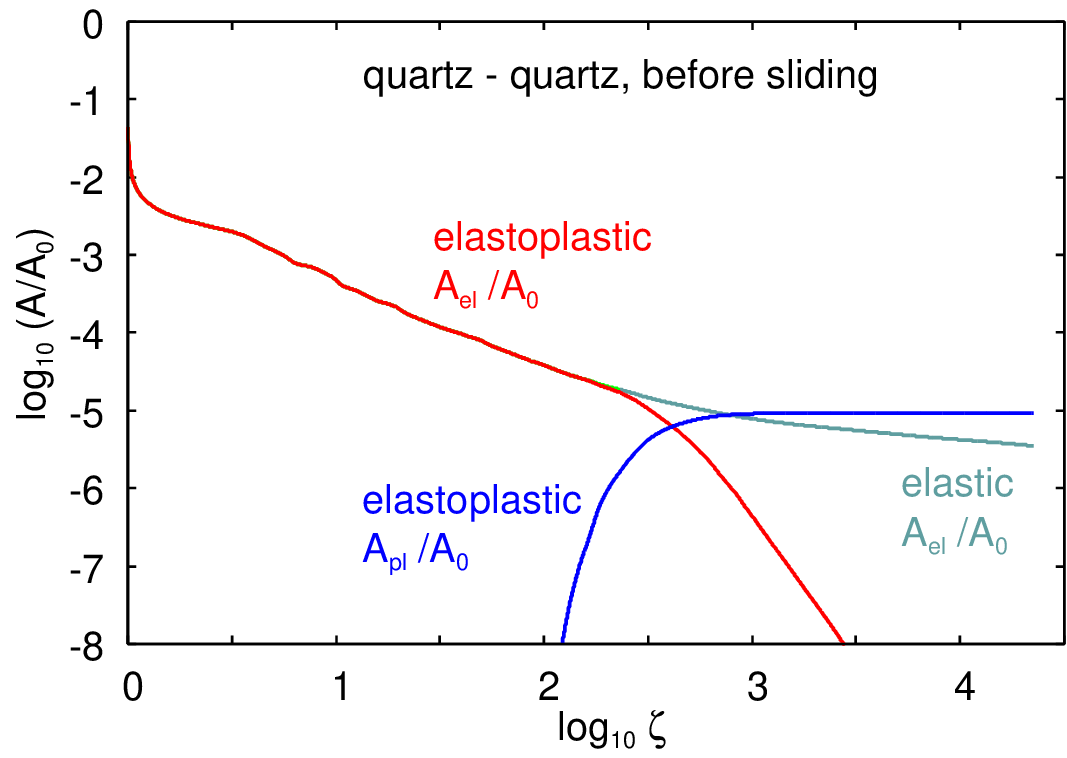}
\caption{
Calculated relative contact area as a function of magnification using the top power spectrum of the original surface (before sliding), scaled by a factor of 2.
The red line shows the elastic contribution, and the blue line the plastically deformed part in an elastoplastic calculation.
The gray line shows the result of a fully elastic calculation.
}
\label{1logZeta.2logArea.Elastoplastic.before.eps}
\end{figure}

\begin{figure}[tbp]
\includegraphics[width=0.45\textwidth,angle=0]{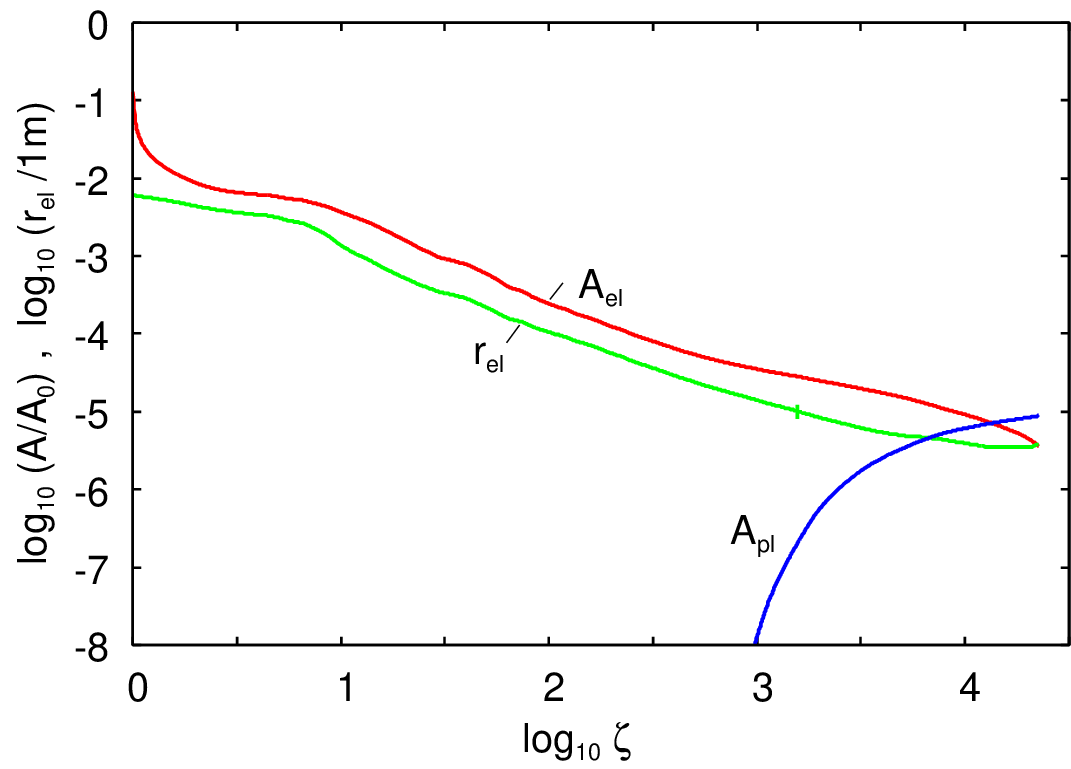}
\caption{
Similar to Fig.~\ref{1logZeta.2logArea.Elastoplastic.before.eps}, but this time using the top power spectrum along the sliding track after a sliding distance of $300.8\,\mathrm{m}$, scaled by a factor of 2.
The characteristic radius of the contact regions is about $3.5\,\upmu\mathrm{m}$ when the magnification is large enough that most contact regions deform plastically.
Nearly identical results are obtained when using the top power spectrum orthogonal to the sliding track.
}
\label{1logZeta.2logAria.logRadie.eps}
\end{figure}

Figs.~\ref{1logZeta.2logArea.Elastoplastic.before.eps} and \ref{1logZeta.2logAria.logRadie.eps} show the evolution of the relative contact area, split into elastically and plastically supported fractions, before and after sliding, respectively.
While, before sliding, the plastically deformed area starts to overtake the elastic contribution when spatial features smaller than $80$~µm are resolved, this cross-over takes place much later after sliding, i.e., near $2.5$~µm.
This is because the mean stress in the elastically supported contact regions scales as $E^* \xi$, where $E^*$ is the contact modulus and $\xi$ 
is the root-mean-square surface gradient, which is reduced after sliding.

\begin{figure}[hbtp]
\includegraphics[width=0.45\textwidth,angle=0]{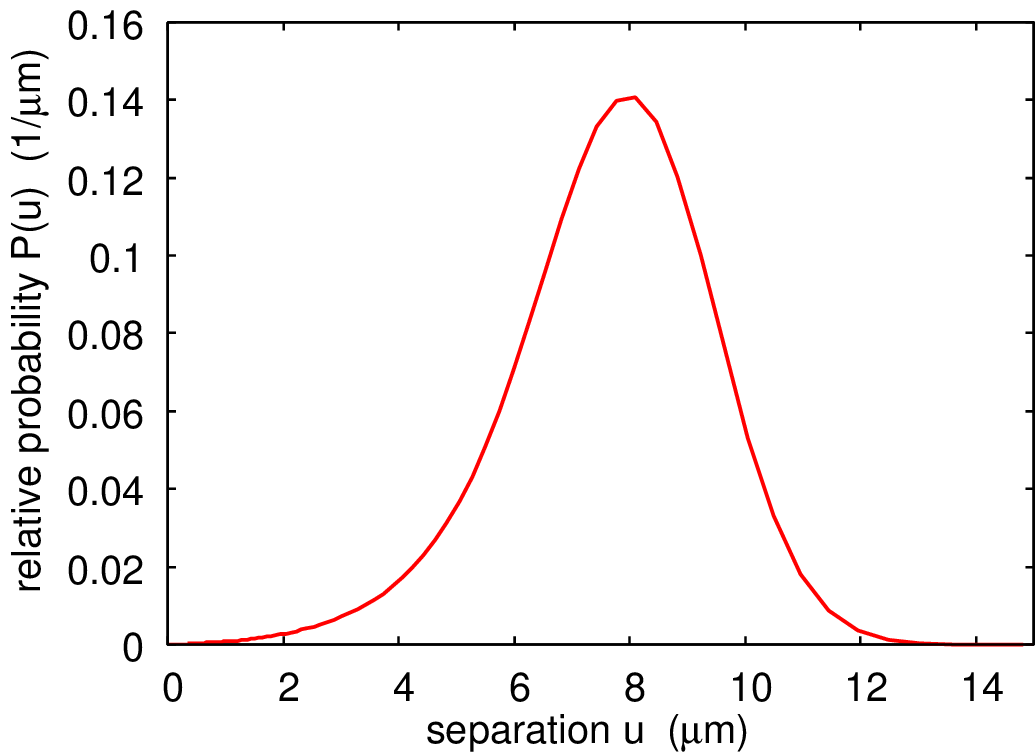}
\caption{
The probability distribution of surface separations. Because of the high elastic modulus
the area of real contact is very small and the surface area where the separation is very small is also very small.
We have used the top power spectrum along the sliding track after sliding $300.8 \ {\rm m}$ scaled by a factor of 2.
}
\label{1u.2Pu.eps}
\end{figure}

Fig.~\ref{1u.2Pu.eps} shows the probability distribution $P(u)$ of surface separations $u$ using the (doubled) top power spectrum.
Because of the high elastic modulus the area of real contact is very small, and the surface area where the 
separation is very small is likewise very small.
Separations are typically well above the size of gouge particles, even without correcting for valleys associated with worn feldspar.
We note that $P(u)$ depends mainly on the long-wavelength roughness components, which deform elastically.
Hence $P(u)$ is only weakly affected by plastic deformation.

Sliding produces frictional heating.
To estimate the resulting flash temperature we first need the characteristic contact size which depends on the magnification $\zeta$.
At low magnification $\zeta$, relatively large and compact contact regions can be observed. An asperity contact region is considered compact if no non-contact 
regions can be observed within it. Generally, an asperity contact region that appears compact at magnification $\zeta$ becomes non-compact 
at the highest magnification $\zeta_1$, often breaking up into several separated regions.

For surfaces with a roll-off region, at low magnification (but not so low that the contact area percolates) and at sufficiently low nominal contact pressure, 
the asperity contact regions are compact and well separated. In this limit, the size of the asperity contact regions is well-defined. 
However, at high magnification it is less clear how to define the size of the asperity contact regions. 
One approach is to use the stress-stress correlation function~\cite{Campana2008PRE,Frerot2018JMPS,Muser2018L,Persson2008JPCMb,Dapp2012PRL,Persson2025JCP_b}.

Consider the stress-stress correlation function
$$g({\bf x},{\bf x}') = 
\langle \sigma ({\bf x}) \sigma ({\bf x}')\rangle - \langle \sigma ({\bf x}) \rangle 
\langle \sigma ({\bf x}') \rangle.$$
For surfaces with isotropic roughness, $g({\bf x},{\bf x}')$ depends only on $r=|{\bf x}-{\bf x}'|$, and we will denote it by $g(r)$.
We define the effective asperity contact radius $r_{\rm c}$ using the condition $g(r_{\rm c}) = \alpha g(0)$, where $\alpha < 1$.
Note that $g(r)$ depends on the range of surface roughness included in the calculation. If $q_0$ and $q_1$ are the smallest and largest roughness 
wavenumbers used in calculating $g(r)$, we include the roughness with wavenumbers $q < \zeta q_0$, where $1 < \zeta < q_1/q_0$. 
Thus, $g(r) = g(r, \zeta)$ will depend on the magnification $\zeta$ and the size of the contact regions will decrease with increasing magnification
Numerical estimates for $r_c$ using this approach~\cite{Persson2025JCP_b} with $\alpha = 0.5$ are included in Fig.~\ref{1logZeta.2logAria.logRadie.eps} 
and yield $r_c \approx 3.5~\upmu{\rm m}$ at the highest resolution.
Using $\sigma_\textrm{p} = 7$~GPa, as suggested by the MD shear simulations, raises the estimated contact radius 
to $r_c \approx 10~\upmu{\rm m}$, in good agreement with experimental observations~\cite{Dieterich1994PAG,Hayward2019JGR}.

\begin{figure} [tbp]
\includegraphics [width=0.25\textwidth,angle=0]{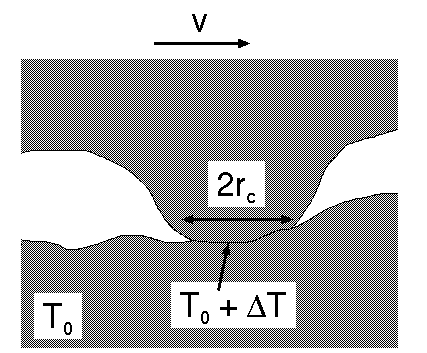}
\caption{
An asperity of the upper solid sliding on the substrate. In the present 
study both material are the same (silica or quartz).
}
\label{HeatSourceTemp.eps}
\end{figure}

To estimate flash temperatures,
consider an asperity on the upper granite block in sliding contact with the lower granite block as in Fig. \ref{HeatSourceTemp.eps}. 
If we assume that the whole frictional energy goes into heat (which is approximately true according to the MD simulations), the maximum temperature in the contact region for the lower solid is given by \cite{Greenwood1991W}
\begin{equation}
\Delta T \approx {s \sigma_{\rm f} v r_{\rm c} \over \kappa} \left(1 + \frac{\pi r_{\rm c} v}{8 D}\right)^{-1/2} ,
\label{eq:greenwood}
\end{equation}
where $\kappa$ is the quartz or silica heat conductivity and $D=\kappa/\rho C_{\rm p}$ the heat  diffusivity  ($\rho$ is the mass density and $C_{\rm p}$ the heat capacity), 
and $\sigma_{\rm f}$ is the frictional shear stress. In ~\eqref{eq:greenwood} $s$ is the fraction of the heat energy $\dot q = \sigma_{\rm f} v$ (energy per unit surface area
and unit time) going into the lower solid. 
For low sliding speed we expect  $s\approx 1/2$ as both materials are equal. 
However, this equation is not valid for high sliding speed as will be shown now.

The maximum surface temperature in the contact region for the upper block is \cite{Greenwood1991W}
\begin{equation}
\Delta T \approx {(1-s) \sigma_{\rm f} v r_{\rm c} \over \kappa}
\label{eq:greenwood_approx}
\end{equation}
If we assume the temperature is continuous in the contact area 
between the solids, as should hold approximately (see Ref.~\cite{Persson2011JPCM}), then using \eqref{eq:greenwood} and \eqref{eq:greenwood_approx}:
\begin{equation}
s \approx {1\over 1+\left (1+{\pi v r_{\rm c} \over 8 D} \right )^{-1/2}}
\label{eq:heat3}
\end{equation}
For quartz $\kappa \approx 11.7 \ {\rm W/mK}$ and for silica $\kappa \approx 1.4 \ {\rm W/mK}$.
The heat capacity and mass densities are similar for quartz and silica and are in the range $C_{\rm p} \approx
650-750 \ {\rm J/kg K}$ and $\rho \approx 2200-2600 \ {\rm kg/m^3}$. Hence, the heat diffusivity
for quartz and silica is $D\approx 10^{-5} \ {\rm m^2/s}$ and $10^{-6} \ {\rm m^2/s}$, respectively. 
Fore sliding speeds $v < 1 \ {\rm cm/s}$ we have $r_{\rm c} v /D < 0.003$ for quartz and $<0.03$
for silica, and hence from \eqref{eq:heat3} we get that $s\approx 1/2$ and half of the 
frictional energy goes into each solid. With $s=1/2$ the maximum temperature increase is
$$\Delta T \approx { \sigma_{\rm f} v r_{\rm c} \over 2 \kappa}$$
Assuming $\sigma_{\rm f} = \mu \sigma_{\rm P} \approx 10 \ {\rm GPa}$ we get
for the highest sliding speed in the experiments $v=1 \ {\rm cm/s}$,
$\Delta T \approx 100 \ {\rm K}$ for silica and $\approx 10 \ {\rm K}$ for quartz.
The temperature increase for silica is non-negligible but does not change the
conclusions presented above. 

The melting temperature of quartz is $\approx 1700^\circ {\rm C}$ while the melting temperature
of silica is not well defined as it softens gradually with increasing temperature.
If the friction coefficient were velocity independent, the temperature in the asperity contact regions
for silica would reach $\approx 1700^\circ {\rm C}$ at the sliding speed 
$\approx 0.2 \ {\rm m/s}$. However, in reality the friction coefficient, and hence the shear stress
$\sigma_{\rm f}$, drops strongly at high temperature, so the actual melting of the silica would occur
at much higher sliding speeds as would be predicted if the friction coefficient were velocity independent.

If two solids slide in contact for a long enough time, the whole interface
will heat up~\cite{Aubry2018GRL}. This effect can be considered as the cumulative effect of the flash
temperature. If a uniform heat source $\dot q_1 $ is turned on at time $t=0$ at the
bottom surface of a semi-infinite solid ($z \geq 0$) then the temperature at the surface
$z=0$ at time $t$ equals
$$\Delta T_1 = {\dot q_1 \over \kappa} \left ({4 D t \over \pi } \right )^{1/2}.$$
Let us compare this to the flash temperature, which we write as
$$\Delta T = {\dot q_0 r_0 \over \kappa}.$$
The heat source $\dot q_0$ acts in the total contact area $A$ while $\dot q_1$ act in the nominal contact area $A_0$. Since the total heat energy is the same in both cases, we get $\dot q_0/\dot q_1 = A_0/A$. 
Assuming that the pressure in all the contact regions is equal to the penetration hardness $\sigma_{\rm P}$ and that the nominally applied pressure is $\sigma_0$ gives $\sigma_0 A_0 = \sigma_{\rm P} A$ or
$\dot q_0/\dot q_1 = \sigma_{\rm P}/\sigma_0$. Hence,
\begin{equation}
    \label{eq:flash_temperature}
{\Delta T\over \Delta T_1} = {\sigma_{\rm P} \over \sigma_0} \left ({\pi r_0^2 \over 4Dt}\right )^{1/2}.
\end{equation}
In earthquake applications $\sigma_{\rm P} \approx 10 \ {\rm GPa}$ and $\sigma_0 \approx 300 \ {\rm MPa}$ and the slip time of order $1 \ {\rm s}$. Using $r_0 = 10 \ {\rm \mu m}$ equation~\eqref{eq:flash_temperature} gives $\Delta T_1 \approx  \Delta T$
Hence in earthquake applications the cumulative effect of the flash temperature is very important and may result in the who sliding interface heating up to a temperature nearly as high as the flash temperature.
In our applications $\sigma_0 \approx 0.2 \ {\rm MPa}$ and even taking into account the longer sliding contact time the cumulative effect of the flash temperature is negligible.

\begin{figure}[tbp]
\includegraphics[width=0.45\textwidth,angle=0]{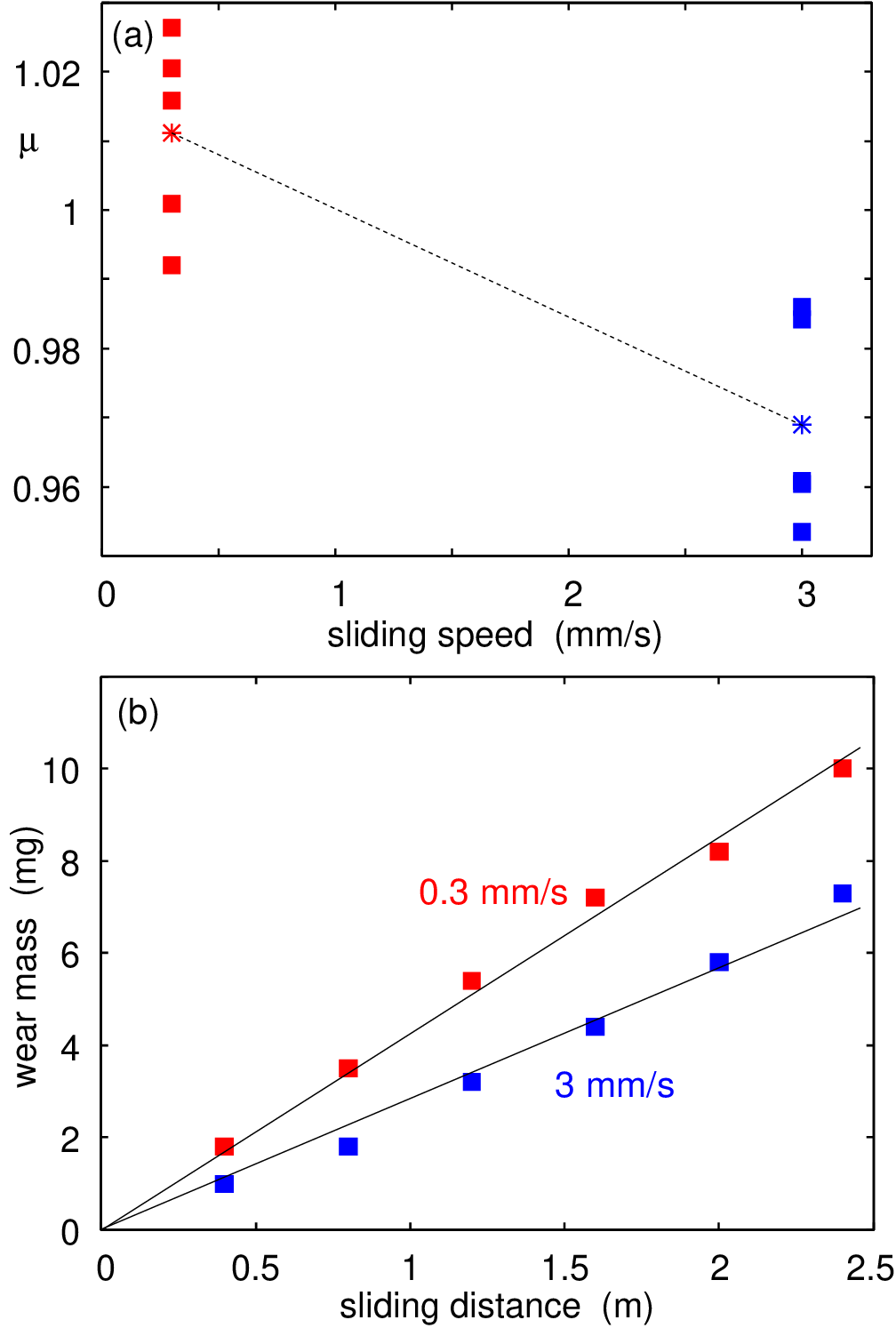}
\caption{
(a) The friction coefficient and (b) the wear rate for six 1-runs at the sliding speeds $v=0.3 \ {\rm mm/s}$ (red squares) and
$3 \ {\rm mm/s}$ (blue squares). In (a) the stars gives the average friction coefficients which are 1.01 and 0.97.
}
\label{1v.2friction.two.velocities.eps}
\end{figure}

As a further test that the flash temperature is not important in the present case, we performed additional friction and wear measurements at $v = 0.3\,\mathrm{mm/s}$,
where $\dot q$, and hence the flash temperature should be nearly 10 times lower than for $v = 3\,\mathrm{mm/s}$.
Figure~\ref{1v.2friction.two.velocities.eps} shows (a) the friction coefficient and (b) the wear rate for six 1-runs at the sliding speeds $v = 0.3\,\mathrm{mm/s}$ (red squares) and $3\,\mathrm{mm/s}$ (blue squares).
The friction coefficient and the wear rate are slightly larger at the lower sliding speed, that is, by $\approx 3\%$ and $\approx 30\%$, respectively. 
The result for the friction coefficient agrees with Fig.~\ref{1time.2Temp.eps}.

In earthquakes the maximum slip velocities are typically on the order of a few m/s, so if the size of the contact regions and friction coefficient were similar to those found above, 
we would have to expect that granite melts in the contact regions~\cite{Hayward2019JGR,Rubin2005JGRSE,Mollon2021JGRSE}.
However, the friction must be expected to drop significantly when quartz approaches its melting temperature, and in that case the contact regions may only soften but not melt.
As a consequence, the flash temperature will be less than expected using a constant friction coefficient. 
This is very similar to friction on ice: when the sliding speed increases, the surface temperature of ice in the contact regions increases due to frictional heating.
On approaching the bulk melting temperature of ice, the ice surface softens (predominantly by sliding-induced amorphization~\cite{Atila2025PRL}) and the friction drops drastically.
As a result the actual (thermodynamic) melting of the ice is shifted to much higher sliding speeds than expected based on the friction coefficient prevailing before frictional heating becomes important.
A similar effect may occur for granite sliding on granite, where the dependence of the friction coefficient on sliding speed, as summarized in Ref.~\cite{DiToro2011N}, is very similar (but shifted to higher sliding speed) to that of ice sliding on ice (see Ref.~\cite{Persson2015JCP,Bore2026PRL}).

\subsection{Elasticity, contact aging, and the limits of rate-and-state friction laws}

During the stationary contact, the asperity contact regions are often assumed to be fully pinned and 
to increase both their real contact area and their local strength over time. 
This \emph{aging} process allows the contact to sustain progressively higher shear force over time. 
When the external shear stress builds up beyond the static threshold, the interface enters the slip phase.
Local contacts rejuvenate, which reduces their resistance to sliding. 
The system comes to rest again once the external driving force has dropped below the kinetic friction. 
Earthquake dynamics are frequently interpreted within this framework---as a sequence of alternating stuck and sliding states~\cite{Dieterich1979JGRSE,Ruina1983JGRSE,Marone1990JGRSE}. 

An increasing body of research challenges the traditional explanation for stick-slip dynamics. 
It has been proposed that the dynamics of stick--slip systems may instead be governed by a kinetic friction coefficient that decreases with sliding velocity at very low speeds, rather than by explicit state evolution~\cite{Baumberger2006AP,Bureau2002EPJE,Rubinstein2007PRL,Yang2008PNAS}.
This raises the question of whether stick-slip dynamics should generally be interpreted as \emph{creep-slip} dynamics 
or if there exist quantitative criteria that determine when the conventional stick-slip description is appropriate. 
Moreover, what conditions cause the breakloose friction not to exceed the kinetic friction?

\begin{figure}
\includegraphics[width=0.3\textwidth,angle=0.0]{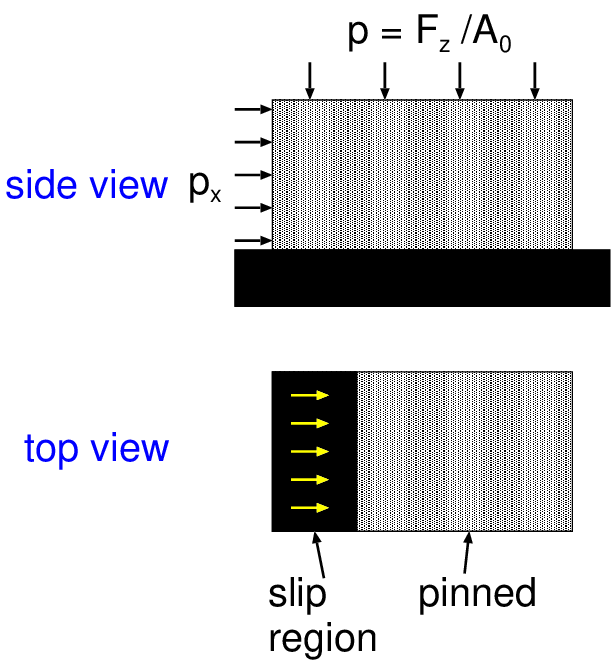}
\caption{\label{ZZZ.ThreeCases1.eps}
Slip at the onset of sliding when the driving force is applied on one side of the solid.
The region where slip occurs can expand gradually and extend over most of the contact before the leading edge starts moving. 
As a result, the breakloose force exceeds the kinetic friction only marginally. 
However, for this to happen, most of the slip region must have moved by a distance exceeding $D$.
}
\end{figure}

A general explanation for why the breakloose friction does not always noticeably exceed the kinetic friction is that a solid block, 
like that shown schematically in Fig.~\ref{ZZZ.ThreeCases1.eps}, does not begin to slide uniformly but through pre-slip. 
Pre-slip refers to the gradual loss of pinning at a fraction of the local contacts before the body as a whole starts to move, or--when applied to a single asperity contact--to the breaking of interfacial molecular bonds before the asperity completely depins. 
Such localized slippage relaxes part of the stored shear stress and thereby reduces the measured static or breakloose friction coefficient, $\mu_{\mathrm{s}}$, below its hypothetical upper bound $\mu_{\mathrm{s}}^*$, which would be attained only if all contact patches were to yield simultaneously.

\begin{figure}[tbp]
\includegraphics[width=0.45\textwidth,angle=0]{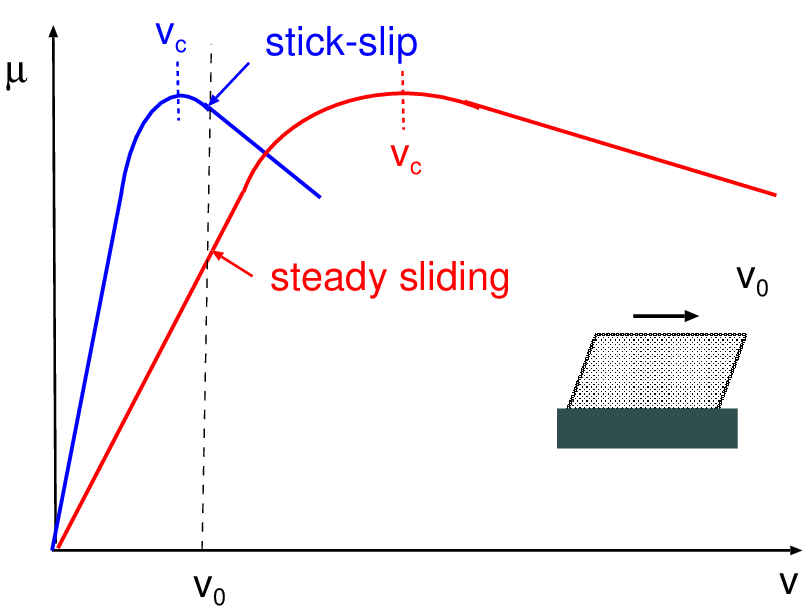}
\caption{
The kinetic friction for two hypothetical cases.
Sliding is stable if the driving velocity $v_0$ is below the maximum of the kinetic friction force at $v_c$ (red curve), while stick-slip occurs when $v_0$ is above it (blue curve).
For most faults the driving velocity (e.g., determined by the continental drift) is very small, $v_0 \approx 10^{-10} \ {\rm m/s}$, and the maximum in the kinetic friction force must occur below this velocity for earthquakes to occur.
For some faults $v_{\rm c} > v_0$ and for such cases only stable sliding occurs.
}
\label{earth.eps}
\end{figure}

A quantitative description of pre-slip in a side-loaded block can be obtained from a continuum version of the Burridge--Knopoff model~\cite{Burridge1967BSSA,Carlson1994RMP}, modified to include a smooth slip-weakening transition~\cite{Lorenz2012JPCM}. 
If the pre-slip displacement $\lambda_{\rm el}$ is larger than the size $D=2r_{\rm c}$ 
of the macroasperity contact regions in most of the nominal contact region when
global slip occurs, then the breakloose friction force may only be marginally larger than the kinetic friction force.
In that continuum framework, which is summarized in Appendix ~\ref{sec:app_theory}, the pre-slip is characterized by the distance
\begin{equation}
    \label{eq:elastic_coherence_length}
    \lambda_{\rm el} = (\sigma_{\rm s} - \sigma_{\rm k}) \frac{L^2}{E h},
\end{equation}
where $E$ is the Young's modulus, $h$ the block height and $L$ its lateral size in the sliding direction, while $\sigma_\textrm{s}$ and $\sigma_\textrm{k}$ are the static and kinetic shear stress, respectively.

Although the model is not a rigorous half-space solution, the qualitative trend — that $\lambda_{\rm el}$ grows with system size for fixed $L/h$ — is expected to remain valid.
This matters because the relative magnitude of $\lambda_{\rm el}$ compared to a microscopic asperity length scale $D$, 
determines the macroscopic breakloose behavior.
If $\lambda_{\rm el} \gg D$, most macroasperity contact regions will be renewed before global slip occurs,
and the breakloose friction force will be close to the kinetic friction force.
That is, the observed static or breakloose friction coefficient $\mu_s$ approaches the kinetic value $\mu_k$. 
Conversely, in stiff or small systems where $\lambda_{\rm el} \ll D$, all macroasperity contact regions are broken at the same time, 
and the full static strength $\mu_s^\ast$ can be reached.
This trend is loosely analogous to how adhesion depends on the system size or the radius of curvature. 
Large, compliant systems fail via interfacial crack propagation. 
In contrast, small, elastically stiff contacts tend to break bonds more uniformly across the interface~\cite{Persson2003W,Muser2022EL}.

Let us apply the above results to the experimental results and to earthquake dynamics.
In the study by McClimon \textit{et al.}~\cite{McClimon2024TL}, a nanoscale silica tip was slid on a silica substrate. 
They observed a breakloose friction force that was, in some cases, 3–4 times higher than the kinetic friction force. 
The experiments were performed in a humid atmosphere, and the aging was attributed to water-catalyzed interfacial \text{Si–O–Si} bond formation.
Due to the small size of the contact region, $\lambda_{\rm el} \ll D$, the full aging effect shows up in the experiments, so that $\mu_{\rm s} = \mu^\ast_{\rm s}$. 
In a perfectly dry atmosphere, McClimon \textit{et al.}~\cite{McClimon2024TL} did not observe any aging.

Next, let us consider earthquake dynamics on larger scales. 
Consider a linear region of length $L$ undergoing slip. 
The driving force for this is the elastic energy stored in a volume element, which extends a distance $L$ orthogonal to the slip plane. 
Using $h=L$ in Eq.~\eqref{eq:elastic_coherence_length} gives:
\begin{equation}
\lambda_{\rm el} \approx (\mu^\ast_{\rm s} - \mu_{\rm k}) \, \frac{\sigma_z L}{E}.
\end{equation}
The normal stress at fault lines is typically $\sigma_z \approx 100 \ {\rm MPa}$. 
Assuming $E = 10^{11} \ {\rm Pa}$, $\mu^\ast_{\rm s} - \mu_{\rm k} \approx 1$, and $D \approx 10 \ {\rm \upmu m}$ (as obtained in Sec.~4), we get $\lambda_{\rm el} \approx D$ if $L \approx 1 \ {\rm cm}$.
Note that an increase in the nominal contact pressure $\sigma_z$ results in an increase in $\lambda_{\rm el}$ which will result in a reduction in $\Delta \mu = \mu_{\rm s} - \mu_{\rm k}$ as observed in Ref.~\cite{Peng2025PRL}.

The friction dynamics in earthquakes is often described using rate-and-state friction models. 
In our study, we find  no memory effects of the type expected from rate-and-state friction laws. 
In the experimental investigations by Dieterich~\cite{Dieterich1979JGRSE}, which were summarized in the theory of Ruina as a rate- and state-dependent law of friction~\cite{Ruina1983JGRSE}, only a few percent change in the friction coefficient was observed. 
In these model experiments, the contact pressure is typically of order $10 \ {\rm MPa}$, and the linear size
of the blocks is on the order of a few centimeters. 
Under these conditions, we get $\lambda_{\rm el} \approx D$, and some (reduced) influence of aging should be observed.

Earthquake faults have roughness on many length scales. 
Integrating out the roughness with wavelength below the linear size $L$ will result in breakloose friction coefficients that approach the kinetic friction coefficient as the length scale $L$ increases. 
From the study above, we expect that there will be no influence of rate-and-state friction when $L$ becomes of order $1 \ {\rm m}$. 
In modeling earthquake dynamics, the long-wavelength roughness may be included explicitly, but the short-wavelength roughness
cannot, as it would result in too many degrees of freedom. 
If earthquake dynamics can be understood using a coarse-grained model with a grid size of order $\sim 1 \ {\rm m}$, then we predict
negligible influence of aging, and rate-and-state description of friction dynamics assuming finite static shear stresses becomes irrelevant. 
If shorter length scales must be included in the modeling, then rate-and-state effects become important. 
However, in this case, the friction law will depend on the coarse-grained length scale $L$. 
Atomic force microscopy measurements~\cite{Tian2017PRL,McClimon2024TL} show significant aging effects at the nanometer length scale. 
If a coarse-grained description with nanometer-sized grids were used to simulate an earthquake of kilometer size, one would end up with a problem  involving $(10^3 / 10^{-9})^3 = 10^{36}$ degrees of freedom!

We believe that the nature of earthquake dynamics depends mainly on the behavior of the kinetic friction coefficient as a function of the sliding speed, $\mu_{\rm k}(v)$, at extremely low sliding speeds, as illustrated in Fig.~\ref{earth.eps}. 
This figure shows the $\mu_{\rm k}(v)$ relation for two hypothetical cases. 
If the driving velocity $v_0$ is below the maximum $v_{\rm c}$ of the kinetic friction force, only stable steady sliding occurs. 
If the driving velocity is above this maximum, then stick-slip occurs. For most faults, the driving velocity,  determined by continental drift (large-scale horizontal movements of continents relative to one another), is very small, $v_0 \approx 10^{-10} \ {\rm m/s}$, and the maximum in the kinetic friction force must occur below  this velocity for earthquakes to occur. 
For some faults, $v_{\rm c} > v_0$, and for such cases only stable sliding occurs.
Understanding the origin of earthquakes must therefore involve understanding the functional form of $\mu_{\rm k}(v)$  at extremely low sliding speeds ($v_0 \sim 10^{-10} \ {\rm m/s}$ or less).
We are not aware of any studies of the $\mu_{\rm k}(v)$ relation for granite-on-granite  at sliding speeds $v < 0.1 \ {\rm \upmu m/s}$.
However, experiments on steel-silicon interfaces at $10^{-5}$~µm/s velocities indicate the absence of true static friction~\cite{Yang2008PNAS}.

\section{Discussion}
\label{sec:discussion}

We wish to strengthen the connections between our experimental, 
theoretical, and simulation results and to further relate them to the existing literature.
To this end, we focus on two key findings: (a) friction in self-mated granite is dominated by adhesion but anti-correlated 
with wear and (b) kinetic friction has a weak dependence on velocity to the extent that the breakloose 
force does not markedly exceed the kinetic friction force in macroscopic systems. 

We first address point (a). If for clean surfaces the 
friction were dominated by plowing or scratching, the friction for 1 run's 
between corundum and quartz would have to exceed that of a self-mated quartz contact, because corundum is harder than 
quartz and the surface roughness bigger and sharper for the corundum paper. 
However, the opposite is observed, which is consistent with the claim that the friction for clean surfaces is caused by adhesion.
Adhesion in corundum--silica contacts (non-passivated or after removal of passivation) will generally be less than in self-mated silica contacts, because a stoichiometric mismatch between anions and cations with two oxygen atoms per cation in SiO$_2$ versus 1.5 in Al$_2$O$_3$ unavoidably leads to frustrated interfacial interactions.
Therefore, silica adhesion with quartz is expected to exceed that of corundum.
In a situation where the interfacial shear is controlled by formation and rupture of interfacial bonds, the higher work of adhesion would lead to larger adhesion-mediated friction of quartz with the gouge compared to silica–corundum contacts.
It is also possible that the corundum asperities have a thin polymer coating (the binder to the fabric), and in that case no cold
welded junctions can form. 
Still, even if this film is removed by wear, 
one expects weaker quartz-corundum interaction than for quartz against quartz for the just-mentioned reasons.

For sliding contacts, in the contact regions, the quartz undergoes 
amorphization~\cite{Pastewka2010NM,Moras2018PRM,Atila2025PRL},
which in tetrahedral network formers is predominantly a mechanically 
rather than thermally induced atom-by-atom or molecule-by-molecule process.
This is similar to ice friction, where sliding-induced amorphization occurs even for temperatures well below the ice melting temperature. 
Nonetheless, silicate interfaces near thermal melting should not be expected to become quite as 
lubricious as those formed by ice, given the vast difference in viscosity, 
e.g., about $\sim 10^{12} \ {\rm Pa s }$ near $1800 {\rm K}$ for pure silica 
versus $1.8\cdot 10^{-3} \ {\rm Pa s}$ for water at $0^\circ$C.

It remains to be understood \emph{why} wear and friction are anti-correlated in our experiments.
To discuss this point, we note that the height of a uniformly spread-out debris layer after a 
typical 1-run is $d \approx 0.45~{\upmu}$m, and after a 100-run $d \approx 4.1~{\upmu}$m.
These values are noticeably smaller than the typical 
spacing of about $8~{\upmu}$m between the two surfaces within the nominal quartz--quartz contacts.
Thus, the debris volume is small enough to be fully accommodated in the surface valleys.
Yet, visual inspection reveals that most of the debris accumulates at the turn-around points, 
indicating that the particles are swept and transported by the regions of true or near-true contact.
This observation suggests that the debris participates are carrying shear, thereby reducing the slip 
at the underlying quartz and consequently mitigating wear.
At the same time, part of the debris may be mechanically softer than quartz, 
which would promote larger and tougher adhesive junctions and thus higher friction.

The next major discussion point revolves around point (b), the weak dependence of solid 
friction on sliding velocity, which we observe even in the molecular dynamics single-asperity sliding contacts.
The generic explanation for why kinetic friction often depends only weakly on 
sliding velocity $v$ is that it is caused by (elastic or plastic) instabilities 
whose nature does not change with~$v$ for small sliding speeds.
This is demonstrated quantitatively in the Prandtl model~\cite{Prandtl1928ZAMM,Popov2012ZAMM,Manzi2021TL}, 
which consists of a mass point (representing either an atom or a more coarse-grained entity, 
such as an atomic-force microscope tip~\cite{Gnecco2000PRL}) dragged by an elastic spring past a 
sinusoidal potential representing the corrugation of a counterface.
Analysis of this model shows that a similar amount of potential energy is dissipated each 
time a degree of freedom pops forward from an unstable position to a new energy minimum, 
so long as the underlying instability mechanism remains unchanged, which is expected for sliding speeds much smaller
than the (usually rapid) slip velocities in the instabilities.
This argument extends far beyond purely elastic systems and has been shown to apply, 
for example, to boundary lubricants~\cite{Muser2002PRL}.
However, our atomistic simulations reveal that this generic picture is too simple.
Thus, the quartz surface undergoes much more massive plastic flow at 0.1~m/s than at 10~m/s.
We conclude that the friction-increasing thermal effects, which enhance both the quality and 
quantity of contacts, are approximately compensated in this system by friction-reducing, 
thermally activated creep that helps to release thermal stress.

It is worth elaborating further on the role of plasticity.
In our simulations of normal indentation, contacts begin to yield plastically at 
normal stresses of about 12 GPa — a result that is surprisingly close to 
experimental estimates~\cite{Whitney2007AM,Strozewski2021JGR}.
This apparent agreement likely benefits from a degree of fortuitous error cancellation, 
as the BKS potential is known to underestimate activation energies for flow in silica 
liquids by around 10\% (about 0.5 eV)~\cite{Horbach1999PRB}.
As a result, our simulations effectively experience an artificial speed-up of more than eight orders of magnitude.
Based on this estimate and setting aside both the elastic anisotropy of quartz and the full structure of the stress 
tensor near the contact, one may argue that yielding occurs when the deviatoric (von Mises) 
stress reaches approximately 12 GPa in normal indentation, where only the normal component $\sigma_{zz}$ is nonzero.
Upon adding shear via $\sigma_{xz} = \sigma_{zx} = \mu \sigma_{zz}$, the same deviatoric stress would 
be reached at $\sigma_{zz} \approx 4.5$ GPa when $\mu = 1$ and at $\sigma_{zz} \approx 7$ GPa when $\mu = 0.8$.
In our simulations under shear, we observe a reduction of the normal stress at yield from 12 GPa to 
approximately 7 GPa, i.e., by 5 GPa — slightly smaller, but still of comparable magnitude to the 7.5 GPa 
reduction predicted by these extremely simplified arguments, which would become smaller if the diagonal 
components of the in-plane stress were not set to zero. 

The last major point of discussion is the relevance of pre-slip for tectonic motion.
In particular, we wish to elaborate in more detail on the approximations made in the 
model~\cite{Lorenz2012JPCM} and the competition between interfacial and elastic interactions, 
which crucially depends on the spatial dimensions of the interface and the elastic body~\cite{Muser2004EPL}.
In their model, Lorenz and Persson~\cite{Lorenz2012JPCM} discretized the moving elastic body into (flat) vertical slices, 
which are thin in the sliding direction but extend macroscopically in the two remaining directions.
In reality, the deformation field will result in curved slices.
Still, we expect the exact result to scale with the parameters in the problem
in the same way as found in the simple model, while prefactors may change in a more detailed model.
Here we note that Ciavarella~\cite{Liang2025TL} has studied a very different model but arrived to the same conclusion as
found in Ref.~\cite{Lorenz2012JPCM}. 

\section{Conclusions}
\label{sec:conclusions}

By combining sliding friction and wear experiments, atomistic simulations, and contact-mechanics theory, 
this work revisits granite friction as a model system for the tribology of rock from a new, systematic angle.
While reproducing key features of faults, such as the size of gouge particles or the anisotropy of surface roughness after sliding, we 
affirm well-established views (e.g., that gouge reduces wear), sharpen and propose new mechanistic interpretations, 
and also challenge some prevailing ideas—most notably by suggesting that irregular tectonic motion is governed by 
creep-slip rather than stick-slip dynamics.

The staged wear-removal experiments reveal that each time debris is brushed off granite surfaces, 
the wear rate increases while friction decreases.
Wear rate and friction remain approximately linear functions of the load, but with different proportionality 
coefficients immediately after debris removal versus after reaching a steady-state-like condition without removal.
The observed anti-correlation between friction and wear implies that asperity interlocking, scratching, 
crushing, and plowing are not dominant contributors to granite friction as frequently assumed.
Instead, friction arises primarily from adhesion in cold-welded junctions, which is traditionally seen as secondary. 
This interpretation is further supported by the finding that hard corundum produces 
more wear but less friction than self-mated granite contacts.
Similarly, adding water to granite–granite interfaces reduces wear 
by an order of magnitude but friction only by about 25\% or less.
Atomistic simulations of a rigid, amorphous silica tip sliding across $\alpha$-quartz—the main and 
presumably load-bearing mineral in granite support and refine this interpretation:
they show friction coefficients of order unity under continuous amorphization of quartz. 
Moreover, three main energy-dissipation mechanisms are identified:
(i) bond-breaking at the trailing edge of the contact, (ii) plastic deformation and amorphization beneath the contact 
center and near the leading edge, and (iii) reversible, stress-induced phase transformations in front and below the leading edge.

The simulations also confirm theoretical flash-temperature estimates, provided that local heating 
does not significantly soften the materials in contact.
The product of the highest sliding velocities $v_\textrm{max}$ and the linear size $D$ of typical asperity contact regions is roughly the same in the MD simulations and the experiments; that is, $v_\textrm{max} \approx 10~\text{m/s}$ and $D = \mathcal{O}(10~\text{nm})$ in the simulations versus $v \approx 1~\text{cm/s}$ and $D = \mathcal{O}(10~\upmu\text{m})$ in the experiments.
According to classical flash-temperature models, heating under these matched conditions is comparable—on the order of 100 K—which does not significantly affect local shear strength.
However, during an earthquake, where slip velocities may exceed 2 m/s at the fault interface, flash heating becomes far more substantial.
Even well below the melting point, we expect pronounced reductions in shear resistance at these elevated temperatures, so that heating will be less substantial in reality than when using interfacial shear stresses determined at low temperature. 
Indeed, our simulations already show a marked decrease in shear stress near the $\alpha$–$\beta$ transition temperature of quartz.
Moreover, the presence of silica polymorphs in post-mortem, which are usually reached from $\alpha$-quartz through heating, turns out to be an insufficient indicator for previous high-temperature:
our simulations produce the low-temperature polymorph OP tridymite underneath the marginally heated leading edge of the tip through a displacive transformation bypassing the usually required, high-temperature intermediates. 

We found a surprising insensitivity of granite friction to temperature and relative sliding speed in the experiments, 
in contrast to expectations from rate–state models.
We rationalized the absence of rate–state effects by the scale dependence of pre-slip: the stress weakening zone, 
in which the interface transitions from perfectly pinned to fully unpinned, can constitute a 
substantial fraction of a small (microscopic) contact but only a marginal fraction of a large (macroscopic) one.

\appendix

\section{Lorenz--Persson theory}
\label{sec:app_theory}

In the domain where the stress transitions from the static to the kinetic shear stress, the displacement field $u(x)$ satisfies the ordinary differential equation
\begin{equation}
Eh\,u'' = \sigma_s - (\sigma_s - \sigma_k)\,\frac{u}{D}.
\end{equation}
Using $\kappa^2 = (\sigma_s - \sigma_k)/(EhD)$, this can be recast as 
\begin{equation}
\label{eq:ODE}
u'' + \kappa^2 u = \frac{\sigma_s}{Eh}.
\end{equation}
We now consider the instant when the leading (right) edge, located at $x = 0$, is about to depin.
In this coordinate system, material points with positions slightly to the left of the edge ($x < 0$) have been displaced to the right and experience a frictional stress between zero and $\sigma_s$.
Thus, the boundary conditions at this moment are
\begin{eqnarray}
u(0) & = & 0, \\
E\,u'(0) & = & -\sigma_s.
\end{eqnarray}
These correspond to the situation where the solid is on the verge of transitioning into global sliding, driven by an adiabatically increasing force applied from the trailing (left) edge.
The solution of Eq.~\eqref{eq:ODE} is 
\begin{equation}
u(x) = \frac{\sigma_s D}{\sigma_s - \sigma_k}
\left[ 1 - \cos(\kappa x) \right]
- \frac{\sigma_s}{E\kappa}\,\sin(\kappa x).
\label{eq:disp_final}
\end{equation}

The hypothetical point $x^*$ to the left of which all material points are completely depinned is the largest negative $x$ for which 
\begin{equation}
\label{eq:depinned}
    |u(x^*)| = D.
\end{equation}
Depending on the parameters, this point may lie inside or outside the block.
The prefactor of the sine term in Eq.~\eqref{eq:disp_final} can be rewritten as
\begin{equation}
\frac{\sigma_s}{E\kappa}
= 
\sqrt{ 
\frac{h}{D}\,
\frac{\sigma_s}{E}\,
\frac{1}{1 - \sigma_k/\sigma_s}
}.
\end{equation}
Except for the height $h$, all terms appearing in this expression are material or loading properties.
Thus, the sine term is negligible in the limit of small $h$ but becomes dominant for large $h$, leading to
\begin{equation}
|x^*| =
\begin{cases}
   \arccos(\sigma_k/\sigma_s) / \kappa & \text{for } h \to 0, \\[6pt]
   D\,E/\sigma_s & \text{for } h \to \infty.
\end{cases}
\end{equation}

Using a Young’s modulus of $E = 70~\text{GPa}$ for silica, a normal stress during sliding of $\sigma_s = 7~\text{GPa}$, and assuming $\sigma_k = \sigma_s/2$, we obtain $|x^*| \approx \sqrt{10\,hD}$ in the small-object limit
In the opposite limit of very large objects, the ratio $|x^*|/L$ becomes negligible, so that the breakloose force exceeds the kinetic friction only marginally.

\section*{Conflict of interest}
The author declares that there are no conflicts of interest.

\section*{Author Contributions}
B.N.J. Persson conceived the project and designed the experiments and the theory.
S.V. Sukhomlinov and M.H. Müser designed the simulations.
S.V. Sukhomlinov conducted the simulations. 
All authors analyzed the data and wrote the manuscript.
The experiments were conducted by Azeddine El Yakini, a technician employed at MultiscaleConsulting.

\section*{Data Availability Statement}
Data supporting the findings of this study are available from the corresponding author upon a reasonable request.

\bibliography{combined}

@article{Afferrante2018TL,
  title = {Elastic Contact Mechanics of Randomly Rough Surfaces: An Assessment of Advanced Asperity Models and Persson’s Theory},
  volume = {66},
  ISSN = {1573-2711},
  url = {http://dx.doi.org/10.1007/s11249-018-1026-x},
  DOI = {10.1007/s11249-018-1026-x},
  number = {2},
  journal = {Tribology Letters},
  publisher = {Springer Science and Business Media LLC},
  author = {Afferrante,  L. and Bottiglione,  F. and Putignano,  C. and Persson,  B. N. J. and Carbone,  G.},
  year = {2018},
  month = may 
}

@article{Aghababaei2016NM,
  title = {Critical length scale controls adhesive wear mechanisms},
  volume = {7},
  ISSN = {2041-1723},
  url = {http://dx.doi.org/10.1038/ncomms11816},
  DOI = {10.1038/ncomms11816},
  number = {1},
  journal = {Nature Communications},
  publisher = {Springer Science and Business Media LLC},
  author = {Aghababaei,  Ramin and Warner,  Derek H. and Molinari,  Jean-Francois},
  year = {2016},
  month = jun 
}

@article{Aghababaei2022MRSB,
  title = {How roughness emerges on natural and engineered surfaces},
  volume = {47},
  ISSN = {1938-1425},
  url = {http://dx.doi.org/10.1557/s43577-022-00469-1},
  DOI = {10.1557/s43577-022-00469-1},
  number = {12},
  journal = {MRS Bulletin},
  publisher = {Springer Science and Business Media LLC},
  author = {Aghababaei,  Ramin and Brodsky,  Emily E. and Molinari,  Jean-Fran\c{c}ois and Chandrasekar,  Srinivasan},
  year = {2022},
  month = dec,
  pages = {1229–1236}
}

@article{Aharonov2019JGRSE,
  title = {The Brittle‐Ductile Transition Predicted by a Physics‐Based Friction Law},
  volume = {124},
  ISSN = {2169-9356},
  url = {http://dx.doi.org/10.1029/2018JB016878},
  DOI = {10.1029/2018jb016878},
  number = {3},
  journal = {Journal of Geophysical Research: Solid Earth},
  publisher = {American Geophysical Union (AGU)},
  author = {Aharonov,  Einat and Scholz,  Christopher H.},
  year = {2019},
  month = mar,
  pages = {2721–2737}
}

@article{Almqvist2011JMPS,
  title = {Interfacial separation between elastic solids with randomly rough surfaces: Comparison between theory and numerical techniques},
  volume = {59},
  ISSN = {0022-5096},
  url = {http://dx.doi.org/10.1016/j.jmps.2011.08.004},
  DOI = {10.1016/j.jmps.2011.08.004},
  number = {11},
  journal = {Journal of the Mechanics and Physics of Solids},
  publisher = {Elsevier BV},
  author = {Almqvist,  A. and Campañá,  C. and Prodanov,  N. and Persson,  B.N.J.},
  year = {2011},
  month = nov,
  pages = {2355–2369}
}

@article{Archard1953JAP,
  title = {Contact and Rubbing of Flat Surfaces},
  volume = {24},
  ISSN = {1089-7550},
  url = {http://dx.doi.org/10.1063/1.1721448},
  DOI = {10.1063/1.1721448},
  number = {8},
  journal = {Journal of Applied Physics},
  publisher = {AIP Publishing},
  author = {Archard,  J. F.},
  year = {1953},
  month = aug,
  pages = {981–988}
}

@article{Atila2025PRL,
  title = {Cold Self-Lubrication of Sliding Ice},
  volume = {135},
  ISSN = {1079-7114},
  url = {http://dx.doi.org/10.1103/1plj-7p4z},
  DOI = {10.1103/1plj-7p4z},
  number = {6},
  journal = {Physical Review Letters},
  publisher = {American Physical Society (APS)},
  author = {Atila,  Achraf and Sukhomlinov,  Sergey V. and M\"{u}ser,  Martin H.},
  year = {2025},
  month = aug,
  pages = {066204}
}

@article{Aubry2018GRL,
  title = {Frictional Heating Processes and Energy Budget During Laboratory Earthquakes},
  volume = {45},
  ISSN = {1944-8007},
  url = {http://dx.doi.org/10.1029/2018GL079263},
  DOI = {10.1029/2018gl079263},
  number = {22},
  journal = {Geophysical Research Letters},
  publisher = {American Geophysical Union (AGU)},
  author = {Aubry,  J. and Passelègue,  F. X. and Deldicque,  D. and Girault,  F. and Marty,  S. and Lahfid,  A. and Bhat,  H. S. and Escartin,  J. and Schubnel,  A.},
  year = {2018},
  month = nov 
}

@article{Baumberger2006AP,
  title = {Solid friction from stick–slip down to pinning and aging},
  volume = {55},
  ISSN = {1460-6976},
  url = {http://dx.doi.org/10.1080/00018730600732186},
  DOI = {10.1080/00018730600732186},
  number = {3–4},
  journal = {Advances in Physics},
  publisher = {Informa UK Limited},
  author = {Baumberger,  Tristan and Caroli,  Christiane},
  year = {2006},
  month = may,
  pages = {279–348}
}

@article{Beeler2008JGR,
  title = {Constitutive relationships and physical basis of fault strength due to flash heating},
  volume = {113},
  ISSN = {0148-0227},
  url = {http://dx.doi.org/10.1029/2007JB004988},
  DOI = {10.1029/2007jb004988},
  number = {B1},
  journal = {Journal of Geophysical Research: Solid Earth},
  publisher = {American Geophysical Union (AGU)},
  author = {Beeler,  N. M. and Tullis,  T. E. and Goldsby,  D. L.},
  year = {2008},
  month = jan 
}

@article{Bore2026PRL,
  title = {Why ice is so slippery},
  journal = {submitted to Physical Review Letters},
  publisher = {American Physical Society (APS)},
  author = {Bore, S. L. and Persson, B.N.J. and Sveinsson, H. A.},
  year = {2026},
}

@book{Bowden2001Book,
  title = {The Friction and Lubrication of Solids},
  ISBN = {9781383021745},
  url = {http://dx.doi.org/10.1093/oso/9780198507772.001.0001},
  DOI = {10.1093/oso/9780198507772.001.0001},
  publisher = {Oxford University PressOxford},
  author = {Bowden,  F P and Tabor,  D},
  year = {2001},
  month = feb 
}

@article{Brodsky2011EPSL,
  title = {Faults smooth gradually as a function of slip},
  volume = {302},
  ISSN = {0012-821X},
  url = {http://dx.doi.org/10.1016/j.epsl.2010.12.010},
  DOI = {10.1016/j.epsl.2010.12.010},
  number = {1–2},
  journal = {Earth and Planetary Science Letters},
  publisher = {Elsevier BV},
  author = {Brodsky,  Emily E. and Gilchrist,  Jacquelyn J. and Sagy,  Amir and Collettini,  Cristiano},
  year = {2011},
  month = feb,
  pages = {185–193}
}

@article{Brown1985JGR,
  title = {Broad bandwidth study of the topography of natural rock surfaces},
  volume = {90},
  ISSN = {0148-0227},
  url = {http://dx.doi.org/10.1029/JB090iB14p12575},
  DOI = {10.1029/jb090ib14p12575},
  number = {B14},
  journal = {Journal of Geophysical Research: Solid Earth},
  publisher = {American Geophysical Union (AGU)},
  author = {Brown,  Stephen R. and Scholz,  Christopher H.},
  year = {1985},
  month = dec,
  pages = {12575–12582}
}

@article{Bureau2002EPJE,
  title = {Rheological aging and rejuvenation in solid friction contacts},
  volume = {8},
  ISSN = {1292-8941},
  url = {http://dx.doi.org/10.1140/epje/i2002-10017-1},
  DOI = {10.1140/epje/i2002-10017-1},
  number = {3},
  journal = {The European Physical Journal E},
  publisher = {Springer Science and Business Media LLC},
  author = {Bureau,  L. and Baumberger,  T. and Caroli,  C.},
  year = {2002},
  month = jun,
  pages = {331–337}
}

@article{Burridge1967BSSA,
  author    = {Burridge, R. and Knopoff, L.},
  title     = {Model and Theoretical Seismicity},
  journal   = {Bulletin of the Seismological Society of America},
  year      = {1967},
  volume    = {57},
  number    = {3},
  pages     = {341--371},
  publisher = {Seismological Society of America}
}

@article{Byerlee1978PAG,
  title = {Friction of rocks},
  volume = {116},
  ISSN = {1420-9136},
  url = {http://dx.doi.org/10.1007/BF00876528},
  DOI = {10.1007/bf00876528},
  number = {4–5},
  journal = {Pure and Applied Geophysics},
  publisher = {Springer Science and Business Media LLC},
  author = {Byerlee,  J.},
  year = {1978},
  month = jul,
  pages = {615–626}
}

@article{Campana2004PRB,
  title = {Irreversibility of the pressure-induced phase transition of quartz and the relation between three hypothetical post-quartz phases},
  volume = {70},
  ISSN = {1550-235X},
  url = {http://dx.doi.org/10.1103/PhysRevB.70.224101},
  DOI = {10.1103/physrevb.70.224101},
  number = {22},
  journal = {Physical Review B},
  publisher = {American Physical Society (APS)},
  author = {Campañá,  Carlos and M\"{u}ser,  Martin H. and Tse,  John S. and Herzbach,  Daniel and Sch\"{o}ffel,  Philipp},
  year = {2004},
  month = dec,
  pages = {224101}
}

@article{Campana2008PRE,
  doi = {10.1103/physreve.78.026110},
  url = {http://dx.doi.org/10.1103/PhysRevE.78.026110},
  year  = {2008},
  month = {aug},
  publisher = {American Physical Society ({APS})},
  volume = {78},
  number = {2},
  author = {Carlos Campa{\~{n}}{\'{a}}},
  title = {Using \protect{G}reen's function molecular dynamics to rationalize the success of asperity models when describing the contact between self-affine surfaces},
  journal = {Physical Review E},
  pages = {026110}
}

@article{Candela2012JGR,
  title = {Roughness of fault surfaces over nine decades of length scales},
  volume = {117},
  ISSN = {0148-0227},
  url = {http://dx.doi.org/10.1029/2011JB009041},
  DOI = {10.1029/2011jb009041},
  number = {B8},
  journal = {Journal of Geophysical Research: Solid Earth},
  publisher = {American Geophysical Union (AGU)},
  author = {Candela,  Thibault and Renard,  Fran\c{c}ois and Klinger,  Yann and Mair,  Karen and Schmittbuhl,  Jean and Brodsky,  Emily E.},
  year = {2012},
  month = aug 
}

@article{Carlson1994RMP,
  title = {Dynamics of earthquake faults},
  volume = {66},
  ISSN = {1539-0756},
  url = {http://dx.doi.org/10.1103/RevModPhys.66.657},
  DOI = {10.1103/revmodphys.66.657},
  number = {2},
  journal = {Reviews of Modern Physics},
  publisher = {American Physical Society (APS)},
  author = {Carlson,  J. M. and Langer,  J. S. and Shaw,  B. E.},
  year = {1994},
  month = apr,
  pages = {657–670}
}

@article{Carpenter1998AM,
  author  = {Carpenter, Michael A. and Salje, Ekhard K. H. and Graeme-Barber, Ann and Wruck, Bernd and Dove, Martin T. and Knight, Kevin S.},
  title   = {Calibration of excess thermodynamic properties and elastic constant variations associated with the $\alpha\leftrightarrow\beta$ phase transition in quartz},
  journal = {American Mineralogist},
  year    = {1998},
  volume  = {83},
  number  = {1-2},
  pages   = {2–22},
  doi     = {10.2138/am-1998-1-206}
}

@article{Cheng2010TL,
  title = {Defining Contact at the Atomic Scale},
  volume = {39},
  ISSN = {1573-2711},
  url = {http://dx.doi.org/10.1007/s11249-010-9682-5},
  DOI = {10.1007/s11249-010-9682-5},
  number = {3},
  journal = {Tribology Letters},
  publisher = {Springer Science and Business Media LLC},
  author = {Cheng,  Shengfeng and Robbins,  Mark O.},
  year = {2010},
  month = aug,
  pages = {329–348}
}

@article{Chester1986PAG,
  title = {Implications for mechanical properties of brittle faults from observations of the Punchbowl fault zone,  California},
  volume = {124},
  ISSN = {1420-9136},
  url = {http://dx.doi.org/10.1007/BF00875720},
  DOI = {10.1007/bf00875720},
  number = {1–2},
  journal = {Pure and Applied Geophysics},
  publisher = {Springer Science and Business Media LLC},
  author = {Chester,  F. M. and Logan,  J. M.},
  year = {1986},
  month = jan,
  pages = {79–106}
}

@article{Dapp2012PRL,
  doi = {10.1103/physrevlett.108.244301},
  url = {http://dx.doi.org/10.1103/PhysRevLett.108.244301},
  year  = {2012},
  month = {jun},
  publisher = {American Physical Society ({APS})},
  volume = {108},
  number = {24},
  author = {Wolf B. Dapp and Andreas L\"{u}cke and Bo N. J. Persson and Martin H. M\"{u}ser},
  title = {Self-Affine Elastic Contacts: Percolation and Leakage},
  journal = {Phys. Rev. Lett.},
  pages = {244301}
}

@article{Dapp2014JPCM,
  title = {Systematic analysis of Persson’s contact mechanics theory of randomly rough elastic surfaces},
  volume = {26},
  ISSN = {1361-648X},
  url = {http://dx.doi.org/10.1088/0953-8984/26/35/355002},
  DOI = {10.1088/0953-8984/26/35/355002},
  number = {35},
  journal = {Journal of Physics: Condensed Matter},
  publisher = {IOP Publishing},
  author = {Dapp,  Wolf B and Prodanov,  Nikolay and M\"{u}ser,  Martin H},
  year = {2014},
  month = jul,
  pages = {355002}
}

@article{Dieterich1979JGRSE,
  title = {Modeling of rock friction: 1. Experimental results and constitutive equations},
  volume = {84},
  ISSN = {0148-0227},
  url = {http://dx.doi.org/10.1029/JB084iB05p02161},
  DOI = {10.1029/jb084ib05p02161},
  number = {B5},
  journal = {Journal of Geophysical Research: Solid Earth},
  publisher = {American Geophysical Union (AGU)},
  author = {Dieterich,  James H.},
  year = {1979},
  month = may,
  pages = {2161–2168}
}

@article{Dieterich1994PAG,
  title = {Direct observation of frictional contacts: New insights for state-dependent properties},
  volume = {143},
  ISSN = {1420-9136},
  url = {http://dx.doi.org/10.1007/BF00874332},
  DOI = {10.1007/bf00874332},
  number = {1–3},
  journal = {Pure and Applied Geophysics},
  publisher = {Springer Science and Business Media LLC},
  author = {Dieterich,  James H. and Kilgore,  Brian D.},
  year = {1994},
  month = mar,
  pages = {283–302}
}

@article{DiToro2011N,
  title = {Fault lubrication during earthquakes},
  volume = {471},
  ISSN = {1476-4687},
  url = {http://dx.doi.org/10.1038/nature09838},
  DOI = {10.1038/nature09838},
  number = {7339},
  journal = {Nature},
  publisher = {Springer Science and Business Media LLC},
  author = {Di Toro,  G. and Han,  R. and Hirose,  T. and De Paola,  N. and Nielsen,  S. and Mizoguchi,  K. and Ferri,  F. and Cocco,  M. and Shimamoto,  T.},
  year = {2011},
  month = mar,
  pages = {494–498}
}

@article{Doremus2002JAP,
  title = {Viscosity of silica},
  volume = {92},
  ISSN = {1089-7550},
  url = {http://dx.doi.org/10.1063/1.1515132},
  DOI = {10.1063/1.1515132},
  number = {12},
  journal = {Journal of Applied Physics},
  publisher = {AIP Publishing},
  author = {Doremus,  Robert H.},
  year = {2002},
  month = dec,
  pages = {7619–7629}
}

@article{Falk1998PRE,
  title = {Dynamics of viscoplastic deformation in amorphous solids},
  volume = {57},
  ISSN = {1095-3787},
  url = {http://dx.doi.org/10.1103/PhysRevE.57.7192},
  DOI = {10.1103/physreve.57.7192},
  number = {6},
  journal = {Physical Review E},
  publisher = {American Physical Society (APS)},
  author = {Falk,  M. L. and Langer,  J. S.},
  year = {1998},
  month = jun,
  pages = {7192–7205}
}

@article{Faulkner2010JSG,
  title = {A review of recent developments concerning the structure,  mechanics and fluid flow properties of fault zones},
  volume = {32},
  ISSN = {0191-8141},
  url = {http://dx.doi.org/10.1016/j.jsg.2010.06.009},
  DOI = {10.1016/j.jsg.2010.06.009},
  number = {11},
  journal = {Journal of Structural Geology},
  publisher = {Elsevier BV},
  author = {Faulkner,  D.R. and Jackson,  C.A.L. and Lunn,  R.J. and Schlische,  R.W. and Shipton,  Z.K. and Wibberley,  C.A.J. and Withjack,  M.O.},
  year = {2010},
  month = nov,
  pages = {1557–1575}
}

@article{Frerot2018JMPS,
  doi = {10.1016/j.jmps.2018.02.015},
  url = {https://doi.org/10.1016/j.jmps.2018.02.015},
  year  = {2018},
  month = {may},
  publisher = {Elsevier {BV}},
  volume = {114},
  pages = {172--184},
  author = {Lucas Fr{\'{e}}rot and Ramin Aghababaei and Jean-Fran{\c{c}}ois Molinari},
  title = {A mechanistic understanding of the wear coefficient: From single to multiple asperities contact},
  journal = {Journal of the Mechanics and Physics of Solids}
}

@article{Fondriest2013G,
  title = {Mirror-like faults and power dissipation during earthquakes},
  volume = {41},
  ISSN = {1943-2682},
  url = {http://dx.doi.org/10.1130/G34641.1},
  DOI = {10.1130/g34641.1},
  number = {11},
  journal = {Geology},
  publisher = {Geological Society of America},
  author = {Fondriest,  M. and Smith,  S. A. F. and Candela,  T. and Nielsen,  S. B. and Mair,  K. and Di Toro,  G.},
  year = {2013},
  month = sep,
  pages = {1175–1178}
}

@article{Gnecco2000PRL,
  title = {Velocity Dependence of Atomic Friction},
  volume = {84},
  ISSN = {1079-7114},
  url = {http://dx.doi.org/10.1103/PhysRevLett.84.1172},
  DOI = {10.1103/physrevlett.84.1172},
  number = {6},
  journal = {Physical Review Letters},
  publisher = {American Physical Society (APS)},
  author = {Gnecco,  E. and Bennewitz,  R. and Gyalog,  T. and Loppacher,  Ch. and Bammerlin,  M. and Meyer,  E. and G\"{u}ntherodt,  H.-J.},
  year = {2000},
  month = feb,
  pages = {1172–1175}
}

@article{Goldsby2002GRL,
  title = {Low frictional strength of quartz rocks at subseismic slip rates},
  volume = {29},
  ISSN = {1944-8007},
  url = {http://dx.doi.org/10.1029/2002GL015240},
  DOI = {10.1029/2002gl015240},
  number = {17},
  journal = {Geophysical Research Letters},
  publisher = {American Geophysical Union (AGU)},
  author = {Goldsby,  David L. and Tullis,  Terry E.},
  year = {2002},
  month = sep,
  pages={25--1-25--4}
}

@article{Graetsch1998AM,
  title = {Characterization of the high-temperature modifications of incommensurate tridymite L3-To(MX-1) from 25 to 250 degrees C},
  volume = {83},
  ISSN = {0003-004X},
  url = {http://dx.doi.org/10.2138/am-1998-7-819},
  DOI = {10.2138/am-1998-7-819},
  number = {7–8},
  journal = {American Mineralogist},
  publisher = {Mineralogical Society of America},
  author = {Graetsch,  Heribert},
  year = {1998},
  month = aug,
  pages = {872–880}
}

@article{Greenwood1991W,
  title = {An interpolation formula for flash temperatures},
  volume = {150},
  ISSN = {0043-1648},
  url = {http://dx.doi.org/10.1016/0043-1648(91)90312-I},
  DOI = {10.1016/0043-1648(91)90312-i},
  number = {1–2},
  journal = {Wear},
  publisher = {Elsevier BV},
  author = {Greenwood,  J.A.},
  year = {1991},
  month = oct,
  pages = {153–158}
}

@article{GronbechJensen2019MP,
  title = {Complete set of stochastic {V}erlet-type thermostats for correct {L}angevin simulations},
  volume = {118},
  ISSN = {1362-3028},
  url = {http://dx.doi.org/10.1080/00268976.2019.1662506},
  DOI = {10.1080/00268976.2019.1662506},
  number = {8},
  journal = {Molecular Physics},
  publisher = {Informa UK Limited},
  author = {Grønbech-Jensen,  Niels},
  year = {2019},
  month = sep,
  pages = {e1662506}
}

@article{Hayward2019JGR,
  title = {Rheological Controls on Asperity Weakening During Earthquake Slip},
  volume = {124},
  ISSN = {2169-9356},
  url = {http://dx.doi.org/10.1029/2019JB018231},
  DOI = {10.1029/2019jb018231},
  number = {12},
  journal = {Journal of Geophysical Research: Solid Earth},
  publisher = {American Geophysical Union (AGU)},
  author = {Hayward,  Kathryn S. and Hawkins,  Rhys and Cox,  Stephen F. and Le Losq,  Charles},
  year = {2019},
  month = dec,
  pages = {12736–12762}
}

@article{He1999S,
  title = {Adsorbed Layers and the Origin of Static Friction},
  volume = {284},
  ISSN = {1095-9203},
  url = {http://dx.doi.org/10.1126/science.284.5420.1650},
  DOI = {10.1126/science.284.5420.1650},
  number = {5420},
  journal = {Science},
  publisher = {American Association for the Advancement of Science (AAAS)},
  author = {He,  Gang and M{\"u}ser,  Martin H. and Robbins,  Mark O.},
  year = {1999},
  month = jun,
  pages = {1650–1652}
}

@article{Herzbach2005JCP,
  title = {Comparison of model potentials for molecular-dynamics simulations of silica},
  volume = {123},
  ISSN = {1089-7690},
  url = {http://dx.doi.org/10.1063/1.2038747},
  DOI = {10.1063/1.2038747},
  number = {12},
  journal = {The Journal of Chemical Physics},
  publisher = {AIP Publishing},
  author = {Herzbach,  Daniel and Binder,  Kurt and M\"{u}ser,  Martin H.},
  year = {2005},
  month = sep,
  pages = {124711}
}

@article{Horbach1999PRB,
  title = {Static and dynamic properties of a viscous silica melt},
  volume = {60},
  ISSN = {1095-3795},
  url = {http://dx.doi.org/10.1103/PhysRevB.60.3169},
  DOI = {10.1103/physrevb.60.3169},
  number = {5},
  journal = {Physical Review B},
  publisher = {American Physical Society (APS)},
  author = {Horbach,  J\"{u}rgen and Kob,  Walter},
  year = {1999},
  month = aug,
  pages = {3169–3181}
}

@article{Horbach2001EPJB,
  title = {High frequency sound and the boson peak in amorphous silica},
  volume = {19},
  ISSN = {1434-6028},
  url = {http://dx.doi.org/10.1007/s100510170299},
  DOI = {10.1007/s100510170299},
  number = {4},
  journal = {The European Physical Journal B},
  publisher = {Springer Science and Business Media LLC},
  author = {Horbach,  J. and Kob,  W. and Binder,  K.},
  year = {2001},
  month = feb,
  pages = {531–543}
}

@article{Ishibashi2018WRR,
  title = {Friction‐Stability‐Permeability Evolution of a Fracture in Granite},
  volume = {54},
  ISSN = {1944-7973},
  url = {http://dx.doi.org/10.1029/2018WR022598},
  DOI = {10.1029/2018wr022598},
  number = {12},
  journal = {Water Resources Research},
  publisher = {American Geophysical Union (AGU)},
  author = {Ishibashi,  Takuya and Elsworth,  Derek and Fang,  Yi and Riviere,  Jacques and Madara,  Benjamin and Asanuma,  Hiroshi and Watanabe,  Noriaki and Marone,  Chris},
  year = {2018},
  month = dec,
  pages = {9901–9918}
}

@article{Jackson2010AM,
  title = {Monoclinic tridymite in clast-rich impact melt rock from the Chesapeake Bay impact structure},
  volume = {96},
  ISSN = {0003-004X},
  url = {http://dx.doi.org/10.2138/am.2011.3589},
  DOI = {10.2138/am.2011.3589},
  number = {1},
  journal = {American Mineralogist},
  publisher = {Mineralogical Society of America},
  author = {Jackson,  J. C. and Horton,  J. W. and Chou,  I.-M. and Belkin,  H. E.},
  year = {2010},
  month = dec,
  pages = {81–88}
}

@article{Jacobs2017STMP,
  MYDOI = {10.1088/2051-672x/aa51f8},
  MYURL = {https://MYDOI.org/10.1088/2051-672x/aa51f8},
  year = {2017},
  month = {jan},
  publisher = {{IOP} Publishing},
  volume = {5},
  number = {1},
  pages = {013001},
  author = {Tevis D B Jacobs and Till Junge and Lars Pastewka},
  title = {Quantitative characterization of surface topography using spectral analysis},
  journal = {Surface Topography: Metrology and Properties}
}

@article{Ji2022EA,
  title = {Friction law for earthquake nucleation: size doesn’t matter},
  url = {http://dx.doi.org/10.31223/X51H0W},
  DOI = {10.31223/x51h0w},
  journal = {EarthArXiv},
  publisher = {California Digital Library (CDL)},
  author = {Ji,  Yuntao and Niemeijer,  André and Baden,  Dawin and Yamashita,  Futoshi and Xu,  Shiqing and Hunfeld,  Luuk and Pijnenburg,  Ronald and Fukuyama,  Eiichi and Spiers,  Christopher},
  year = {2022},
  month = jun 
}

@article{Kanamori2001PT,
  title = {The Physics of Earthquakes},
  volume = {54},
  ISSN = {1945-0699},
  url = {http://dx.doi.org/10.1063/1.1387590},
  DOI = {10.1063/1.1387590},
  number = {6},
  journal = {Physics Today},
  publisher = {AIP Publishing},
  author = {Kanamori,  Hiroo and Brodsky,  Emily E.},
  year = {2001},
  month = jun,
  pages = {34–40}
}

@article{Kanamori2004RPP,
  title = {The physics of earthquakes},
  volume = {67},
  ISSN = {1361-6633},
  url = {http://dx.doi.org/10.1088/0034-4885/67/8/R03},
  DOI = {10.1088/0034-4885/67/8/r03},
  number = {8},
  journal = {Reports on Progress in Physics},
  publisher = {IOP Publishing},
  author = {Kanamori,  Hiroo and Brodsky,  Emily E},
  year = {2004},
  month = jul,
  pages = {1429–1496}
}

@article{Kihara1977ZKCM,
  title = {An orthrohombic superstructure of tridymite existing between about 105 and 180°C},
  volume = {146},
  ISSN = {2196-7105},
  url = {http://dx.doi.org/10.1524/zkri.1977.146.4-6.185},
  DOI = {10.1524/zkri.1977.146.4-6.185},
  number = {4–6},
  journal = {Zeitschrift f\"{u}r Kristallographie - Crystalline Materials},
  publisher = {Walter de Gruyter GmbH},
  year = {1977},
  month = feb,
  pages = {185–203},
  author = {Kihara, K}
}

@article{Lee2017NG,
  title = {Quasi-equilibrium melting of quartzite upon extreme friction},
  volume = {10},
  ISSN = {1752-0908},
  url = {http://dx.doi.org/10.1038/NGEO2951},
  DOI = {10.1038/ngeo2951},
  number = {6},
  journal = {Nature Geoscience},
  publisher = {Springer Science and Business Media LLC},
  author = {Lee,  Sung Keun and Han,  Raehee and Kim,  Eun Jeong and Jeong,  Gi Young and Khim,  Hoon and Hirose,  Takehiro},
  year = {2017},
  month = may,
  pages = {436–441}
}

@article{Lei2022G,
  title = {Study on the Effect of Nanoindentation Test Method on Micromechanical Properties of Granite Minerals},
  volume = {2022},
  ISSN = {1468-8115},
  url = {http://dx.doi.org/10.1155/2022/5834979},
  DOI = {10.1155/2022/5834979},
  journal = {Geofluids},
  publisher = {Wiley},
  author = {Lei,  Man and Dang,  Fa-Ning and Xue,  Haibin and Xing,  Jin and Ding,  Weihua},
  editor = {Hou,  Peng},
  year = {2022},
  month = may,
  pages = {1–9}
}

@article{Li2011N,
  title = {Frictional ageing from interfacial bonding and the origins of rate and state friction},
  volume = {480},
  ISSN = {1476-4687},
  url = {http://dx.doi.org/10.1038/nature10589},
  DOI = {10.1038/nature10589},
  number = {7376},
  journal = {Nature},
  publisher = {Springer Science and Business Media LLC},
  author = {Li,  Qunyang and Tullis,  Terry E. and Goldsby,  David and Carpick,  Robert W.},
  year = {2011},
  month = nov,
  pages = {233–236}
}

@article{Li2014TL,
  title = {Effects of Interfacial Bonding on Friction and Wear at Silica/Silica Interfaces},
  volume = {56},
  ISSN = {1573-2711},
  url = {http://dx.doi.org/10.1007/s11249-014-0425-x},
  DOI = {10.1007/s11249-014-0425-x},
  number = {3},
  journal = {Tribology Letters},
  publisher = {Springer Science and Business Media LLC},
  author = {Li,  Ao and Liu,  Yun and Szlufarska,  Izabela},
  year = {2014},
  month = oct,
  pages = {481–490}
}

@article{Li2020PRL,
  title = {Length Scale Effect in Frictional Aging of Silica Contacts},
  volume = {125},
  ISSN = {1079-7114},
  url = {http://dx.doi.org/10.1103/PhysRevLett.125.215502},
  DOI = {10.1103/physrevlett.125.215502},
  number = {21},
  journal = {Physical Review Letters},
  publisher = {American Physical Society (APS)},
  author = {Li,  Shen and Zhang,  Shuai and Chen,  Zhe and Feng,  Xi-Qiao and Li,  Qunyang},
  year = {2020},
  month = nov 
}

@article{Li2025GRL,
  title = {Nanoscale Plastic Wear of $\alpha$-Quartz Asperities During Shear Sliding: Insights From Molecular Dynamics Simulations},
  volume = {52},
  ISSN = {1944-8007},
  url = {http://dx.doi.org/10.1029/2025GL116288},
  DOI = {10.1029/2025gl116288},
  number = {20},
  journal = {Geophysical Research Letters},
  publisher = {American Geophysical Union (AGU)},
  author = {Li,  Sheng and Fukuyama,  Eiichi},
  year = {2025},
  month = oct 
}

@article{Liang2025TL,
  title = {A JKR/Griffith Model for the Inception of Slip in the Contact Between Nominally Flat Rough Surfaces},
  volume = {73},
  ISSN = {1573-2711},
  url = {http://dx.doi.org/10.1007/s11249-025-02075-z},
  DOI = {10.1007/s11249-025-02075-z},
  number = {4},
  journal = {Tribology Letters},
  publisher = {Springer Science and Business Media LLC},
  author = {Liang,  X. M. and Ciavarella,  M.},
  year = {2025},
  month = oct 
}

@article{Lorenz2012JPCM,
  title = {On the origin of why static or breakloose friction is larger than kinetic friction,  and how to reduce it: the role of aging,  elasticity and sequential interfacial slip},
  volume = {24},
  ISSN = {1361-648X},
  url = {http://dx.doi.org/10.1088/0953-8984/24/22/225008},
  DOI = {10.1088/0953-8984/24/22/225008},
  number = {22},
  journal = {Journal of Physics: Condensed Matter},
  publisher = {IOP Publishing},
  author = {Lorenz,  B and Persson,  B N J},
  year = {2012},
  month = may,
  pages = {225008}
}

@article{Mair1999JGR,
  title = {Friction of simulated fault gouge for a wide range of velocities and normal stresses},
  volume = {104},
  ISSN = {0148-0227},
  url = {http://dx.doi.org/10.1029/1999JB900279},
  DOI = {10.1029/1999jb900279},
  number = {B12},
  journal = {Journal of Geophysical Research: Solid Earth},
  publisher = {American Geophysical Union (AGU)},
  author = {Mair,  Karen and Marone,  Chris},
  year = {1999},
  month = dec,
  pages = {28899–28914}
}

@article{Majumdar1990Wear,
  MYDOI = {10.1016/0043-1648(90)90154-3},
  MYURL = {https://MYDOI.org/10.1016/0043-1648(90)90154-3},
  year = {1990},
  month = {mar},
  publisher = {Elsevier {BV}},
  volume = {136},
  number = {2},
  pages = {313--327},
  author = {A. Majumdar and C.L. Tien},
  MYtitle = {Fractal characterization and simulation of rough surfaces},
  journal = {Wear}
}

@article{Manzi2021TL,
  title = {Prandtl–{T}omlinson-Type Models for Molecular Sliding Friction},
  volume = {69},
  ISSN = {1573-2711},
  url = {http://dx.doi.org/10.1007/s11249-021-01523-w},
  DOI = {10.1007/s11249-021-01523-w},
  number = {4},
  journal = {Tribology Letters},
  publisher = {Springer Science and Business Media LLC},
  author = {Manzi,  Sergio Javier and Carrera,  Sebastian Eduardo and Furlong,  Octavio Javier and Kenmoe,  Germaine Djuidje and Tysoe,  Wilfred T.},
  year = {2021},
  month = oct 
}

@article{Marone1990JGRSE,
  title = {Frictional behavior and constitutive modeling of simulated fault gouge},
  volume = {95},
  ISSN = {0148-0227},
  url = {http://dx.doi.org/10.1029/JB095iB05p07007},
  DOI = {10.1029/jb095ib05p07007},
  number = {B5},
  journal = {Journal of Geophysical Research: Solid Earth},
  publisher = {American Geophysical Union (AGU)},
  author = {Marone,  Chris and Raleigh,  C. Barry and Scholz,  C. H.},
  year = {1990},
  month = may,
  pages = {7007–7025}
}

@article{Martinelli2020M,
  title = {Fracture Analysis of $\alpha$-Quartz Crystals Subjected to Shear Stress},
  volume = {10},
  ISSN = {2075-163X},
  url = {http://dx.doi.org/10.3390/min10100870},
  DOI = {10.3390/min10100870},
  number = {10},
  journal = {Minerals},
  publisher = {MDPI AG},
  author = {Martinelli,  Giovanni and Plescia,  Paolo and Tempesta,  Emanuela and Paris,  Enrico and Gallucci,  Francesco},
  year = {2020},
  month = sep,
  pages = {870}
}

@article{McClimon2024TL,
  title = {The Effects of Humidity on the Velocity-Dependence and Frictional Ageing of Nanoscale Silica Contacts},
  volume = {72},
  ISSN = {1573-2711},
  url = {http://dx.doi.org/10.1007/s11249-024-01904-x},
  DOI = {10.1007/s11249-024-01904-x},
  number = {4},
  journal = {Tribology Letters},
  publisher = {Springer Science and Business Media LLC},
  author = {McClimon,  J. Brandon and Li,  Zhuohan and Baral,  Khagendra and Goldsby,  David and Szlufarska,  Izabela and Carpick,  Robert W.},
  year = {2024},
  month = aug,
  pages = {105}
}

@article{Mollon2021JGRSE,
  title = {Simulating Melting in 2D Seismic Fault Gouge},
  volume = {126},
  ISSN = {2169-9356},
  url = {http://dx.doi.org/10.1029/2020JB021485},
  DOI = {10.1029/2020jb021485},
  number = {6},
  journal = {Journal of Geophysical Research: Solid Earth},
  publisher = {American Geophysical Union (AGU)},
  author = {Mollon,  Guilhem and Aubry,  Jér\^ome and Schubnel,  Alexandre},
  year = {2021},
  month = jun 
}

@article{Moras2018PRM,
  title = {Shear melting of silicon and diamond and the disappearance of the polyamorphic transition under shear},
  volume = {2},
  ISSN = {2475-9953},
  url = {http://dx.doi.org/10.1103/PhysRevMaterials.2.083601},
  DOI = {10.1103/physrevmaterials.2.083601},
  number = {8},
  journal = {Physical Review Materials},
  publisher = {American Physical Society (APS)},
  author = {Moras,  Gianpietro and Klemenz,  Andreas and Reichenbach,  Thomas and Gola,  Adrien and Uetsuka,  Hiroshi and Moseler,  Michael and Pastewka,  Lars},
  year = {2018},
  month = aug 
}

@article{Muser2001PCM,
  title = {Molecular dynamics study of the $\alpha-\beta$ transition in quartz: elastic properties,  finite size effects,  and hysteresis in the local structure},
  volume = {28},
  ISSN = {1432-2021},
  url = {http://dx.doi.org/10.1007/s002690100203},
  DOI = {10.1007/s002690100203},
  number = {10},
  journal = {Physics and Chemistry of Minerals},
  publisher = {Springer Science and Business Media LLC},
  author = {M\"{u}ser,  M. H. and Binder,  K.},
  year = {2001},
  month = nov,
  pages = {746–755}
}

@article{Muser2001PRL,
  title = {Simple Microscopic Theory of Amontons’s Laws for Static Friction},
  volume = {86},
  ISSN = {1079-7114},
  url = {http://dx.doi.org/10.1103/PhysRevLett.86.1295},
  DOI = {10.1103/physrevlett.86.1295},
  number = {7},
  journal = {Physical Review Letters},
  publisher = {American Physical Society (APS)},
  author = {M\"{u}ser,  Martin H. and Wenning,  Ludgar and Robbins,  Mark O.},
  year = {2001},
  month = feb,
  pages = {1295–1298}
}

@article{Muser2002PRL,
  title = {Nature of Mechanical Instabilities and Their Effect on Kinetic Friction},
  volume = {89},
  ISSN = {1079-7114},
  url = {http://dx.doi.org/10.1103/PhysRevLett.89.224301},
  DOI = {10.1103/physrevlett.89.224301},
  number = {22},
  journal = {Physical Review Letters},
  publisher = {American Physical Society (APS)},
  author = {M\"{u}ser,  Martin H.},
  year = {2002},
  month = nov 
}

@article{Muser2004EPL,
  title = {Structural lubricity: Role of dimension and symmetry},
  volume = {66},
  ISSN = {1286-4854},
  url = {http://dx.doi.org/10.1209/epl/i2003-10139-6},
  DOI = {10.1209/epl/i2003-10139-6},
  number = {1},
  journal = {Europhysics Letters (EPL)},
  publisher = {IOP Publishing},
  author = {M\"{u}ser,  M. H},
  year = {2004},
  month = apr,
  pages = {97–103}
}

@article{Muser2017TL,
  title = {Meeting the Contact-Mechanics Challenge},
  volume = {65},
  ISSN = {1573-2711},
  url = {http://dx.doi.org/10.1007/s11249-017-0900-2},
  DOI = {10.1007/s11249-017-0900-2},
  number = {4},
  journal = {Tribology Letters},
  publisher = {Springer Science and Business Media LLC},
  author = {M\"{u}ser,  Martin H. and Dapp,  Wolf B. and Bugnicourt,  Romain and Sainsot,  Philippe and Lesaffre,  Nicolas and Lubrecht,  Ton A. and Persson,  Bo N. J. and Harris,  Kathryn and Bennett,  Alexander and Schulze,  Kyle and Rohde,  Sean and Ifju,  Peter and Sawyer,  W. Gregory and Angelini,  Thomas and Ashtari Esfahani,  Hossein and Kadkhodaei,  Mahmoud and Akbarzadeh,  Saleh and Wu,  Jiunn-Jong and Vorlaufer,  Georg and Vernes,  András and Solhjoo,  Soheil and Vakis,  Antonis I. and Jackson,  Robert L. and Xu,  Yang and Streator,  Jeffrey and Rostami,  Amir and Dini,  Daniele and Medina,  Simon and Carbone,  Giuseppe and Bottiglione,  Francesco and Afferrante,  Luciano and Monti,  Joseph and Pastewka,  Lars and Robbins,  Mark O. and Greenwood,  James A.},
  year = {2017},
  month = aug 
}

@article{Muser2018L,
  title = {Contact-Patch-Size Distribution and Limits of Self-Affinity in Contacts between Randomly Rough Surfaces},
  volume = {6},
  ISSN = {2075-4442},
  url = {http://dx.doi.org/10.3390/lubricants6040085},
  DOI = {10.3390/lubricants6040085},
  number = {4},
  journal = {Lubricants},
  publisher = {MDPI AG},
  author = {Müser,  Martin H. and Wang,  Anle},
  year = {2018},
  month = sep,
  pages = {85}
}

@article{Muser2019TL,
  title = {Elasticity Does Not Necessarily Break Down in Nanoscale Contacts: Comparing Stresses from Atomistic Simulations to Continuum Theory},
  volume = {67},
  ISSN = {1573-2711},
  url = {http://dx.doi.org/10.1007/s11249-019-1170-y},
  DOI = {10.1007/s11249-019-1170-y},
  number = {2},
  journal = {Tribology Letters},
  publisher = {Springer Science and Business Media LLC},
  author = {M\"{u}ser,  Martin H.},
  year = {2019},
  month = apr 
}

@article{Muser2021TL,
  title = {Elastic Contacts of Randomly Rough Indenters with Thin Sheets,  Membranes Under Tension,  Half Spaces,  and Beyond},
  volume = {69},
  ISSN = {1573-2711},
  url = {http://dx.doi.org/10.1007/s11249-020-01383-w},
  DOI = {10.1007/s11249-020-01383-w},
  number = {1},
  journal = {Tribology Letters},
  publisher = {Springer Science and Business Media LLC},
  author = {M\"{u}ser,  Martin H.},
  year = {2021},
  month = feb 
}

@article{Muser2022EL,
  title = {Crack and pull-off dynamics of adhesive,  viscoelastic solids},
  volume = {137},
  ISSN = {1286-4854},
  url = {http://dx.doi.org/10.1209/0295-5075/ac535c},
  DOI = {10.1209/0295-5075/ac535c},
  number = {3},
  journal = {Europhysics Letters},
  publisher = {IOP Publishing},
  author = {M\"{u}ser,  Martin H. and Persson,  Bo N. J.},
  year = {2022},
  month = feb,
  pages = {36004}
}

@article{Noaki2023NCM,
  title = {Development of the reactive force field and silicon dry/wet oxidation process modeling},
  volume = {9},
  ISSN = {2057-3960},
  url = {http://dx.doi.org/10.1038/s41524-023-01112-6},
  DOI = {10.1038/s41524-023-01112-6},
  number = {1},
  journal = {npj Computational Materials},
  publisher = {Springer Science and Business Media LLC},
  author = {Noaki,  Junichi and Numazawa,  Satoshi and Jeon,  Joohyun and Kochi,  Shuntaro},
  year = {2023},
  month = sep,
  pages = {161}
}

@article{Pastewka2010NM,
  title = {Anisotropic mechanical amorphization drives wear in diamond},
  volume = {10},
  ISSN = {1476-4660},
  url = {http://dx.doi.org/10.1038/nmat2902},
  DOI = {10.1038/nmat2902},
  number = {1},
  journal = {Nature Materials},
  publisher = {Springer Science and Business Media LLC},
  author = {Pastewka,  Lars and Moser,  Stefan and Gumbsch,  Peter and Moseler,  Michael},
  year = {2010},
  month = nov,
  pages = {34–38}
}

@article{Peng2025PRL,
  title = {Decrease of Static Friction Coefficient with Interface Growth from Single to Multiasperity Contact},
  volume = {134},
  ISSN = {1079-7114},
  url = {http://dx.doi.org/10.1103/PhysRevLett.134.176202},
  DOI = {10.1103/physrevlett.134.176202},
  number = {17},
  journal = {Physical Review Letters},
  publisher = {American Physical Society (APS)},
  author = {Peng,  Liang and Roch,  Thibault and Bonn,  Daniel and Weber,  Bart},
  year = {2025},
  month = apr 
}

@article{Persson2001JCP,
 author = {Persson, B N J},
 doi = {10.1063/1.1388626},
 journal = {J. Chem. Phys.},
 number = {8},
 pages = {3840--3861},
 publisher = {AIP Publishing},
 title = {Theory of rubber friction and contact mechanics},
 volume = {115},
 year = {2001}
}

@article{Persson2001PRL,
  title = {Elastoplastic Contact between Randomly Rough Surfaces},
  volume = {87},
  ISSN = {1079-7114},
  url = {http://dx.doi.org/10.1103/PhysRevLett.87.116101},
  DOI = {10.1103/physrevlett.87.116101},
  number = {11},
  journal = {Physical Review Letters},
  publisher = {American Physical Society (APS)},
  author = {Persson,  B. N. J.},
  year = {2001},
  month = aug 
}

@article{Persson2003W,
  title = {Nanoadhesion},
  volume = {254},
  ISSN = {0043-1648},
  url = {http://dx.doi.org/10.1016/S0043-1648(03)00233-3},
  DOI = {10.1016/s0043-1648(03)00233-3},
  number = {9},
  journal = {Wear},
  publisher = {Elsevier BV},
  author = {Persson,  B.N.J},
  year = {2003},
  month = may,
  pages = {832–834}
}

@article{Persson2008JPCMb,
 doi = {10.1088/0953-8984/20/31/312001},
  url = {http://dx.doi.org/10.1088/0953-8984/20/31/312001},
  year  = {2008},
  month = {jun},
  publisher = {{IOP} Publishing},
  volume = {20},
  number = {31},
  pages = {312001},
  author = {B. N. J. Persson},
  title = {On the elastic energy and stress correlation in the contact between elastic solids with randomly rough surfaces},
  journal = {Journal of Physics: Condensed Matter}
}

@article{Persson2011JPCM,
  title = {Phononic heat transfer across an interface: thermal boundary resistance},
  volume = {23},
  ISSN = {1361-648X},
  url = {http://dx.doi.org/10.1088/0953-8984/23/4/045009},
  DOI = {10.1088/0953-8984/23/4/045009},
  number = {4},
  journal = {Journal of Physics: Condensed Matter},
  publisher = {IOP Publishing},
  author = {Persson,  B N J and Volokitin,  A I and Ueba,  H},
  year = {2011},
  month = jan,
  pages = {045009}
}

@article{Persson2014TL,
  MYtitle={On the fractal dimension of rough surfaces},
  author={Persson, BNJ},
  journal={Tribology Letters},
  volume={54},
  number={1},
  pages={99--106},
  year={2014},
  publisher={Springer}
}

@article{Persson2015JCP,
  title = {Ice friction: Role of non-uniform frictional heating and ice premelting},
  volume = {143},
  ISSN = {1089-7690},
  url = {http://dx.doi.org/10.1063/1.4936299},
  DOI = {10.1063/1.4936299},
  number = {22},
  journal = {The Journal of Chemical Physics},
  publisher = {AIP Publishing},
  author = {Persson,  B. N. J.},
  year = {2015},
  month = dec,
  pages = {224701}
}

@article{Persson2022MRS,
  title = {Functional properties of rough surfaces from an analytical theory of mechanical contact},
  volume = {47},
  ISSN = {1938-1425},
  url = {http://dx.doi.org/10.1557/s43577-022-00472-6},
  DOI = {10.1557/s43577-022-00472-6},
  number = {12},
  journal = {MRS Bulletin},
  publisher = {Springer Science and Business Media LLC},
  author = {Persson,  B. N. J.},
  year = {2022},
  month = dec,
  pages = {1211–1219}
}

@article{Persson2025JCP_a,
  title = {Rubber friction: Theory,  mechanisms,  and challenges},
  volume = {163},
  ISSN = {1089-7690},
  url = {http://dx.doi.org/10.1063/5.0293616},
  DOI = {10.1063/5.0293616},
  number = {14},
  journal = {The Journal of Chemical Physics},
  publisher = {AIP Publishing},
  author = {Persson,  B. N. J. and Xu,  R.},
  year = {2025},
  month = oct,
  pages = {141001}
}

@article{Persson2025JCP_b,
  title = {Rubber wear: Experiment and theory},
  volume = {162},
  ISSN = {1089-7690},
  url = {http://dx.doi.org/10.1063/5.0248199},
  DOI = {10.1063/5.0248199},
  number = {7},
  journal = {The Journal of Chemical Physics},
  publisher = {AIP Publishing},
  author = {Persson,  B. N. J. and Xu,  R. and Miyashita,  N.},
  year = {2025},
  month = feb,
  pages = {074704}
}

@article{Popov2012ZAMM,
  doi = {10.1002/zamm.201200097},
  url = {https://doi.org/10.1002/zamm.201200097},
  year = {2012},
  month = jul,
  publisher = {Wiley},
  volume = {92},
  number = {9},
  pages = {683--708},
  author = {V.L. Popov and J.A.T. Gray},
  title = {{Prandtl-Tomlinson model: History and applications in friction,  plasticity,  and nanotechnologies}},
  journal = {{ZAMM} - Journal of Applied Mathematics and Mechanics / Zeitschrift f\"{u}r Angewandte Mathematik und Mechanik}
}

@article{Power1987GRL,
  title = {Roughness of natural fault surfaces},
  volume = {14},
  ISSN = {1944-8007},
  url = {http://dx.doi.org/10.1029/GL014i001p00029},
  DOI = {10.1029/gl014i001p00029},
  number = {1},
  journal = {Geophysical Research Letters},
  publisher = {American Geophysical Union (AGU)},
  author = {Power,  W. L. and Tullis,  T. E. and Brown,  S. R. and Boitnott,  G. N. and Scholz,  C. H.},
  year = {1987},
  month = jan,
  pages = {29–32}
}

@article{Prandtl1928ZAMM,
author = {Prandtl, L.},
title = {{Ein Gedankenmodell zur kinetischen Theorie der festen Körper}},
journal = {Zeitschrift für Angewandte Mathematik und Mechanik},
volume = {8},
number = {2},
pages = {85-106},
doi = {10.1002/zamm.19280080202},
url = {https://onlinelibrary.wiley.com/doi/abs/10.1002/zamm.19280080202},
year = {1928}
}

@book{Rabinowicz1995Book,
  author = {Rabinowicz, Ernest},
  title = {Friction and Wear of Materials},
  edition = {2nd},
  year = {1995},
  publisher = {Wiley-Interscience, John Wiley \& Sons, Inc.},
  address = {New York, NY},
  isbn = {978-0471830849}
}

@article{Rathbun2013JGRSE,
  title = {Symmetry and the critical slip distance in rate and state friction laws},
  volume = {118},
  ISSN = {2169-9356},
  url = {http://dx.doi.org/10.1002/jgrb.50224},
  DOI = {10.1002/jgrb.50224},
  number = {7},
  journal = {Journal of Geophysical Research: Solid Earth},
  publisher = {American Geophysical Union (AGU)},
  author = {Rathbun,  Andrew P. and Marone,  Chris},
  year = {2013},
  month = jul,
  pages = {3728–3741}
}

@article{Rice2006JGR,
  title = {Heating and weakening of faults during earthquake slip},
  volume = {111},
  ISSN = {0148-0227},
  url = {http://dx.doi.org/10.1029/2005JB004006},
  DOI = {10.1029/2005jb004006},
  number = {B5},
  journal = {Journal of Geophysical Research: Solid Earth},
  publisher = {American Geophysical Union (AGU)},
  author = {Rice,  James R.},
  year = {2006},
  month = may 
}

@article{Rubin2005JGRSE,
  title = {Earthquake nucleation on (aging) rate and state faults},
  volume = {110},
  ISSN = {0148-0227},
  url = {http://dx.doi.org/10.1029/2005JB003686},
  DOI = {10.1029/2005jb003686},
  number = {B11},
  journal = {Journal of Geophysical Research: Solid Earth},
  publisher = {American Geophysical Union (AGU)},
  author = {Rubin,  A. M. and Ampuero,  J.‐P.},
  year = {2005},
  month = nov 
}

@article{Rubinstein2007PRL,
  title = {Dynamics of Precursors to Frictional Sliding},
  volume = {98},
  ISSN = {1079-7114},
  url = {http://dx.doi.org/10.1103/PhysRevLett.98.226103},
  DOI = {10.1103/physrevlett.98.226103},
  number = {22},
  journal = {Physical Review Letters},
  publisher = {American Physical Society (APS)},
  author = {Rubinstein,  S. M. and Cohen,  G. and Fineberg,  J.},
  year = {2007},
  month = jun 
}

@article{Ruina1983JGRSE,
  title = {Slip instability and state variable friction laws},
  volume = {88},
  ISSN = {0148-0227},
  url = {http://dx.doi.org/10.1029/JB088iB12p10359},
  DOI = {10.1029/jb088ib12p10359},
  number = {B12},
  journal = {Journal of Geophysical Research: Solid Earth},
  publisher = {American Geophysical Union (AGU)},
  author = {Ruina,  Andy},
  year = {1983},
  month = dec,
  pages = {10359–10370}
}

@article{Scholz1998N,
  title = {Earthquakes and friction laws},
  volume = {391},
  ISSN = {1476-4687},
  url = {http://dx.doi.org/10.1038/34097},
  DOI = {10.1038/34097},
  number = {6662},
  journal = {Nature},
  publisher = {Springer Science and Business Media LLC},
  author = {Scholz,  Christopher H.},
  year = {1998},
  month = jan,
  pages = {37–42}
}

@book{Scholz2018Book,
  title = {The Mechanics of Earthquakes and Faulting},
  ISBN = {9781316615232},
  url = {http://dx.doi.org/10.1017/9781316681473},
  DOI = {10.1017/9781316681473},
  publisher = {Cambridge University Press},
  author = {Scholz,  Christopher H.},
  year = {2018},
  month = dec 
}

@article{Sharma1996PMS, 
    title={Pressure induced amorphization of materials}, 
    volume={40}, ISSN={0079-6425}, 
    url={http://dx.doi.org/10.1016/0079-6425(95)00006-2}, 
    DOI={10.1016/0079-6425(95)00006-2}, 
    number={1}, 
    journal={Progress in Materials Science}, 
    publisher={Elsevier BV}, 
    author={Sharma, Surinder M. and Sikka, S.K.},
    year={1996}, 
    month=jan, 
    pages={1–77} 
}

@article{Strozewski2021JGR,
  title = {Viscoplastic Rheology of $\alpha$‐Quartz Investigated by Nanoindentation},
  volume = {126},
  ISSN = {2169-9356},
  url = {http://dx.doi.org/10.1029/2021JB022229},
  DOI = {10.1029/2021jb022229},
  number = {9},
  journal = {Journal of Geophysical Research: Solid Earth},
  publisher = {American Geophysical Union (AGU)},
  author = {Strozewski,  Benjamin and Sly,  Michael K. and Flores,  Katharine M. and Skemer,  Philip},
  year = {2021},
  month = sep 
}

@article{Stukowski2009MSMSE,
  title = {Visualization and analysis of atomistic simulation data with OVITO–the Open Visualization Tool},
  volume = {18},
  ISSN = {1361-651X},
  url = {http://dx.doi.org/10.1088/0965-0393/18/1/015012},
  DOI = {10.1088/0965-0393/18/1/015012},
  number = {1},
  journal = {Modelling and Simulation in Materials Science and Engineering},
  publisher = {IOP Publishing},
  author = {Stukowski,  Alexander},
  year = {2009},
  month = dec,
  pages = {015012}
}

@article{Tada2025JCP,
  title = {On the origin of the breakloose friction force},
  volume = {162},
  ISSN = {1089-7690},
  url = {http://dx.doi.org/10.1063/5.0266065},
  DOI = {10.1063/5.0266065},
  number = {17},
  journal = {The Journal of Chemical Physics},
  publisher = {AIP Publishing},
  author = {Tada,  T. and Persson,  B. N. J.},
  year = {2025},
  month = may,
  pages = {174709}
}

@article{Thompson2022CPC,
  title = {LAMMPS - a flexible simulation tool for particle-based materials modeling at the atomic,  meso,  and continuum scales},
  volume = {271},
  ISSN = {0010-4655},
  url = {http://dx.doi.org/10.1016/j.cpc.2021.108171},
  DOI = {10.1016/j.cpc.2021.108171},
  journal = {Computer Physics Communications},
  publisher = {Elsevier BV},
  author = {Thompson,  Aidan P. and Aktulga,  H. Metin and Berger,  Richard and Bolintineanu,  Dan S. and Brown,  W. Michael and Crozier,  Paul S. and in ’t Veld,  Pieter J. and Kohlmeyer,  Axel and Moore,  Stan G. and Nguyen,  Trung Dac and Shan,  Ray and Stevens,  Mark J. and Tranchida,  Julien and Trott,  Christian and Plimpton,  Steven J.},
  year = {2022},
  month = feb,
  pages = {108171}
}

@article{Tian2017PRL,
  title = {Load and Time Dependence of Interfacial Chemical Bond-Induced Friction at the Nanoscale},
  volume = {118},
  ISSN = {1079-7114},
  url = {http://dx.doi.org/10.1103/PhysRevLett.118.076103},
  DOI = {10.1103/physrevlett.118.076103},
  number = {7},
  journal = {Physical Review Letters},
  publisher = {American Physical Society (APS)},
  author = {Tian,  Kaiwen and Gosvami,  Nitya N. and Goldsby,  David L. and Liu,  Yun and Szlufarska,  Izabela and Carpick,  Robert W.},
  year = {2017},
  month = feb 
}

@article{Togo2024STAMM,
  title = {Spglib: a software library for crystal symmetry search},
  volume = {4},
  ISSN = {2766-0400},
  url = {http://dx.doi.org/10.1080/27660400.2024.2384822},
  DOI = {10.1080/27660400.2024.2384822},
  number = {1},
  journal = {Science and Technology of Advanced Materials: Methods},
  publisher = {Informa UK Limited},
  author = {Togo,  Atsushi and Shinohara,  Kohei and Tanaka,  Isao},
  year = {2024},
  month = oct 
}

@article{Tuononen2016SR,
  title = {Onset of frictional sliding of rubber–glass contact under dry and lubricated conditions},
  volume = {6},
  ISSN = {2045-2322},
  url = {http://dx.doi.org/10.1038/srep27951},
  DOI = {10.1038/srep27951},
  number = {1},
  journal = {Scientific Reports},
  publisher = {Springer Science and Business Media LLC},
  author = {Tuononen,  Ari J.},
  year = {2016},
  month = jun 
}

@article{Tse1991PRL,
  title = {Mechanical instability of $\alpha$-quartz: A molecular dynamics study},
  volume = {67},
  ISSN = {0031-9007},
  url = {http://dx.doi.org/10.1103/PhysRevLett.67.3559},
  DOI = {10.1103/physrevlett.67.3559},
  number = {25},
  journal = {Physical Review Letters},
  publisher = {American Physical Society (APS)},
  author = {Tse,  John S. and Klug,  Dennis D.},
  year = {1991},
  month = dec,
  pages = {3559–3562}
}

@article{vanBeest1990PRL,
  title = {Force fields for silicas and aluminophosphates based onab initiocalculations},
  volume = {64},
  ISSN = {0031-9007},
  url = {http://dx.doi.org/10.1103/PhysRevLett.64.1955},
  DOI = {10.1103/physrevlett.64.1955},
  number = {16},
  journal = {Physical Review Letters},
  publisher = {American Physical Society (APS)},
  author = {van Beest,  B. W. H. and Kramer,  G. J. and van Santen,  R. A.},
  year = {1990},
  month = apr,
  pages = {1955–1958}
}

@article{vanDuin2003JPC,
  title = {ReaxFF$_{\rm SiO}$ Reactive Force Field for Silicon and Silicon Oxide Systems},
  volume = {107},
  ISSN = {1520-5215},
  url = {http://dx.doi.org/10.1021/jp0276303},
  DOI = {10.1021/jp0276303},
  number = {19},
  journal = {The Journal of Physical Chemistry A},
  publisher = {American Chemical Society (ACS)},
  author = {van Duin,  Adri C. T. and Strachan,  Alejandro and Stewman,  Shannon and Zhang,  Qingsong and Xu,  Xin and Goddard,  William A.},
  year = {2003},
  month = apr,
  pages = {3803–3811}
}

@article{Vollmayr1996PRB,
  title = {Cooling-rate effects in amorphous silica: A computer-simulation study},
  volume = {54},
  ISSN = {1095-3795},
  url = {http://dx.doi.org/10.1103/PhysRevB.54.15808},
  DOI = {10.1103/physrevb.54.15808},
  number = {22},
  journal = {Physical Review B},
  publisher = {American Physical Society (APS)},
  author = {Vollmayr,  Katharina and Kob,  Walter and Binder,  Kurt},
  year = {1996},
  month = dec,
  pages = {15808–15827}
}

@article{Wang2018M,
  title = {Structural and Electronic Properties of Different Terminations for Quartz (001) Surfaces as Well as Water Molecule Adsorption on It: A First-Principles Study},
  volume = {8},
  ISSN = {2075-163X},
  url = {http://dx.doi.org/10.3390/min8020058},
  DOI = {10.3390/min8020058},
  number = {2},
  journal = {Minerals},
  publisher = {MDPI AG},
  author = {Wang,  Xianchen and Zhang,  Qin and Li,  Xianbo and Ye,  Junjian and Li,  Longjiang},
  year = {2018},
  month = feb,
  pages = {58}
}

@article{Whitney2007AM,
  title = {Hardness,  toughness,  and modulus of some common metamorphic minerals},
  volume = {92},
  ISSN = {0003-004X},
  url = {http://dx.doi.org/10.2138/am.2007.2212},
  DOI = {10.2138/am.2007.2212},
  number = {2–3},
  journal = {American Mineralogist},
  publisher = {Mineralogical Society of America},
  author = {Whitney,  D. L. and Broz,  M. and Cook,  R. F.},
  year = {2007},
  month = feb,
  pages = {281–288}
}

@article{Woo2023SR,
  title = {Frictional melting mechanisms of rocks during earthquake fault slip},
  volume = {13},
  ISSN = {2045-2322},
  url = {http://dx.doi.org/10.1038/s41598-023-39752-9},
  DOI = {10.1038/s41598-023-39752-9},
  number = {1},
  journal = {Scientific Reports},
  publisher = {Springer Science and Business Media LLC},
  author = {Woo,  Sangwoo and Han,  Raehee and Oohashi,  Kiyokazu},
  year = {2023},
  month = aug 
}

@article{Xu2024IJSS,
  title = {Persson’s theory of purely normal elastic rough surface contact: A tutorial based on stochastic process theory},
  volume = {290},
  ISSN = {0020-7683},
  url = {http://dx.doi.org/10.1016/j.ijsolstr.2024.112684},
  DOI = {10.1016/j.ijsolstr.2024.112684},
  journal = {International Journal of Solids and Structures},
  publisher = {Elsevier BV},
  author = {Xu,  Yang and Li,  Xiaobao and Chen,  Qi and Zhou,  Yunong},
  year = {2024},
  month = mar,
  pages = {112684}
}

@article{Xu2025TL,
  title = {Sliding Wear: Role of Plasticity},
  volume = {73},
  ISSN = {1573-2711},
  url = {http://dx.doi.org/10.1007/s11249-025-02044-6},
  DOI = {10.1007/s11249-025-02044-6},
  number = {3},
  journal = {Tribology Letters},
  publisher = {Springer Science and Business Media LLC},
  author = {Xu,  R. and Persson,  B. N. J.},
  year = {2025},
  month = jul,
  pages = {109}
}

@article{Yang2008PNAS,
  title = {Dynamics of static friction between steel and silicon},
  volume = {105},
  ISSN = {1091-6490},
  url = {http://dx.doi.org/10.1073/pnas.0806174105},
  DOI = {10.1073/pnas.0806174105},
  number = {36},
  journal = {Proceedings of the National Academy of Sciences},
  publisher = {Proceedings of the National Academy of Sciences},
  author = {Yang,  Zhiping and Zhang,  H. P. and Marder,  M.},
  year = {2008},
  month = sep,
  pages = {13264–13268}
}

\end{document}